\documentclass[apj,twocolumn,twocolappendix,numberedappendix]{openjournal}
\usepackage{apjfonts,amsfonts,amsmath,amssymb,bm,enumitem,multirow,ctable,verbatim}
\usepackage[T1]{fontenc}
\usepackage[breaklinks,colorlinks,citecolor=blue,urlcolor=blue]{hyperref}
\usepackage{pifont}
\newcommand{\xmark}{\ding{55}}%

\newcommand{\Alf}{{Alfv\'en}}
\newcommand{\bhat}{{\bf b}}

\newcommand{\orcidauthor}[3]{\author{\href{http://orcid.org/#1}{#2$^{#3}$}}}

\shorttitle{CRs on Galaxy Scales}
\shortauthors{Hopkins}

\begin{document}
\title{\vspace{-0.8cm}Cosmic Rays on Galaxy Scales: Progress and Pitfalls for CR-MHD Dynamical Models\vspace{-1.5cm}}
\orcidauthor{0000-0003-3729-1684}{Philip F. Hopkins}{1*}
\affiliation{$^{1}$TAPIR, Mailcode 350-17, California Institute of Technology, Pasadena, CA 91125, USA}
\thanks{$^*$E-mail: \href{mailto:phopkins@caltech.edu}{phopkins@caltech.edu}},

\begin{abstract}
Recent years have seen many arguments for cosmic rays (CRs) as an important influence on galactic and circum-galactic medium (CGM) physics, star and galaxy formation. We present a pedagogical overview of state-of-the-art modeling of CR-magnetohydrodynamics (CR-MHD) on macro scales ($\gtrsim$\,kpc), highlighting their fundamental dependence on the micro ($\lesssim$\,au) scales of CR gyro orbits and meso ($\sim$\,pc) scales of CR mean-free-paths, intended to connect the extragalactic, Galactic, and plasma physics CR transport modeling communities. 
We note the pitfalls and systematic errors that arise from older assumptions in CR modeling, including: 
use of a simple Fokker-Planck equation or ad-hoc two-moment formalisms for CR transport; 
assumption of leaky boxes or plane-parallel or shear-periodic boundaries for comparison to local interstellar medium (LISM) CR observations; 
ignoring detailed LISM constraints on CR spectra (e.g.\ focusing only on extragalactic observables or spectrally integrated models); and/or 
assuming CR transport is mediated by classical models of advection, streaming from ``self-confinement'' (super-\Alf{ic}/damped or \Alf{ic}/saturated), or ``extrinsic turbulence.''
We emphasize recent progress addressing all of these:
development of rigorously-derived CR-MHD equations; 
use of global, 3D galaxy+disk+halo models for LISM comparisons; 
new methods for full-spectrum CR dynamics; 
and
novel models for intermittent CR scattering and/or new scattering drivers. 
We compile new extragalactic+LISM observations to show how $\sim$\,GeV CR transport is being rapidly constrained in the CGM, and present simple phenomenological models which can be used in future simulations. 
We conclude by highlighting critical open questions for micro, meso, and macro-scale CR-MHD simulations.
\end{abstract}
\maketitle

\section{Introduction}
\label{sec:intro}

Cosmic rays are of fundamental interest for high-energy astro-particle physics, galaxy and star and planet formation and evolution, astro-chemistry, dark matter physics, and ISM and space plasma physics. Many excellent reviews have been written in the last decade, both comprehensive and focused on various aspects of CRs including e.g.\ CR acceleration \citep{bell:2013.cr.acceleration.review}, propagation and transport \citep{strong:2007.cr.propagation.interactions.review,2018AdSpR..62.2731A,hanasz:2021.cr.propagation.sims.review}, interactions with the ISM \citep{grenier:2015.cr.galaxies.review,gabici:2022.low.energy.cr.ionization.review}, non-thermal emission \citep{kronecki:2022.cosmic.ray.gamma.ray.review.not.calorimeters}, ``feedback'' on galaxy and star formation \citep{zweibel:cr.feedback.review,owen:2023.cr.review.galaxies.feedback,ruszkowski.pfrommer:cr.review.broad.cr.physics}, and open questions \citep{gabici:2019.cr.paradigm.challenges.mostly.well.explained.in.galaxy.sims,kachelries:2019.cr.microphysics.review.obs.broad.range.anomolies.local.sources}. 
In recent years, there has been an explosion of work connecting galaxy-scale processes and $\gtrsim$\,kpc or macro-scale galaxy formation models to the microphysics of CR scattering and transport through dynamical, fully-coupled CR-MHD simulations \citep[see reviews above and e.g.][]{ensslin:2011.cr.transport.clusters,salem:2013.cosmic.ray.outflows,Wien13,Sale16,simpson:2016.cr.pressure.driven.outflows.idealized.sims,wiener:2017.cr.streaming.winds,Buts18,gaches:2018.protostellar.cr.acceleration,Buts18,butsky:2020.cr.fx.thermal.instab.cgm,butsky:2022.cr.kappa.lower.limits.cgm,butsky:2023.cosmic.ray.scattering.patchy.ism.structures,chan:2018.cosmicray.fire.gammaray,chan:2021.cosmic.ray.vertical.balance,su:2018.stellar.fb.fails.to.solve.cooling.flow,su:turb.crs.quench,su:2021.agn.jet.params.vs.quenching,kraft2023lineemissionmapperlem,su:2025.crs.at.shock.fronts.from.jets.injection,bustard:2020.crs.multiphase.ism.accel.confinement,ji:fire.cr.cgm,ji:20.virial.shocks.suppressed.cr.dominated.halos,hopkins:cr.mhd.fire2,hopkins:2020.cr.outflows.to.mpc.scales,hopkins:cr.transport.constraints.from.galaxies,su:turb.crs.quench,armillotta:2021.cr.streaming.vs.environment.multiphase,wehahn:2021.gamma.rays,werhahn:2021.cr.calorimetry.simulated.galaxies,peschken:2022.crs.naab.sims.outflows.similar.to.fire,thomas:2022.self-confinement.non.eqm.dynamics,wellons:2022.smbh.growth,ponnada:fire.magnetic.fields.vs.obs,ponnada:2023.fire.synchrotron.profiles,ponnada:2023.synch.signatures.of.cr.transport.models.fire,ponnada:2024.fire.fir.radio.from.crs.constraints.on.outliers.and.transport,byrne:2023.fire.elliptical.galaxies.with.agn.feedback,martin.alvarez:radiation.crs.galsim.similar.conclusions.fire,wijers:2024.neviii.failure.of.simulations,brugaletta:2024.cr.sims.ionization.heating.important.in.some.dwarfs,weber:2025.cr.thermal.instab.cgm.fx.dept.transport.like.butsky.study,farcy:2025.cr.feedback.eor.galaxies.crs.increased.with.sne.decreased.lower.escape.fraction.but.similar.stellar.masses,dome:2025.no.burstiness.change.w.bfields.crs.other.physics.if.feedback.present.in.dwarfs.highz.galaxies}. Information flows in both directions: what we know about galaxies (for example, the fact that they all have large magnetized, turbulent gaseous halos extending to hundreds of kpc from the central galaxy) and extra-galactic observables inform fundamental assumptions needed to model CR transport and constrain its microphysics, and that microphysics in turn informs how CRs can exert feedback effects on galaxies in the interstellar or circumgalactic medium (ISM/CGM) through their energy deposition and pressure. 

\begin{footnotesize}
\ctable[caption={{\small Spatial Scales Of Key Importance\vspace{-0.2cm}}\label{tbl:scales}},center]{l l l}{
}{
\hline
Key Scale (\S~\ref{sec:background}) & Physical Units & Transport Formalism (\S~\ref{sec:transport}) \\
\hline
Micro & $\sim 0.2\,{\rm au}\,(R/{\rm GV})\,B_{\rm \mu G}^{-1}$ & Vlasov-Poisson (analytic) \\ 
\ \ (Gyro Scale) & \ \ ($\ell_{\rm g}$; Eq.~\ref{eqn:micro}) & MHD-PIC (numerical) \\
\hline
Meso & $\sim 10\,{\rm pc}\,(R/{\rm GV})^{1/2}$ & Gyro-Averaged (Eq.~\ref{eqn:mu}) or \\
\ \ (Mean Free Path) & \ \  ($\ell_{\rm mfp}$; Eq.~\ref{eqn:meso}) & Two-Moment (Eq.~\ref{eqn:twomoment}) \\
\hline
Macro & $\gtrsim$\,kpc & Single-Moment (analytic) \\
\ \ (CR Gradient) & \ \  ($\ell_{\rm \nabla}$; Eq.~\ref{eqn:macro}) & Two-Moment (numerical) \\
\hline
}
\end{footnotesize}

In this manuscript, we present a pedagogical overview of this connection, emphasizing and collecting developments in the last few years, to highlight the major challenges and goals, especially for CR-MHD and MHD-PIC simulations interested in the plasma physics of CRs in the ISM/CGM. 
Our goal, as distinct from the reviews above, is to connect and present recent insights and constraints from CR transport modeling from the extragalactic, Galactic, and plasma physics communities. All of these communities involve active research on these connected topics, but often communication between the communities is more limited and, as such, outdated assumptions from ``external'' communities are often adopted within each. 
We do not discuss the important, but distinct questions of CR acceleration, high and ultra-high energy CRs, or their role as tests of high-energy particle physics. 

\S~\ref{sec:background} introduces various definitions and scales, including the crucial concept of the micro/meso/macro-scale in CRs. 
In \S~\ref{sec:transport}, we then review the progress on deriving the effective equations for CR transport and coupled CR-MHD dynamics (including numerical implementations) on micro (\S~\ref{sec:transport.micro}), meso (\S~\ref{sec:transport.meso}), and macro (\S~\ref{sec:transport.macro}) scales, going well beyond state-of-the-art approaches just a few years ago. 
\S~\ref{sec:obs.lism} gives an overview of what we measure (\S~\ref{sec:obs.spectra}) and what we have robustly learned (\S~\ref{sec:lessons}) from Galactic CR transport models aimed at reproducing detailed Solar-neighborhood CR spectra, emphasizing the need for global Galactic models and extended CR scattering halos (\S~\ref{sec:halos}), as well as remaining open questions where dynamical CR-MHD approaches can play a major role (\S~\ref{sec:dyn.models}). 
In \S~\ref{sec:extrasolar} we discuss indirect CR constraints from outside the Solar system and in other galaxies, discussing the different measurements available and their caveats (\S~\ref{sec:extrasolar.obs}), what they tell us about CR transport in the ISM of the Milky Way and other galaxies (\S~\ref{sec:xgal.ism}), and what we have learned (mostly in just the last couple years) about CR transport in the CGM of Milky Way-like galaxies (\S~\ref{sec:obs.cgm}). 
\S~\ref{sec:fx} reviews what these and related constraints tell us about the \textit{dynamical} importance of CRs to the near environments around AGN (\S~\ref{sec:fx:agn}); the neutral ISM, GMCs, and starburst galaxies (\S~\ref{sec:fx.gmc}); the warm and hot ISM (\S~\ref{sec:fx.ism}); and the CGM/IGM (\S~\ref{sec:fx:cgm}); as well as what parts of this are sensitive to CR transport uncertainties. 
With this in mind, \S~\ref{sec:uncertainty} discusses the major uncertainties in our physical understanding of CR transport in these different environments, reviewing how different physical CR scattering models predict widely divergent dependence on local plasma properties (\S~\ref{sec:uncertainty:plasma.props}), and how ``classic'' models for CR transport are fundamentally incompatible with the observed rigidity-dependence of CR scattering rates from $\sim 1-1000$\,GeV (\S~\ref{sec:uncertainty:rigidity}). This includes advection/convection/saturated self-confinement, \S~\ref{sec:advective}; damped/un-saturated self-confinement, \S~\ref{sec:super.alfvenic}; extrinsic turbulence from \Alf\ or slow modes, \S~\ref{sec:et.alfven}; extrinsic turbulence from fast modes, \S~\ref{sec:et.fast}; and scattering from highly non-gyro-resonant structures, \S~\ref{sec:no.resonant}. In \S~\ref{sec:intermittency} we discuss how this motivates newly-developed models for alternative drivers or intermittent CR scattering (\S~\ref{sec:intermittency}). 
We conclude in \S~\ref{sec:conclusions} with an admittedly biased summary of the most important questions new efforts can address for these linked problems on micro, meso, and macro scales.

We note that while this is primarily structured as a pedagogical presentation and review of recent work, much of what we show and discuss here has not been presented or collected before -- including almost all of the Figures, and the data/model compilations in Tables~\ref{tbl:spectral.fits}, \ref{tbl:scattering.models} \&\ Figures~\ref{fig:bc.be.epos}, \ref{fig:be.halo}, \ref{fig:cgm.kappa.obs}, and the streamlined derivation+equations of the ``solution collapse'' problem in \S~\ref{sec:super.alfvenic}.

\begin{footnotesize}
\ctable[caption={{\small Acronyms and Key Terms\vspace{-0.2cm}}\label{tbl:acronyms}},center]{l l}{
}{
\hline
Acronym/Term & Definition \\
\hline
CR & Cosmic Ray \\
MHD & Magneto-Hydrodynamics \\
ISM & Interstellar Medium \\
CGM & Circum-Galactic Medium \\
IGM & Inter-Galactic Medium \\
LISM & Local Interstellar Medium (Solar neighborhood) \\
PIC & Particle-in-Cell (simulation method) \\
MHD-PIC & Hybrid MHD + kinetic CR particles \\
SNe/SNR & Supernova(e)/Supernova Remnant \\
GMC & Giant Molecular Cloud \\
QLT & Quasi-Linear Theory (\S~\ref{sec:uncertainty:rigidity}) \\
IC & Inverse Compton \\
Scattering rate ($\bar{\nu}$) & Effective pitch-angle scattering rate (\S~\ref{sec:transport.meso}) \\
Self-confinement & CR scattering by self-excited waves (\S~\ref{sec:super.alfvenic}) \\
Extrinsic turbulence & CR scattering by external MHD turbulence (\S~\ref{sec:et.alfven}) \\
\hline
}
\end{footnotesize}

\begin{footnotesize}
\ctable[caption={{\small Variables Used Throughout\vspace{-0.2cm}}\label{tbl:variables}},center]{l l l}{
}{
\hline
Name & Definition & Units \\
\hline
$R_{\rm cr}$ & CR Rigidity $ p_{\rm cr} c/Z_{\rm cr} e$ (momentum $p_{\rm cr}$)  & $10^{-3}$-$10^{3}\,$GV \\
$v_{\rm cr}$, $\beta$ & CR speed $ \beta_{\rm cr} c$ & $\sim$\,c \\
$E^{\rm tot}_{\rm cr}$, $\gamma$ & CR total energy $ \gamma m c^{2}$ & MeV-TeV \\
$T^{\rm kin}_{\rm cr}/{\rm nuc}$ & CR kinetic-energy per nucleon & MeV-TeV \\ 
\hline
$f$, $\bar{f}_{0}$ & CR phase-space density $d{\rm N}_{\rm cr} / d^{3}{\bf x}\,d^{3}{\bf p}$ & (\S~\ref{sec:transport}) \\
$\mu$, $\langle \mu\rangle$ & CR pitch-angle $ \hat{\bf p} \cdot \bhat$, $\langle \mu \rangle \equiv v_{\rm drift}/v_{\rm cr}$ & -- \\
$e_{\rm cr}$ & CR kinetic energy density $\int d^{3}{\bf p}\, T_{\rm kin} f$ & $\sim {\rm eV\,cm^{-3}}$ \\
$P_{\rm cr}$ & CR pressure $\mathbb{P}_{\rm cr} \equiv e_{\rm cr} \mathbb{D}_{\rm cr} = \int d^{3}{\bf p}\, {\bf p} {\bf v} f$ & $\sim e_{\rm cr}/3$ \\
\hline
$\nu$, $\bar{\nu}$ & CR pitch-angle scattering rate ($\bar{\nu} \sim D_{\mu\mu}({\bf p},\,{\bf x},\,t)$) & $10^{-(11-8)}\,{\rm s^{-1}}$ \\
$\kappa_{\|}$, $\kappa_{\rm eff}$ & Diffusivity $v_{\rm cr}^{2}/3\bar{\nu}$, ``effective'' (net flux) & $10^{28-31}\,{\rm cm^{2}\,s^{-1}}$ \\
$\bar{v}_{A}$, $v_{\rm st, eff}$ & Streaming speed, ``effective'' ($\sim\kappa_{\rm eff}/\ell_{\nabla}$) & $10-300\,{\rm km\,s^{-1}}$ \\
$\Delta t_{\rm res}$ & CR ``Residence Time'' in disk/halo & $\sim\,$Myr ($\propto R_{\rm cr}^{-\delta}$) \\ 
\hline
$X_{\rm gramm}$ & CR Grammage ($\int \rho\, v_{\rm cr}\, d t$) & $\sim {\rm g\,cm^{-2}}$ \\ 
$\mathcal{R}$ & CR Loss (radiative, collisional) rate & $\sim {\rm Gyr}^{-1}$ \\
$\ell$, $k_{\|}$ & Spatial scale ($\ell_{\rm g,\,mfp,\,\nabla}$; Table~\ref{tbl:scales}), $k \sim 1/\ell$ & $\sim$\,au-kpc \\
$r$, $H_{\rm halo}$ & Galacto-centric radius $r$, halo ``size'' & $10$-$1000$\,kpc \\
\hline
$\rho$, $n$ & Gas density $n=\rho/\mu m_{p}$ & $\sim {\rm cm^{-3}}$ \\
$B$, $\bhat$ & Magnetic field strength/direction ${\bf B} \equiv B\,\bhat$ & $\sim {\rm \mu G}$ \\
$v_{A}$ & \Alf\ speed $v_{A} \equiv B/\sqrt{4\pi\,\rho_{\rm ion}}$ & $1$-$100\,{\rm km\,s^{-1}}$ \\
${\bf u}$, ${\bf v}_{e}$ & Gas velocity ${\bf u}$, effective CR frame ${\bf v}_{e} \equiv {\bf u} + \bar{v}_{A}\bhat$ & $1$-$10^{3}\,{\rm km\,s^{-1}}$ \\
\hline
$\delta$ & CR transport energy scaling ($\kappa_{\rm eff} \propto R_{\rm cr}^{\delta}$) & $\sim 0.5$ (0.4-0.7)\\
$\bar{\nu}_{0}$ & LISM mean CR $\bar{\nu}$, around spectral peak & $\sim 10^{-9}\,{\rm s^{-1}}$ \\
\hline
}
\end{footnotesize}

\begin{figure*}
	\centering
	\includegraphics[width=0.48\textwidth]{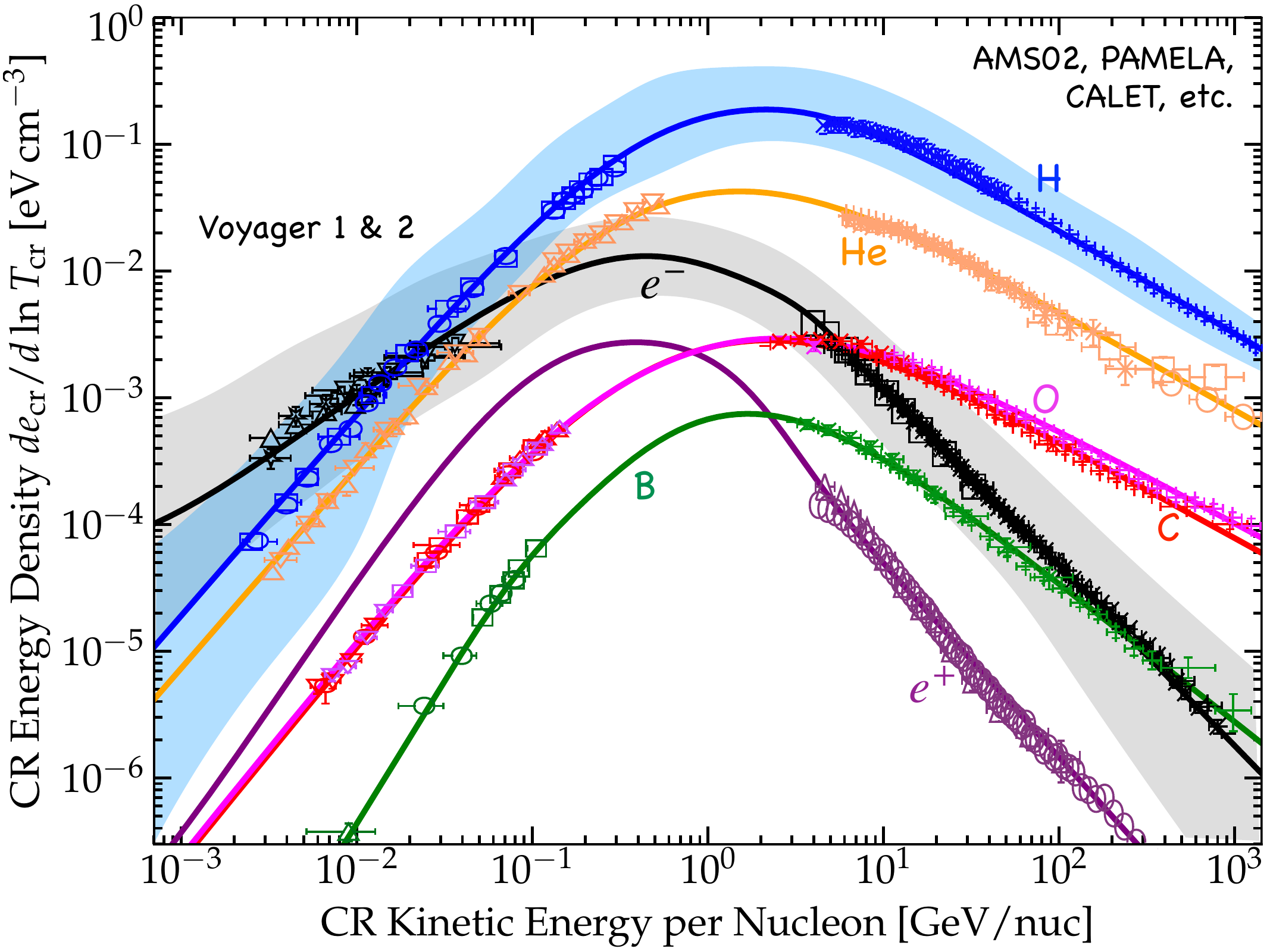}
	\includegraphics[width=0.48\textwidth]{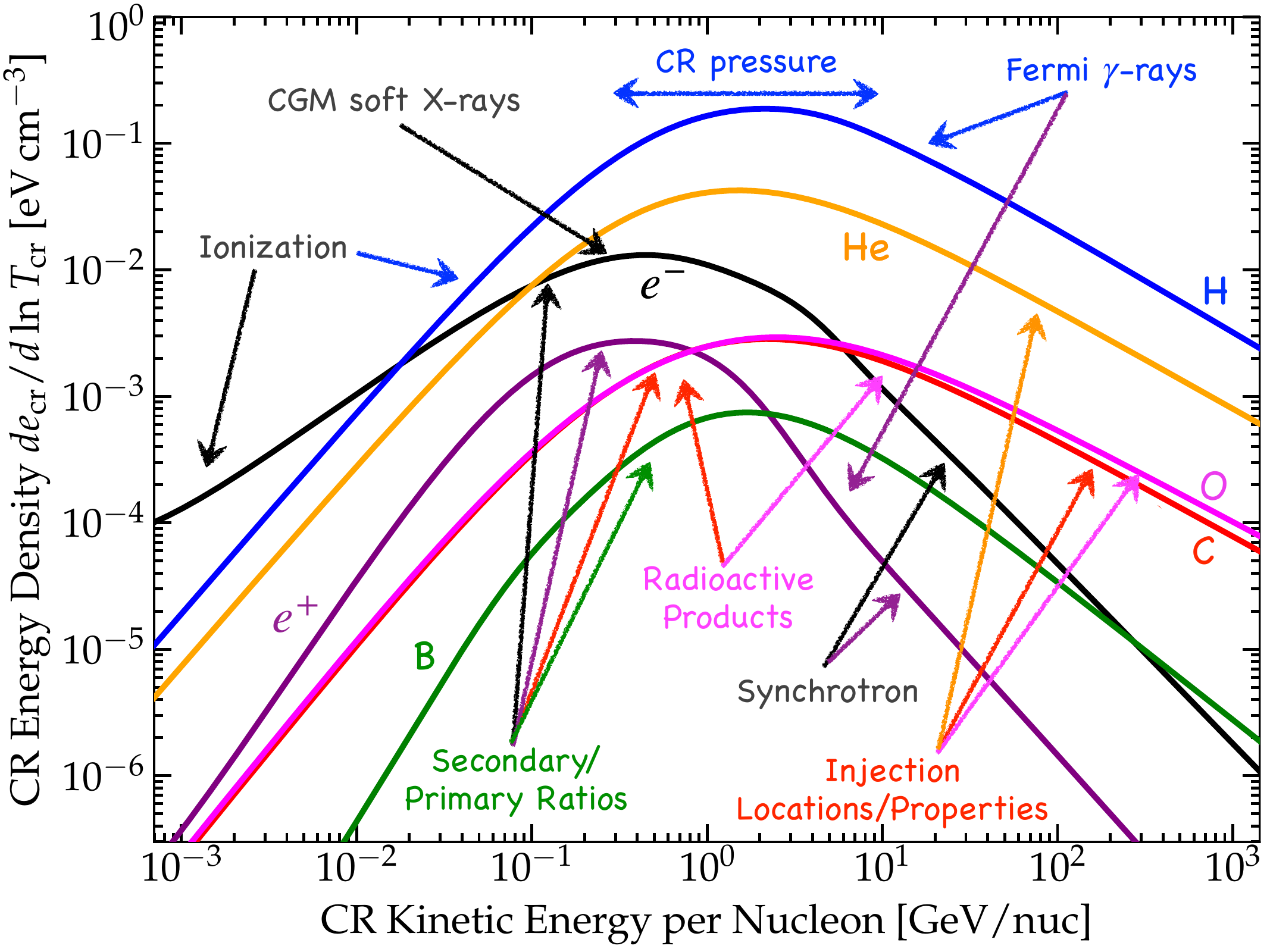}
	\caption{Local ISM (LISM) CR spectra. 
	{\em Left:} We show the best-fit modulation-corrected spectra from time-static CR transport codes like GALPROP ({\em lines}) with data ({\em points}, matched color) compiled from \citet{cosmic.ray.database.2023.version.citation} -- primarily AMS-02, PAMELA, CALET, Voyager 1 \&\ 2 -- at energies where the modulation correction is small ($\lesssim 10\%$; all Voyager, plus Solar system experiments at $\gtrsim 4\,$GeV). We also show ({\em shaded}) the range predicted about the mean proton (H) and electron $e^{-}$ spectra, for different specific Solar locations at $R \sim 7-9\,$kpc from the Galactic center and LISM gas density $\sim 0.1-3\,{\rm cm^{-3}}$, in dynamical CR-MHD simulations \citep{hopkins:2021.sc.et.models.incompatible.obs,ponnada:2023.fire.synchrotron.profiles}. While ``CR weather'' can systematically influence spectral shapes (e.g.\ losses being stronger/weaker in more/less dense regions), the CR spectra are well-known and phenomenological models for CR transport are well-constrained. 
	{\em Right:} Importance of different species and energy ranges for different physical processes. 
	Most of the CR energy/pressure comes from $\sim$\,GeV protons, but the entire MeV-TeV range for many species are critical to different processes and observations.
	\label{fig:specdemo}}
\end{figure*} 

\section{Definitions and Scales}
\label{sec:background}

For ISM/galaxy/CGM science, the CRs of greatest interest are low-energy $\sim$\,GeV (or more broadly $\sim$\,MeV-TeV) CRs, especially protons and electrons, as these (discussed below) dominate the effects of CRs on gas (ionization, heating, pressure) and extra-Solar observables. 
This is obvious from their LISM spectra (Fig.~\ref{fig:specdemo}). 

There are three especially salient scales (Table~\ref{tbl:scales}): (1) the micro scales of order the CR gyro radii, (2) the meso scales of order the CR scattering mean free path (pitch angle deflection length), and (3) macro scales of order the CR pressure gradient scale on which they can act on the ambient medium and globally evolve. Quantitatively:
\begin{align}
\label{eqn:micro} \ell_{\rm micro} &\sim \ell_{\rm g} \equiv \frac{p_{\rm cr}}{Z_{\rm cr} e B} \sim 0.2\,{\rm au} \left( \frac{R_{\rm cr}}{\rm GV} \right) \left( \frac{\rm \mu G }{B} \right) \ ,  \\ 
\label{eqn:meso} \ell_{\rm meso} &\sim \ell_{\rm mfp,\,\|} \equiv \frac{v_{\rm cr}}{\nu[R_{\rm cr},\,...]} \sim 10\,{\rm pc}\,\left( \frac{R_{\rm cr}}{\rm GV} \right)^{1/2}  \ , \\ 
\label{eqn:macro} \ell_{\rm macro} &\sim \ell_{\rm cr,\,\nabla} \equiv \frac{P_{\rm cr}}{|\nabla_{\|} P_{\rm cr}|} \gtrsim \,{\rm kpc} \ .
\end{align}
Variables are defined in Table~\ref{tbl:variables}. 
Here $v_{\rm cr}$, $p_{\rm cr}$, $Z_{\rm cr}$, $R_{\rm cr}$ are individual CR velocities, momenta, charge, and rigidity, $B=|{\bf B}|$ is the magnetic field strength, $\nu$ the CR scattering rate, and $P_{\rm cr} \sim e_{\rm cr}/3$ for ultra-relativistic species the CR (scalar) pressure.\footnote{We define the scalings above in terms of the CR rigidity $R_{\rm cr} \equiv p_{\rm cr}\,c/Z_{\rm cr}\,e$ -- this is most often used in the plasma literature because for fixed $B$, CRs of the same $R_{\rm cr}$ have the same gyro radii and should follow essentially the same scattering/transport physics. However, in the CR detection literature, CRs are often compared at fixed CR kinetic energy-per-nucleon $T^{\rm kin}_{\rm cr}/N_{\rm nuc}$, while in other literature sometimes just the kinetic energy $T_{\rm cr}$ or total energy $E^{\rm tot}_{\rm cr} \equiv T^{\rm kin}_{\rm cr} + m_{\rm cr}\,c^{2}$ are used. For ultra-relativistic singly-charged species, i.e.\ $e^{\pm}$ at $\gg$\,MeV and $p$, $\bar{p}$ at $\gg$\,GeV energies, these are trivially related, with $E^{\rm tot}_{\rm cr} \approx T^{\rm kin}_{\rm cr} \approx p_{\rm cr}\,c \approx {\rm GeV}\,(R_{\rm cr}/{\rm GV})$, so we will generally treat them synonymously except where needed.}

\section{CR Transport Formalism (Equations-of-Motion)}
\label{sec:transport}

Any modeling of CRs and their effects requires asking: what are the actual transport equations of CRs? While some analytic scalings were developed in the 1960s and 70s, until quite recently there was no rigorously-derived numerical CR-MHD transport theory relevant for meso and macro scales, nor was there a tractable method for e.g.\ numerical simulations of most transport physics on micro-scales. There has been tremendous progress on these fronts in the last decade. Given the huge scale-separation in the problem, we discuss each scale in turn.

\subsection{Micro-Scale}
\label{sec:transport.micro}

On micro-scales, $\ell_{\rm g}/\ell_{\rm micro} \sim \mathcal{O}(1)$ by definition, and if there are interesting variations in ${\bf B}$ on the same scales, gyro orbits need to be explicitly modeled -- i.e.\ one integrates directly the Lorentz force ${\bf F} \approx d_{t} {\bf p} = (q/c)\,({\bf E} + {\bf v}\times {\bf B}$) along individual CR trajectories (other non-Lorentz forces can safely be ignored, to leading order). 
This means that on micro-scales, CR scattering rates $\bar{\nu}$ are predicted {\em directly}, not assumed.
Other distance scales like the average separation between CRs ($\sim 10$\,meters in the ISM) are tiny, and so the CR population can be evolved in terms of a 3-dimensional distribution function $f_{3} \equiv d {\rm N}_{\rm cr} /  (d^{3} {\bf p}_{\rm cr} \, d^{3}{\bf x})$ obeying the Vlasov equation $d_{t} f_{3} + {\bf v}_{\rm cr} \cdot \nabla_{\bf x} f_{3} + {\bf F} \cdot \nabla_{\bf p} f_{3} = d_{t} f_{3} |_{\rm coll}$ in terms of gradients in position ${\bf x}$ and momentum ${\bf p}$ space and collision/loss operators. 

Older work used this to develop quasi-linear theory of collisionless CR scattering \citep{jokipii:1966.cr.propagation.random.bfield,Voelk1973,1975MNRAS.172..557S,cesarsky.kulsrud:1981.cr.confinement.damping.hot.gas}, whereby fluctuations in the magnetic fields ${\bf B}$ (especially gyro-resonant fluctuations with wavelength $\lambda \sim \ell_{\rm g}$) deflect the CR pitch angle $\mu \equiv \cos{\theta} \equiv {\bf p}_{\rm cr} \cdot \bhat / p_{\rm cr}$, isotropizing the pitch-angle distribution so that $|\langle \mu \rangle | \ll 1$, i.e.\ the bulk CR drift/streaming/diffusive propagation speed $v_{\rm st} = |\langle \mu \rangle|\,v_{\rm cr} \ll v_{\rm cr}$ is very small compared to $c$. But many interesting, and especially non-linear cases can only be studied by numerical particle-in-cell (PIC) methods. Pure PIC methods evolved over the very large spatial/timescales relevant for most CR transport problems, however, are intractable for GeV CRs in the ISM, owing to the enormous mismatch between gyro-radii of non-relativistic electrons ($\sim 10^{4}\,{\rm cm}$) and CRs ($\sim 10^{12}\,{\rm cm}$; see review in e.g.\ \citealt{holcolmb.spitkovsky:saturation.gri.sims}). 

However, more recently this has been turned to an advantage. Utilizing the facts that (1) CR gyro radii are much larger than gyro/Debye scales of the background plasma, (2) the CRs carry a very small fraction of the total charge, and (3) the background is non-relativistic, new hybrid MHD-PIC methods have been developed, in which the background plasma obeys the fluid/MHD equations, with just CRs integrated as PIC particles including a back-reaction force term from their current. These have been developed for Eulerian \citep{bai:2015.mhd.pic,mignone:2018.mhd.pic} and Lagrangian codes \citep{ji:2021.cr.mhd.pic.dust.sims}, extended to $\delta-f$ formalisms for more accurate integration of small perturbations \citep{bai:2019.cr.pic.streaming}, optimized by use of rigorously-derived reduced-speed-of-light formulations (that ensure converged steady-state results while allowing larger timesteps; \citealt{ji:2021.mhd.pic.rsol}), and applied to a variety of conditions to study the CR gyro-resonant streaming instability \citep{bai:2019.cr.pic.streaming}, damping of scattering modes in partially-neutral environments \citep{plotnikov:2021.cr.mhd.pic.sims.streaming.ion.neutral.strong.damping.deconfines,bambic:2021.mhd.pic.transport.inhomogeneous.ionization.effective.confinement.very.different.from.length.corr.of.turb.much.similar.to.cr.fluid.dynamics.models}, and hybrid interactions with other large-gyro-radii species like dust grains \citep{ji:2021.cr.mhd.pic.dust.sims}.

These constitute an extremely promising tool for CR transport, though pure-PIC simulations are still necessary to study e.g.\ initial acceleration (since ions begin non-relativistic) and certain plasma damping processes \citep[][and references therein]{holcolmb.spitkovsky:saturation.gri.sims,lemmers:2024.cr.pic.modified.with.landau.closure.to.model.certain.plasma.effects.on.waves,schroer:2025.nll.damping.sims.pic.but.weak.and.non.expected.saturation}, and there are still limitations (most often that the CR gyro times, and hence timesteps, are vastly shorter than e.g.\ the global growth timescales of relevant linear instabilities on these scales).

\begin{table*}
\begin{center}
\begin{footnotesize}
\caption{\label{tbl:summary}What We Learn from Local CR Data (constrained by data \checkmark; unconstrained \xmark )}
\begin{tabular}{ lllll }
\hline\hline
Data & Injection Spectrum & Scattering Rate $\bar{\nu}_{0}$ & Energy-Dependence $\delta$ & Losses \\
\hline\hline
Spectra & \checkmark & \checkmark (normalization) & \checkmark & hadrons: low-E (Coulomb+ionization) \\
(primaries) & $\sim (dN/dtdE)_{\rm inj} \Delta t_{\rm res}$ & normalization ($\Delta t_{\rm res}$) at $\sim$\,GeV & high-E hadron shape & leptons: high-E (synchrotron+IC) \\
 \hline
B/C & \xmark & \checkmark (normalization) & \checkmark & sub-GeV (Coulomb/ionization) \\
(sec/primary) &  & GeV grammage $X\sim\langle n_{\rm gas} c \Delta t_{\rm res} \rangle$ & shape &  \\
 \hline
$^{10}$Be/$^{9}$Be & \xmark & \checkmark (shape) & \checkmark & sub-GeV (Coulomb/ionization) \\
(radioactive) &  & $\langle\Delta t_{\rm res}\rangle$ after secondary & high-E shape &  \\
 \hline
$e^{+}/e^{-}$ & \checkmark (low-E shape) & \checkmark (normalization) & \checkmark & high-E (synchrotron+IC) \\
 (leptonic) & (through $p$-vs-$e^{-}$ injection) & grammage & & very low-E (Coulomb+ionization)  \\
 \hline
$\bar{p}/p$ & \checkmark (normalization) & \checkmark (normalization) & \checkmark & weak \\
(other $\bar{A}/A$) & ($T/n$ not conserved) & grammage & high-E shape &   \\
 \hline 
A/H & \checkmark (universality \&\ sources) & \xmark & \xmark & \xmark \\
(CNO,Fe,...) & (where/when around SNe)  &  &   &   \\
 \hline 
Summary & from SNe at $r \sim$\,few kpc  & $\sim 10^{-9}\,{\rm s^{-1}}$  & $\sim 0.5$  & $\gtrsim10\,$GeV leptons: synchrotron+IC \\
 & $dN/dp \propto p^{-(2.2-2.4)}$ & (factor $\sim2-3$ uncertainty) & (range $0.4-0.7$) &  $\lesssim 0.1\,$GeV: Coulomb+ionization \\
 & in reverse shocks & $X$ vs $\Delta t_{\rm res}$ requires few-kpc halo & high-E, smaller $\Delta t_{\rm res}$ & $\sim 0.1$-$10\,$GeV and higher hadrons: escape\\
\hline
\end{tabular}
\end{footnotesize}
\end{center}
\end{table*}

\subsection{Meso-Scale}
\label{sec:transport.meso}

On meso-scales, one can take $\ell_{\rm micro}$ to be small, and formally treat it as an expansion parameter for the dynamics equations, {\em averaging over} the gyro radii and orbits in a non-relativistic background medium, like with kinetic MHD. For a gyro-averaged distribution function $f_{\mu}$ defined by $d {\rm N}_{\rm cr} \equiv f_{\mu} 2\pi p_{\rm cr}^{2} dp_{\rm cr}\, d\mu\, d^{3}{\bf x}$, the evolution equation can be simplified to the focused CR equation for free transport \citep{skilling:1971.cr.diffusion,1975MNRAS.172..557S,1997JGR...102.4719I,2001GeoRL..28.3831L,2005ApJ...626.1116L,zank:2014.book,2015ApJ...801..112L} plus slab scattering terms \citep[references above and][]{schlickeiser:89.cr.transport.scattering.eqns}. Together, after expansion in $\ell_{\rm g}/\ell_{\rm meso}$ and $|{\bf u}|/c$, this gives \citep{hopkins:m1.cr.closure} for each CR species: 
\begin{align}
\label{eqn:mu} D_{t} f_{\mu}  &+  \nabla \cdot (\mu\,v_{\rm cr} f_{\mu} \,\bhat) =  j_{\mu} +  \\
\nonumber & \frac{\partial}{\partial\mu}\left[ \chi_{\mu} \left\{ f_{\mu} v_{\rm cr} \nabla\cdot\bhat + {\nu}\left( \frac{\partial f_{\mu}}{\partial \mu} + \frac{\bar{v}_{A}}{v_{\rm cr}} \frac{\partial f_{\mu}}{\partial \ln{p_{\rm cr}}} \right)  \right\} \right] + \\
\nonumber & \frac{1}{p_{\rm cr}^{2}} \frac{\partial }{\partial p_{\rm cr}}\Bigl[ p_{\rm cr}^{3} \Bigl\{f_{\mu} \left(\mathcal{R} +  \mathbb{D}_{\mu}:\nabla {\bf u} \right) + \\ 
\nonumber & \ \ \ \ \ \ \ \ \ \ \ \ \ \ \ \ \  {\nu\,\chi_{\mu}}\left( \frac{\bar{v}_{A}}{v_{\rm cr}}\, \frac{\partial f_{\mu}}{\partial \mu} + \frac{v_{A}^{2}}{v_{\rm cr}^{2}}\,\frac{\partial f_{\mu}}{\partial \ln{p_{\rm cr}}} \right) \Bigr\} \Bigr] \ , 
\end{align}
where $D_{t} X \equiv (\partial X/\partial t) + \nabla \cdot ({\bf u} X)$ is a convective derivative, $\bhat\equiv {\bf B}/B$ is the magnetic field direction, $j_{\mu}$ represents sources and catastrophic losses, $\mathcal{R}$ radiative losses, $\mathbb{D}_{\mu} \equiv \chi_{\mu} \mathbb{I} + (1-3\chi_{\mu})\bhat\bhat$, $\chi_{\mu} \equiv (1-\mu^{2})/2$, $\nu=\nu(p_{\rm cr},\,\mu,\,{\bf x},\,...)$, $v_{A} \equiv B/\sqrt{4\pi\,\rho_{\rm ion}}$ is the phase velocity of \Alf\ waves with wavelength equal to the gyro radius (so depends on the {\em ion} density, distinct from the ideal MHD \Alf\ speed relevant for much longer-wavelength modes), and $\bar{v}_{A} \equiv v_{A}\,(\nu_{+}-\nu_{-})/(\nu_{+}+\nu_{-})$ in terms of the scattering rate contributed by forward ($\nu_{+}$) versus backward ($\nu_{-}$) propagating magnetic fluctuations (${\nu} \equiv \nu_{+}+\nu_{-}$).

Note that all quantities needed or evolved here are defined on meso-scales, except for the scattering rates $\nu$, which by definition come from micro-scales. On meso-scales, $\nu$ must be assumed to follow some simplified scaling
(most commonly parameterized as $\bar{\nu} \sim \beta_{\rm cr}\,\bar{\nu}_{0} (R_{\rm cr}/R_{0})^{-\delta}$)\footnote{Equivalent (in the diffusive limit) to an isotropic diffusivity as often parameterized in the isotropic Fokker-Planck equations on macro-scales in \S~\ref{sec:transport.macro} of $\kappa_{\rm iso} \sim \kappa_{\|}/3 \sim v_{\rm cr}^{2} / 9\,\bar{\nu} \sim \beta_{\rm cr}\,D_{0}\,(R_{\rm cr}/R_{0})^{\delta}$ with $D_{0} \sim c^{2}/9\bar{\nu}_{0}$.} or evolution equation ($\partial_{t} \nu = ...$, or the equivalent in terms of the wave energies/amplitudes driving scattering, see \S~\ref{sec:uncertainty} below and e.g.\ \citealt{Zwei13,thomas.pfrommer.18:alfven.reg.cr.transport}).

While discarding some information, the dimensionality of the problem can be reduced by taking moments of this rather than evolving the pitch-angle distribution explicitly. This gives a pair of equations for $\bar{f}_{0} \equiv \langle f_{\mu}\rangle$ (so $d {\rm N}_{\rm cr} \equiv \bar{f}_{0} 4\pi p_{\rm cr}^{2} dp_{\rm cr} d^{3}{\bf x}$) and $\bar{f}_{1} \equiv \langle \mu f_{\mu} \rangle = \langle \mu\rangle \bar{f}_{0}$, from \citet{hopkins:m1.cr.closure}:
\begin{align}
\label{eqn:twomoment} D_{t}& \bar{f}_{0}  + \nabla \cdot  (v_{\rm cr} \bhat\,\bar{f}_{1})  
=  j_{0} + 
\\
\nonumber &  
\frac{1}{p_{\rm cr}^{2}}\frac{\partial}{\partial p_{\rm cr}}\left[ p^{3}_{\rm cr}\,\left\{  \left( \mathcal{R} + \mathbb{D}:\nabla{\bf u} + \bar{\nu}  \frac{v_{A}^{2}}{v_{\rm cr}^{2}} \chi \psi_{p} \right) \bar{f}_{0}  
+ \bar{\nu} \frac{\bar{v}_{A}}{v_{\rm cr}}\,\bar{f}_{1}
\right\} \right] \ ,  \\
\nonumber 
D_{t} & \bar{f}_{1} +  
v_{\rm cr} \bhat \cdot \nabla \cdot (\mathbb{D} \bar{f}_{0}) = -\bar{\nu} \left[ \bar{f}_{1} +   \frac{\bar{v}_{A}}{v_{\rm cr}} \chi \psi_{p} \bar{f}_{0}  \right] + j_{1}  \ ,
\end{align}
where $\bar{\nu}$ is now pitch-angle averaged, $\mathbb{D} \equiv \chi\,\mathbb{I} + ( 1-3\,\chi )\,\bhat\bhat$, $\psi_{p} \equiv \partial \ln{\bar{f}_{0}}/\partial \ln p$, $\chi \equiv (1-\langle \mu^{2} \rangle)/2$, $\langle \mu \rangle \equiv \bar{f}_{1}/\bar{f}_{0}$, and $\langle \mu^{2} \rangle \approx (3+4\,\langle \mu \rangle^{2})/(5 + 2\,[4-3\,\langle \mu \rangle^{2}]^{1/2})$ is given by some closure.\footnote{Different closure assumptions are discussed and derived in \citet{hopkins:m1.cr.closure} and \citet{thomas:2021.compare.cr.closures.from.prev.papers}. Both conclude that so long as the closure permits a physical distribution function for any realizeable pair of $\langle \mu\rangle$ and $\langle\mu^{2} \rangle$ the differences are minimal. This is much simpler than in the M1 radiation hydrodynamics case because the closure here only averages out one degree of freedom ($\mu$), not three dimensions, and because situations like anti-parallel streams allowed for radiation are, for CRs, microphysically highly unstable and isotropize extremely quickly.}

\begin{figure*}
	\centering
	\includegraphics[width=0.33\textwidth]{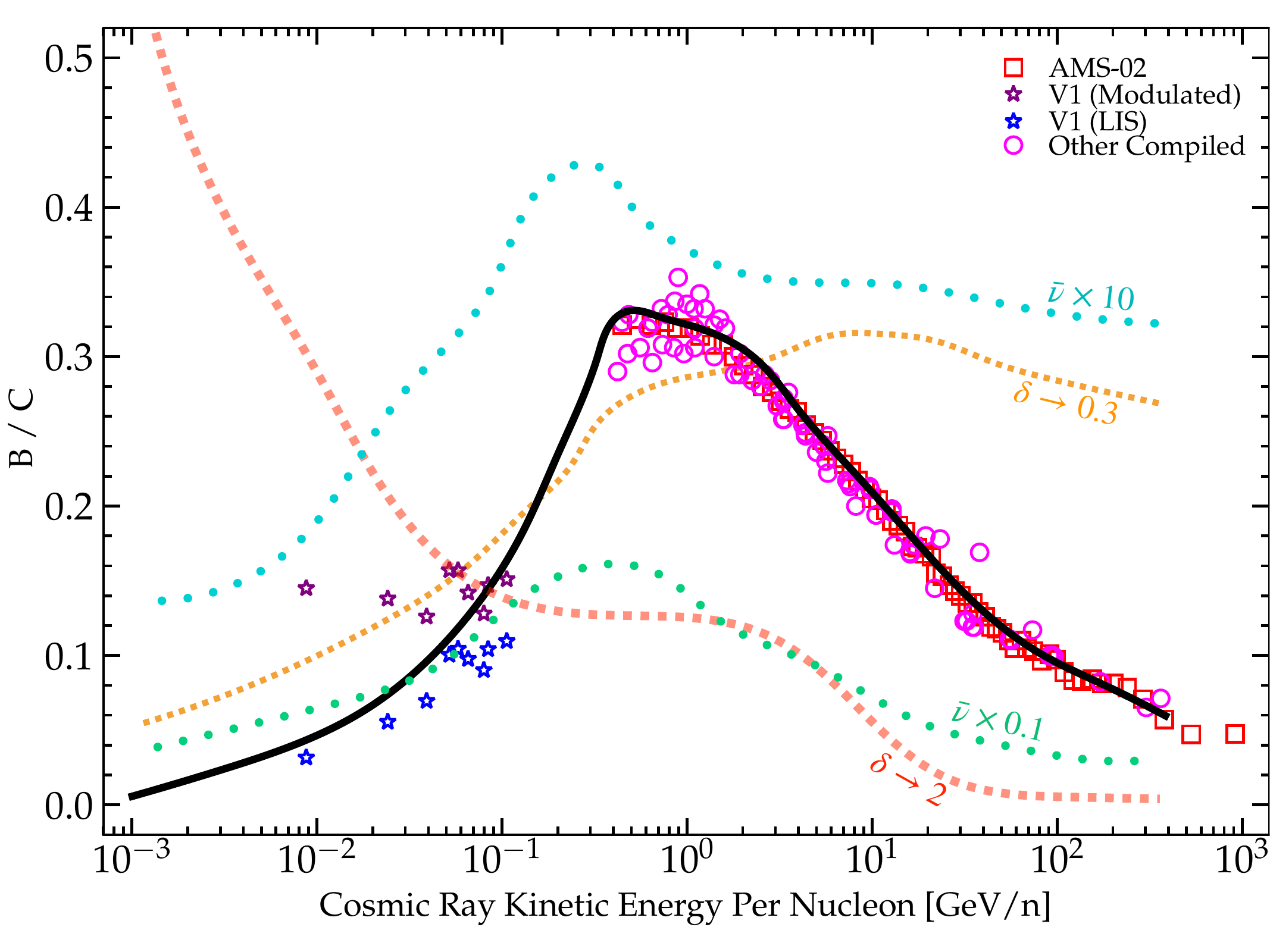}
	\includegraphics[width=0.33\textwidth]{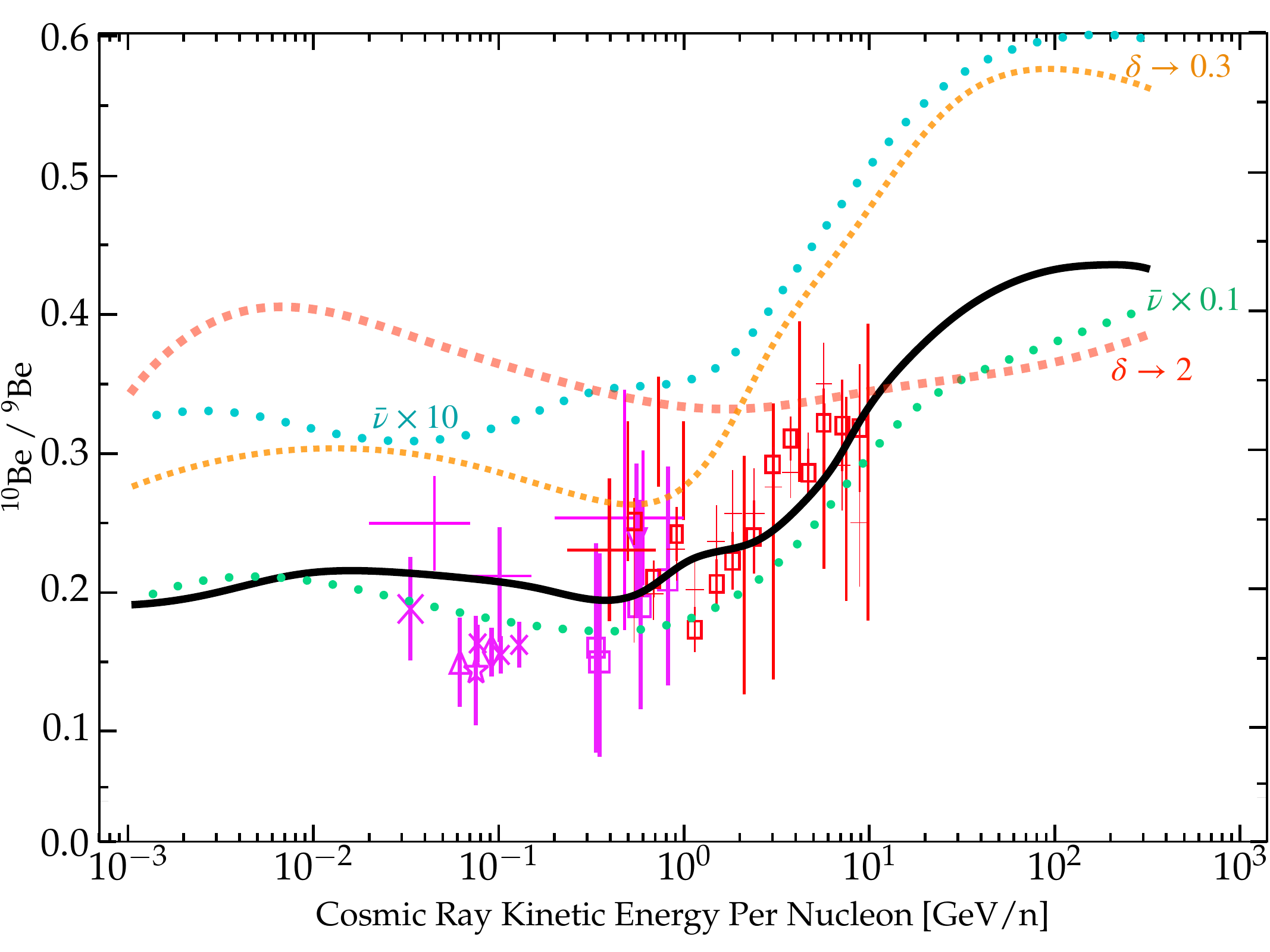}
	\includegraphics[width=0.33\textwidth]{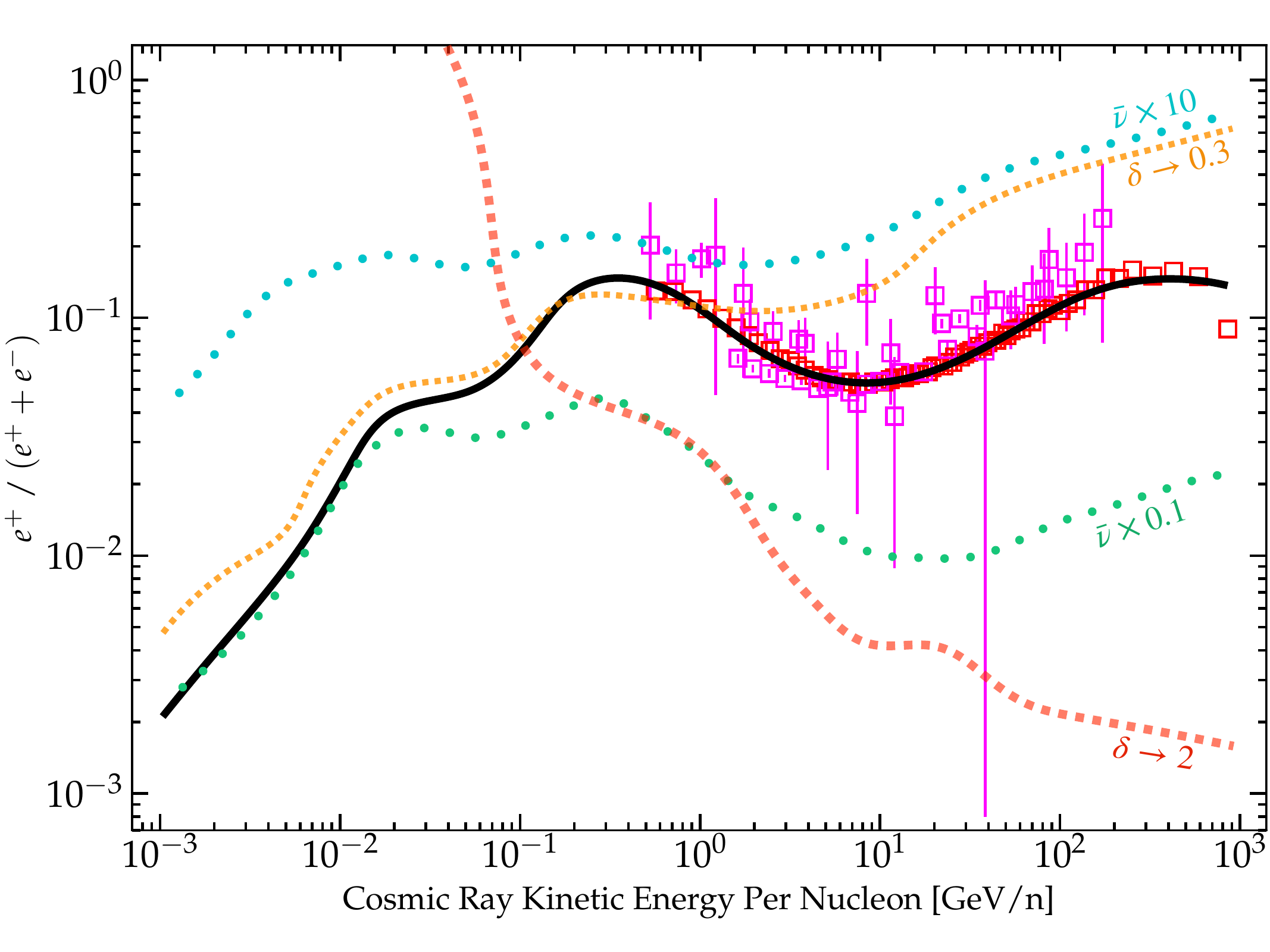}
	\caption{Illustration of constraints from LISM CR spectra on transport coefficients. A baseline modern full-galaxy model ({\em solid black line}) is assumed with uniform scattering rate $\bar{\nu} = \beta_{\rm cr} \bar{\nu}_{0} \,(R_{\rm cr}/{\rm GV})^{-\delta}$, $\bar{\nu}_{0} = 10^{-9}\,{\rm s^{-1}}$, $\delta=1/2$, and these parameters are varied and compared to (1) B/C, a hadronic secondary-to-primary ratio sensitive to grammage; (2) $^{10}$Be/$^{9}$Be, a ratio of radioactive-to-stable secondaries (produced in the same processes) sensitive to residence time; and (3) positron-to-electron ratio, with $e^{+}$ produced primarily from CR H, and annihilating, so loss-rate-sensitive (note the $e^{+}/(e^{+}+e^{-})$ data only begin around $\sim$\,GeV owing to experimental limitations; lower-energy measurements would be valuable but face strong Solar modulation and small positron fluxes). All are compared to various CR observations compiled as in Fig.~\ref{fig:specdemo}. LISM CR transport parameters are strongly constrained, in global Galactic transport models.
	\label{fig:bc.be.epos}}
\end{figure*}

For some applications, the spectrally-integrated equations, for total CR energy $e_{\rm cr} \equiv \int E_{\rm cr}(p_{\rm cr}) \bar{f}_{0} 4\pi p_{\rm cr}^{2} d p_{\rm cr}$ and energy flux $F_{\rm cr}\equiv \int E_{\rm cr}(p_{\rm cr}) v_{\rm cr} \bar{f}_{1} 4\pi p_{\rm cr}^{2} d p_{\rm cr}$, are useful. Assuming ultra-relativistic CRs, this gives: 
\begin{align}
\nonumber D_{t} e_{\rm cr} +& \nabla \cdot ( F_{\rm cr} \bhat ) = S_{\rm net} - \mathbb{P}:\nabla {\bf u}  \\ 
\nonumber & - \frac{\hat{\nu} }{c^{2}}\left[  \bar{v}_{A} F_{\rm cr} - 3\hat{\chi}\,v_{A}^{2}(e_{\rm cr}+P_{\rm cr}) \right] \ , \\ 
\label{eqn:energy} D_{t} F_{\rm cr} +& c^{2}\bhat\cdot(\nabla \cdot \mathbb{P}) = - \hat{\nu} \left[  F_{\rm cr} - 3\hat{\chi}\,\bar{v}_{A}(e_{\rm cr} +P_{\rm cr} ) \right] \ , 
\end{align}
where $\mathbb{P} \equiv e_{\rm cr}\hat{\mathbb{D}} \equiv \sum_{s} \int  4\pi\,p^{2}\,\nabla \cdot (\mathbb{D}\,p\,v\,\bar{f}_{0})\, {\rm d} p$ is the total CR pressure tensor, and $\hat{\chi}$, $\hat{\nu}$, $\hat{\mathbb{D}}$ are some appropriately-weighted averages. 
Of course, this discards a huge amount of information from the CR spectra and species, requires imposing an assumption of a universal spectral shape and effectively grey scattering rates and anisotropy, incorrectly describes sub-relativistic CRs, and cannot correctly capture certain effects (for example, the effect of diffusive reacceleration broadening the spectrum is lost, while catastrophic and radiative losses are conflated into a single $S_{\rm net}$).\footnote{The gyro-averaged force of CRs on {\em gas}, at a given point, is given by \citet{hopkins:m1.cr.closure}:
\begin{align}
\label{eqn:cr.gas.force} D_{t} (\rho\, {\bf u}) &= -\nabla \cdot \mathbb{P} -  \bhat\,\sum_{\rm species} \int 4 \pi p^{3} D_{t} \bar{f}_{1}  dp \\ 
\nonumber &\approx -\nabla \cdot \mathbb{P} + \bhat \left[\bhat\cdot\nabla\cdot\mathbb{P} + 
\frac{\hat{\nu}}{c^{2}}
\left\{
F_{\rm cr} - 3\hat{\chi} \bar{v}_{A}\,(e_{\rm cr} + P_{\rm cr} )
\right\} 
\right] 
\end{align}
where the latter is the spectral-integrated version, along with the appropriate expressions for energy transfer to the gas, radiation, and magnetic fields from e.g.\ acceleration/work, thermalized catastrophic losses, ionization, Coulomb scattering, synchrotron and inverse Compton and bremsstrahlung \citep[collected from various references in the appendices of ][]{hopkins:cr.multibin.mw.comparison}.}

Note that as shown in \citet{hopkins:m1.cr.closure} and \citet{thomas:2021.compare.cr.closures.from.prev.papers} (building on earlier work by \citealt{Zwei13,zweibel:cr.feedback.review,thomas.pfrommer.18:alfven.reg.cr.transport}), Eq.~\ref{eqn:twomoment} is the correct two-moment set of equations to leading order in $\ell_{g}/\ell_{\rm meso}$ and $|{\bf u}|/c$, and quasi-linear scattering theory. This is not a diffusion-advection equation nor telegraph nor Fokker-Planck equation, nor is it functionally the same as the two-moment radiation transport equations (though there are some similarities). Earlier simulations in e.g.\ \citet{jiang.oh:2018.cr.transport.m1.scheme,chan:2018.cosmicray.fire.gammaray} attempted to formulate {\em ad-hoc} two moment equations for the CR energy Eq.~\ref{eqn:energy} (motivated heuristically by M1 radiation transport). While useful for some insights, we stress that the ad-hoc formulations are missing important terms and contain several spurious terms (shown explicitly in Appendix~\ref{sec:chang.jo.note}), and will give the wrong answer (sometimes qualitatively, i.e.\ give the incorrect sign of $D_{t} e_{\rm cr}$ or $D_{t} F_{\rm cr}$ or even producing CRs propagating in the opposite direction) in certain common scenarios, for example if there are gradients in $\bhat$ on scales smaller than the CR scattering mean-free path, or the CRs are not in flux steady state or have significant anisotropy or comparable diffusion and streaming speeds.\footnote{Specifically (see \S~\ref{sec:chang.jo.note}), the equations in \citet{jiang.oh:2018.cr.transport.m1.scheme,chan:2018.cosmicray.fire.gammaray} are incorrect, in any limit, for the evolution of non-zero $\partial_{t} F_{\rm cr}$, and are only correct in the first-moment equation $\partial_{t} e_{\rm cr}$ in the limit of nearly-isotropic CRs, with $|\bar{v}_{A}|=v_{A}$, zero perpendicular diffusion, $\bhat \cdot \nabla P_{\rm cr} \ne 0$, vanishing scattering mean-free-path, and $\partial_{t} F_{\rm cr} = 0$.}
Fortunately, numerically implementing Eq.~\ref{eqn:energy} in stable fashion is actually simpler than any of the earlier schemes in \citet{jiang.oh:2018.cr.transport.m1.scheme,chan:2018.cosmicray.fire.gammaray,thomas.pfrommer.18:alfven.reg.cr.transport}, and has been demonstrated in many papers \citep{hopkins:2020.cr.transport.model.fx.galform,hopkins:2020.cr.outflows.to.mpc.scales,ji:20.virial.shocks.suppressed.cr.dominated.halos,su:2021.agn.jet.params.vs.quenching,hopkins:m1.cr.closure,ji:2021.cr.mhd.pic.dust.sims,chan:2021.cosmic.ray.vertical.balance,thomas:2021.compare.cr.closures.from.prev.papers,thomas:2022.self-confinement.non.eqm.dynamics,sike:2024.cr.winds.pfrommer.model.launch.warm.gas,weber:2025.cr.thermal.instab.cgm.fx.dept.transport.like.butsky.study}. Meanwhile breakthroughs in \citet{girichidis:cr.spectral.scheme} and subsequent work \citep{hanasz:2021.cr.propagation.sims.review,girichidis:2021.cr.transport.w.spectral.reconnection.hack,ogrodnik:2021.spectral.cr.electron.code} have enabled simulations of the full spectral Eq.~\ref{eqn:twomoment} for multiple species \citep{hopkins:cr.multibin.mw.comparison,baldacchino.jordan:2025.piernik.cr.transport.multi.species.sims} generalized to arbitrary scattering models \citep{hopkins:2021.sc.et.models.incompatible.obs}, with exact energy and momentum conservation ensured by use of the updated scheme in e.g.\ \citet{hopkins:cr.spectra.accurate.integration}, and these have been generalized to numerical reduced-speed-of-light methods with converged solutions in \citet{hopkins:m1.cr.closure}. We also stress that Eqs.~\ref{eqn:mu}-\ref{eqn:energy} do \textit{not} assume the mean-free-path is small: they are valid on scales $\gg \ell_{\rm micro} \sim \ell_{\rm g}$ even if $\ell_{\rm mfp} \rightarrow \infty$ (e.g.\ with zero scattering, they reduce correctly to ballistic transport along field lines).

\begin{figure*}
	\centering
	\includegraphics[width=0.9\textwidth]{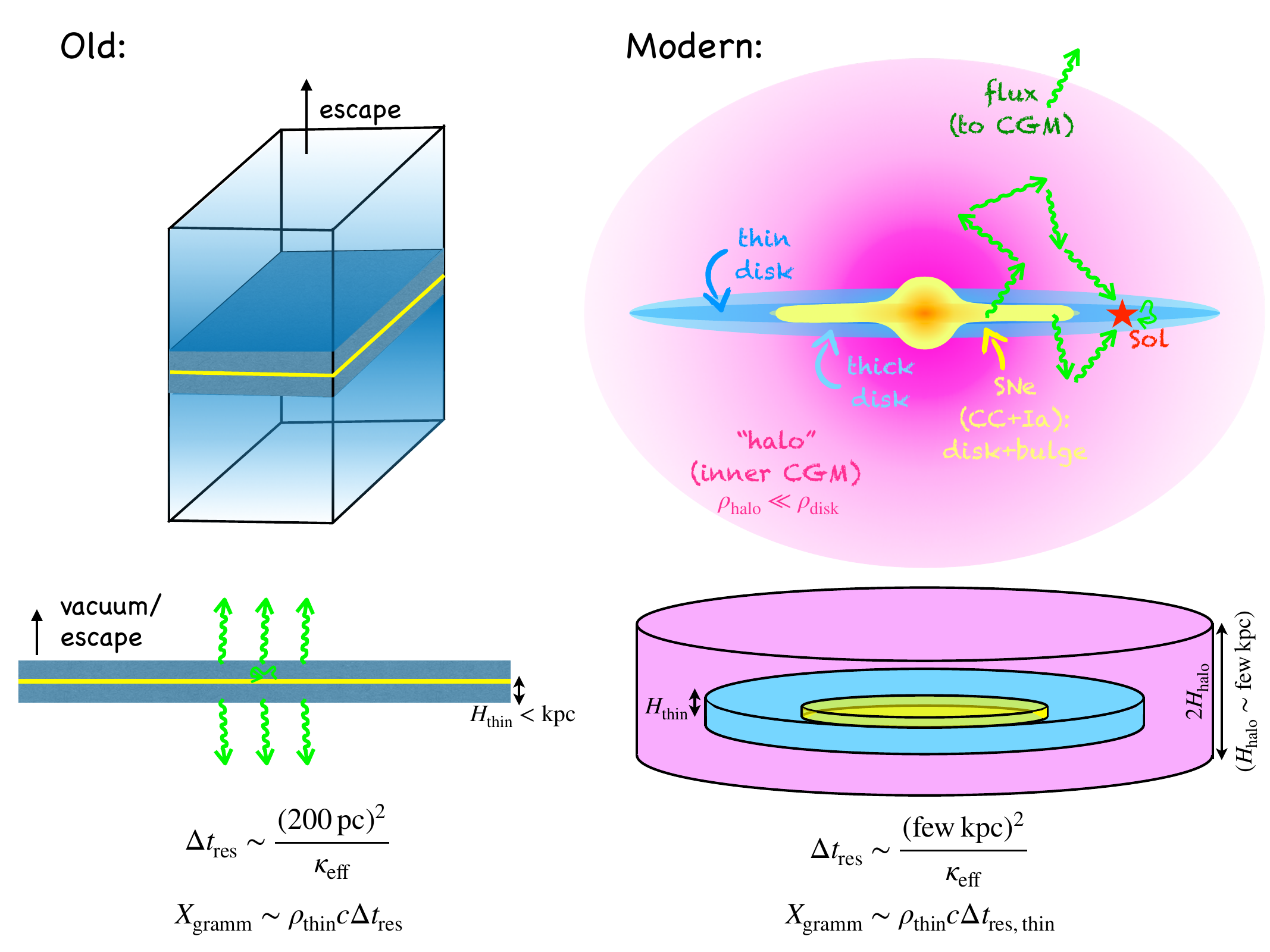}
	\caption{Galactic CR transport model setups: older (nested leaky or shear-periodic boxes; {\em left}) vs.\ modern (global CR-MHD or flat-halo-diffusion; {\em right}). 
	{\em Left:} Boxes assume a plane-parallel source layer in the midplane of a thin disk (height $H_{\rm thin} \sim 200\,$pc), and upper boundary above which CRs are ``removed.'' CR transport is effectively 1D along $\hat{z}$, and majority of CRs at the midplane in steady-state come from a region $\Delta x \sim H_{\rm thin}$ with residence time and grammage scaling as shown. This biases $\kappa_{\rm eff}$ to much smaller values and cannot reproduce the observed $\Delta t_{\rm res}$ and $X_{\rm gramm}$ simultaneously.
	{\em Right:} Modern global, 3D models use either full-Galaxy modeling (simulation or empirical) of bulge+thin+thick stellar+gas disks and halo, or simplified embedded cylindrical volumes with source density a function of $R,\,z,\,\phi$ (tracing observed SNe rates) and flux-steady boundaries at the (lower-density) CR scattering halo size $H_{\rm halo} \sim $\,few kpc. Here $\Delta t_{\rm res} \sim ({\rm few\,kpc})^{2}/\kappa_{\rm eff}$ as CRs scatter through halo to the LISM, while $X_{\rm gramm}$ is dominated by thin-disk residence time.
	\label{fig:transport.old.vs.new}}
\end{figure*}

\subsection{Macro-Scale} 
\label{sec:transport.macro}

On macro scales $\ell \gg \ell_{\rm meso} \sim \ell_{\rm mfp,\,\|} \sim c/\bar{\nu}$, it is common to simplify Eq.~\ref{eqn:twomoment} by assuming the flux equation has reached equilibrium everywhere ($D_{t} \bar{f}_{1} \rightarrow 0$, equivalent to taking $\bar{\nu}\rightarrow \infty$ and $v_{\rm cr} \sim c \rightarrow \infty$), and that the CRs are very close to isotropic, so $|\langle \mu \rangle| \rightarrow 0$ and $\langle \mu^{2} \rangle \rightarrow 1/3$, $\mathbb{D} \rightarrow \mathbb{I}/3$, which then gives a single equation for $\bar{f}_{0}$ or $e_{\rm cr}$:
\begin{align}
\label{eqn:onemoment} \frac{\partial \bar{f}_{0}}{\partial t} &\rightarrow   j_{0} + \nabla \cdot  \left[ \kappa_{\|} \bhat \nabla_{\|} \bar{f}_{0} - {\bf v}_{e} \bar{f}_{0} \right]
 + 
\\
\nonumber &  
\frac{1}{p_{\rm cr}^{2}}\frac{\partial }{\partial p_{\rm cr}}\left[ p_{\rm cr}^{3} \left( \mathcal{R} 
+ \frac{\nabla \cdot {\bf v}_{e}}{3} \right) \bar{f}_{0} 
+ \frac{(v_{A}^{2} - \bar{v}_{A}^{2})\,p_{\rm cr}^{4}}{9 \kappa_{\|}}   \frac{\partial \bar{f}_{0}}{\partial {p_{\rm cr}}}
\right] \ ,  \\
\nonumber \frac{\partial e_{\rm cr}}{\partial t} &\rightarrow   S_{\rm net} + \nabla \cdot  \left[ \hat{\kappa}_{\|}\bhat \nabla_{\|} e_{\rm cr} - {\bf v}_{e} e_{\rm cr} \right] \\ 
\nonumber & \ \ \ \ \ \ \ \ \ - P\, \nabla \cdot {\bf v}_{e} + \frac{v_{A}^{2} - \bar{v}_{A}^{2}}{3 \hat{\kappa}_{\|}}\,(e_{\rm cr}+P_{\rm cr})  \ , \\ 
\nonumber {\bf v}_{e} &\equiv \bar{v}_{A} \bhat + {\bf u} \  \ \ \ , \ \ \  \ \kappa_{\|} \equiv \frac{v_{\rm cr} ^{2}}{3\bar{\nu}} \ .
\end{align}
This gives the classical ``convective'' (${\bf u}$), ``streaming'' ($\bar{v}_{A} \bhat$), and ``diffusive'' ($\kappa_{\|}$)-like flux terms , as well as the  ``adiabatic'' ($\nabla \cdot {\bf u}$), ``streaming loss'' ($\nabla \cdot \bar{v}_{A}\bhat$), and ``turbulent reacceleration'' ($(v_{A}^{2}-\bar{v}_{A}^{2})/9\kappa_{\|}$) loss/gain terms. 

On these scales the streaming $\bar{v}_{A}$ (imbalance of $\nu_{\pm}$) is taken to point down the direction of the CR number gradient, $\bar{v}_{A} \rightarrow -f_{\rm st}\,v_{A}\,\nabla_{\|} \bar{f}_{0}/|\nabla_{\|} \bar{f}_{0}|$ (with $0 \le f_{\rm st} \equiv |\nu_{+}-\nu_{-}|/|\nu_{+}+\nu_{-}| \le 1$); this plus its pointing along $\bhat$ mean that it can be mathematically added to an effective diffusion coefficient, and conversely $\kappa_{\|}$ can be added to an effective streaming speed: 
\begin{align}
\kappa_{\|,\,\rm eff} &\rightarrow \kappa_{\|} + (f_{\rm st} v_{A} +u_{\|})\,\ell_{f,\,\nabla} \\ 
v_{\rm st,\,eff} &\rightarrow \left( f_{\rm st} v_{A} + u_{\|} + \kappa_{\|}/\ell_{f,\,\nabla}  \right)\,\nabla_{\|} \bar{f}_{0}/|\nabla_{\|} \bar{f}_{0}|  
\end{align}
with $\ell_{f,\,\nabla} \equiv \bar{f}_{0}/|\nabla_{\|} \bar{f}_{0}|$.

We stress Eq.~\ref{eqn:onemoment} is still quite distinct from the Fokker-Planck equation usually referred to in older CR studies and semi-analytic equilibrium codes like GALPROP, DRAGON, PICARD, USINE and others (discussed below). To obtain that equation from Eq.~\ref{eqn:onemoment}, we must further (1) drop all streaming terms/corrections ($\bar{v}_{A} \rightarrow 0$), (2) replace parallel diffusion with isotropic diffusion, so $\nabla \cdot (\bhat \kappa_{\|} \nabla_{\|} \bar{f}_{1d}) \rightarrow \nabla (D_{xx} \nabla \bar{f}_{1d})$ with an assumed (isotropically tangled on all scales) $D_{xx} \sim \kappa_{\|}/3$, and (3) assume steady-state ($\partial \bar{f}_{0} / \partial t \rightarrow 0$). On top of this usually (4) any small scale (e.g.\ local turbulent) structure in ${\bf u}$ or ${\bf B}$ or loss terms $\mathcal{R}$ (which depend on $\rho$, $|{\bf B}|$, $e_{\rm rad}$, and neutral and ionized fractions) are neglected, and (5) terms like $v_{A}$ (in e.g.\ the turbulent reacceleration) and $\langle \nabla \cdot {\bf u}\rangle$ are either dropped or assumed to be universal constants.

While useful for e.g.\ intuition-building, and analytic steady-state models, many studies have shown that Eq.~\ref{eqn:onemoment} is \textit{almost never} appropriate for use in dynamical applications (e.g.\ CR-MHD numerical simulations), for many reasons (both physical and numerical). This includes: (1) it breaks down, and can give unphysical answers, if there are gradients in ${\bf B}$, or $e_{\rm cr}$, or $\bar{\nu}$, or $v_{A}$, on scales comparable to the CR scattering mean-free-path, which can exceed $>100\,$pc for $\sim$\,TeV CRs; (2) it fails very near sources (distance $r \rightarrow 0$), for similar reasons; (3) it fails far from sources ($r\rightarrow \infty$), by failing to account correctly for finite-travel-time effects; (4) it features infinite signal speeds; (5) it fails on \textit{any} scale (giving incorrect or unphysical or undefined/non-unique solutions) if there are certain types of un-resolved discontinuities (e.g.\ shocks), or if $\nabla_{\|} \bar{f}_{0} = \bhat \cdot \nabla \bar{f}_{0}$ (or $\nabla_{\|} P_{\rm cr}$) changes sign, which happens ubiquitously because it occurs not just at extrema of $\bar{f}_{0}$ or $P_{\rm cr}$ but anywhere that $\bhat$ changes direction;\footnote{A simple example is if $\bhat$ changes direction on small scales in a background with a large-scale coherent $\nabla P_{\rm cr}$, causing $\nabla \cdot {\bf v}_{e}$ to diverge; other examples in even the simplest pure-streaming single-direction limit are discussed in \citet{sharma.2010:cosmic.ray.streaming.timestepping,chan:2018.cosmicray.fire.gammaray,thomas.pfrommer.18:alfven.reg.cr.transport} and in shocks in \citet{gupta:2021.cosmic.ray.numerical.nonuniqueness.with.streaming.only}.} (6) like the infamous telegraph equation, it has (for certain initial/boundary conditions) solutions with negative CR number and/or energy \citep{hopkins:m1.cr.closure,thomas:2021.compare.cr.closures.from.prev.papers}. All of these are resolved by using the correct two-moment equations (Eq.~\ref{eqn:twomoment}) instead. And in most CR-MHD simulations, these two-moment methods are actually numerically cheaper to evolve than single-moment methods. Hence the majority of recent macro-scale simulations have adopted either the spectrally-integrated two-moment Eq.~\ref{eqn:energy}, or full-spectrum Eq.~\ref{eqn:twomoment} \citep[references above and e.g.][]{ponnada:2024.fire.fir.radio.from.crs.constraints.on.outliers.and.transport,ponnada:2023.fire.synchrotron.profiles,ponnada:2023.synch.signatures.of.cr.transport.models.fire,fitzaxen:2024.cr.transport.into.gmcs.suppressed.starforge,su:2025.crs.at.shock.fronts.from.jets.injection}.

\section{Modern Galactic CR Transport: The Fundamental Need for Global Models with Halos}
\label{sec:obs.lism}

\begin{figure*}
	\centering
	\includegraphics[width=0.8\textwidth]{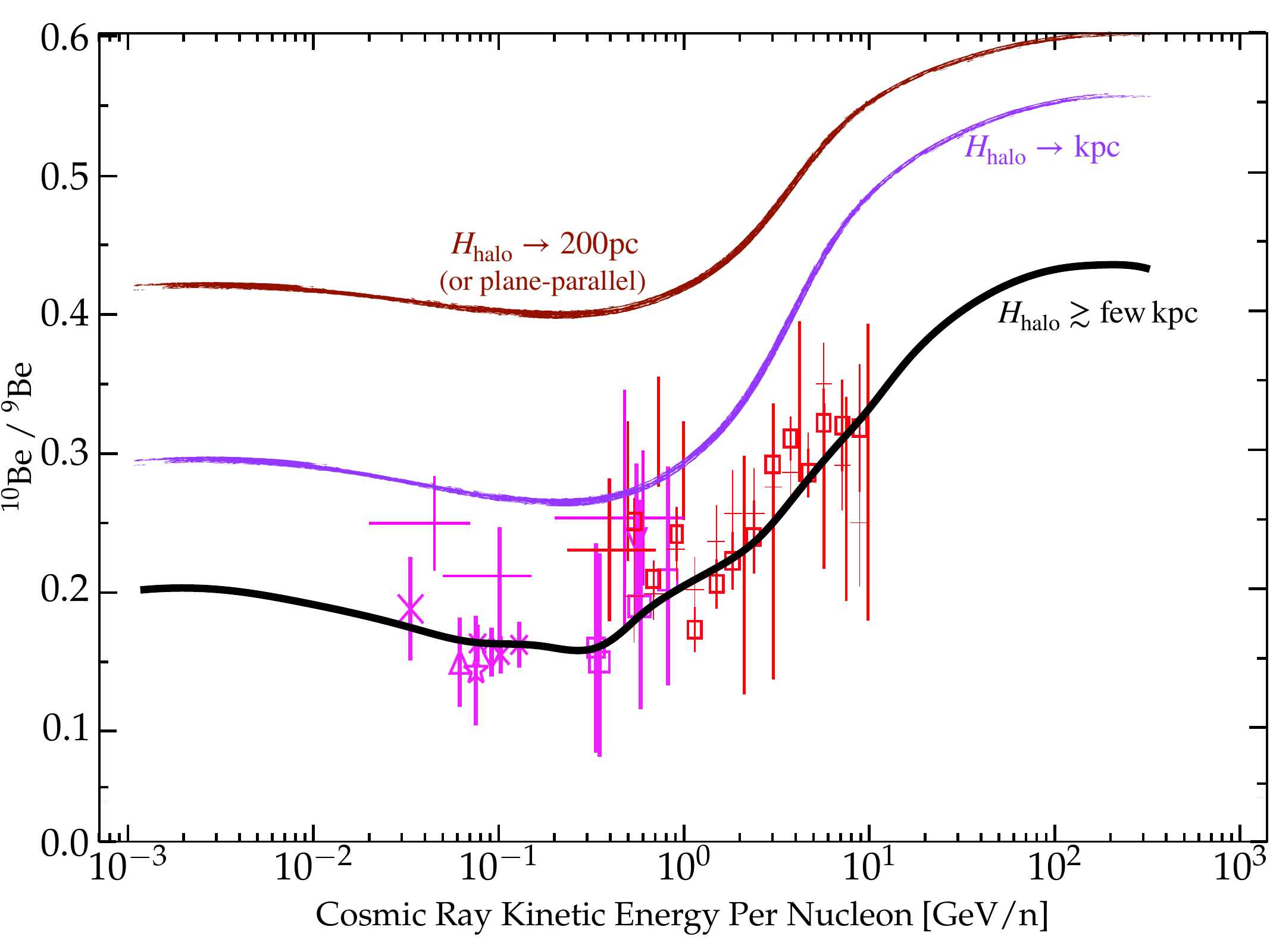}
	\caption{Illustration of the LISM radioactive-to-stable ratios ($^{10}$Be/$^{9}$Be here) dependence on CR scattering halo size. Each line is a model from static CR transport code GALPROP (other codes give very similar results; Table~\ref{tbl:spectral.fits}), marginalizing over all CR transport parameters, loss rate parameters, and free parameters like streaming/advection speeds, source rate distributions and source spectra (within tolerance allowed by direct SNe observations), magnetic field strength, etc., fitting to all CR spectra {\em except} Be. The models are then used to predict $^{10}$Be/$^{9}$Be. 
	Without an extended CR scattering halo, or in plane-parallel or leaky box (non-global) models, fitting the grammage $X \sim n_{\rm gas}\,c \Delta t_{\rm res}$ requires production of $^{10}$Be continuously alongside $^{9}$Be within the disk between source and LISM, predicting much-too-large $^{10}$Be/$^{9}$Be. In global models with a halo, secondaries produced in the disk at $R < R_{\odot}$ take longer excursions into the halo on their way to the LISM, decaying away the $^{10}$Be while preserving B/C. So long as the halo size is $\gtrsim$\,few kpc and the model is global, the prediction becomes insensitive to its size and agrees well with measurements.
	\label{fig:be.halo}}
\end{figure*}

\begin{figure}
	\centering
	\includegraphics[width=0.98\columnwidth]{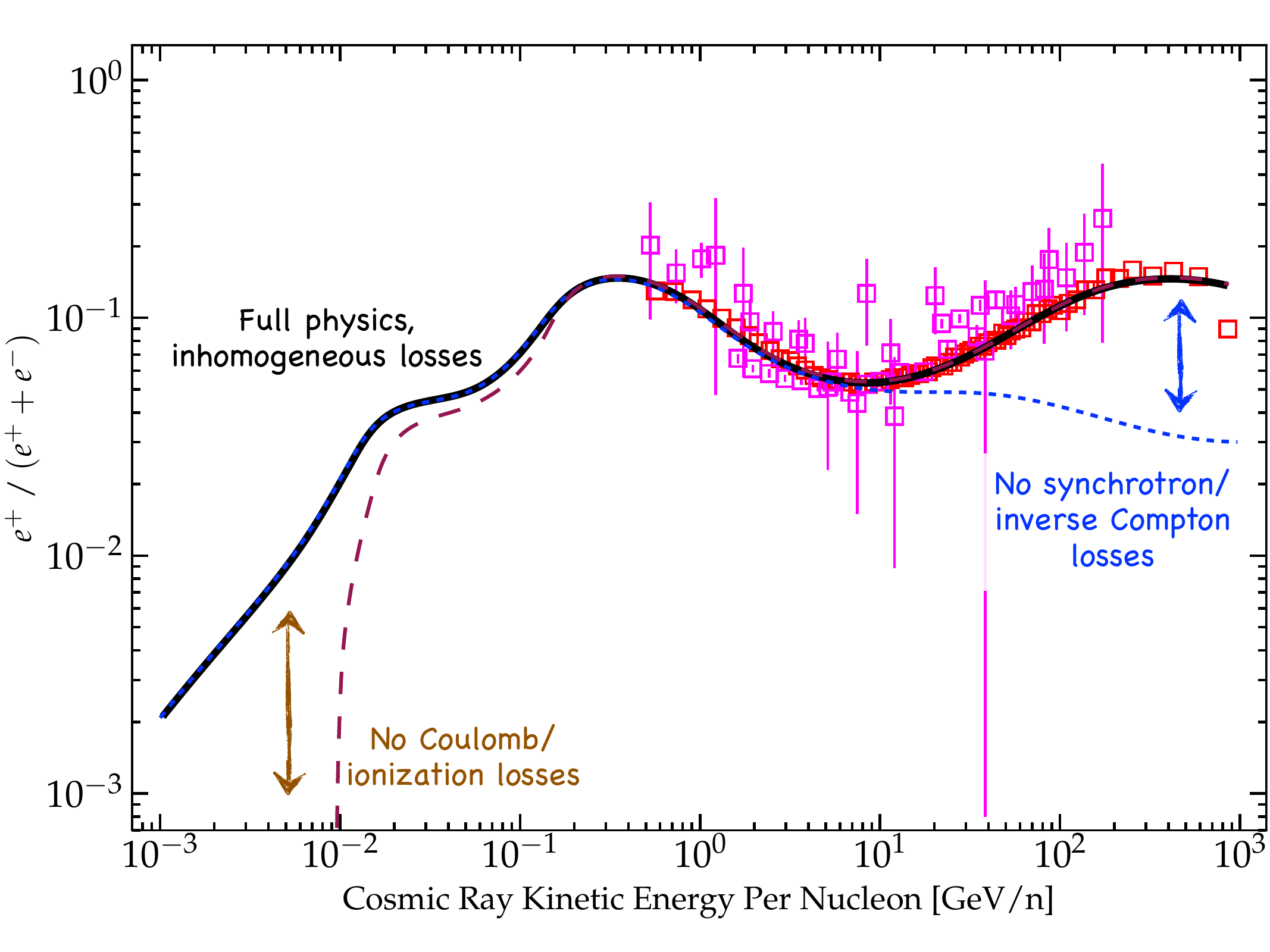}
	\caption{Illustration of sensitivity of CR spectra to losses, here using positron-to-electron ratios. Because $e^{+}$ are produced as secondaries in flight by CR proton interactions, they have shorter lifetimes/residence times before arrival at the LISM, hence are less suppressed by radiative losses, so ignoring those losses preferentially increases $e^{-}$ and lowers $e^{+}/(e^{-}+e^{+})$. The salient loss terms are synchrotron+inverse Compton (these scale identically with CR properties here) at high lepton energies $\gg 10\,$GeV, and Coulomb+ionization interactions at low energies $\ll 0.1\,$GeV.
	\label{fig:epos.losses}}
\end{figure}

\begin{footnotesize}
\ctable[caption={{\normalsize Fits to LISM CR Data\vspace{-0.2cm}}\label{tbl:spectral.fits}},center,star]{| l l c c c c r |}{\tnote{Recent fits to Galactic+LISM+Solar system CR datasets. (1) Study. (2) Code/method: GALPROP/DRAGON2/PICARD solve steady-state Fokker-Planck in a 3D global galaxy model; USINE/Custom 1D/Weighted Slab use semi-analytic 1D transfer with halo+corrections for global terms.; GIZMO uses 3D CR-MHD dynamics in an evolving galaxy (includes small-scale turbulent ${\bf u}$, ${\bf B}$, \&\ streaming $\bar{v}_{A}$). (3) Estimated injection slope at $\gtrsim$\,GV, $dN_{\rm cr}/d {p}_{\rm cr} \propto p_{\rm cr}^{-\alpha_{\rm inj}}$ (best-fits for set of models considered; $^{\ast}$ values fixed from previous studies). (4) Effective spectrally-integrated, CR path-integrated parallel diffusivity $\kappa_{\rm eff,\,\|}$, in units of $10^{29}\,{\rm cm^{2}\,s^{-1}}$. (5) Effective $\delta$, for $\Delta t_{\rm escape} \propto \kappa_{\rm eff}^{-1} \propto R_{\rm cr}^{-\langle \delta \rangle}$ from $\sim 1-100$\,GV. (6) CR scattering halo ``size'' above which CRs free-escape, in kpc. (7) Additional notes.}
}{
\hline
Reference & Code & $\alpha_{\rm inj}$ & $\kappa_{\rm eff,\,\|}^{29}$ & $\langle \delta \rangle$ & $H_{\rm halo}$ & Notes \\
\hline
\citet{dimauro:2023.cr.diff.constraints.updated.galprop.very.similar.our.models.but.lots.of.interp.re.selfconfinement.that.doesnt.mathematically.work} & GALPROP & 1.8-2.4 & 1-14 & 0.4-0.7 & 4.2$^{\ast}$ & $\kappa$ varies with height $z$ \\ 
\citet{silver:2024.cr.propagation.low.energies.new.data} & GALPROP & 2.00-2.37 & 1.5-3 & 0.4-0.5 & 4$^{\ast}$ & Varying injection+losses below GV \\ 
\citet{korsmeier:2022.cr.fitting.update.ams02} & GALPROP & 2.3-2.4 & 3.3-4.3 & 0.44-0.5 & 4$^{\ast}$ & Non-negligible convective term \\ 
\citet{zhao:2021.spatially.dependent.cr.propagation.disk.halo.models} & GALPROP &  2.3-2.36 & 1-1.5 & 0.6-0.7 & 8 (4.8-11.4) & Different disk/halo $\kappa$ allowed \\
\citet{korsmeier:2021.light.element.requires.halo.but.upper.limit.unconfined} & GALPROP & 2.1-2.4 & 1.2-3 & 0.43-0.55 & 7 (4-9) & Full (LiBeBCNO) fits \\ 
\citet{delaTorre:2021.dragon2.methods.new.model.comparison} & GALPROP & 2.0-2.4 & 2.0 & 0.44 & 6.9 & Fits with Be \\
\hline
\citet{delatorre.luque:2024.gas.models.of.galaxy.key.for.scale.height.but.need.halo.for.crs} & DRAGON2 & 2.0-2.4 & 3.1 & 0.5 & 8 (6-12) & Spiral arms (3D) $\rightarrow$ larger $\kappa$, $H_{\rm halo}$ \\
\citet{tovar:2024.inhomogeneous.diffusion.cr.spectra} & DRAGON2 & 2.0-2.4 & 2.6-2.9 & 0.49-0.53 & 4.72$^{\ast}$ & (Weak) $\kappa$ variation with $R_{\rm gal}$ \\ 
\citet{delaTorre:2021.dragon2.methods.new.model.comparison} & DRAGON2 & 2.0-2.4 & 2.3 & 0.42 & 6.8 & Fits with Be \\
\citet{delatorreluque:2021.proton.antiproton.cr.modeling} & DRAGON2 & 2.0-2.4 & 2.2 & 0.43-0.55 & 7.1 & Sub-GV inversions in $\alpha_{\rm inj}$, $\kappa$ allowed \\ 
\hline
\citet{ramirez:2024.3d.struct.galaxy.gas.influences.cr.fitting.params.vs.2d} & PICARD & 1.9-2.4 & 1.5 & 0.4 & 4$^{\ast}$ & 3D gas models important at low-$E_{\rm cr}$ \\
\hline
\citet{hopkins:cr.multibin.mw.comparison} & GIZMO & 2.0-2.4 & 2-7 & 0.5-0.7 & $\infty$ & CR-MHD cosmological dynamics \\ 
\hline
\citet{weinrich:2020.usine.halo.size.modeling} & USINE & 2.3$^{\ast}$ & 1.2-2.4 & 0.45-0.55 & 4-17 & Favored $H_{\rm halo} \sim 6$ depends on dataset \\ 
\citet{weinrich:2020.usine.combined.analysis.cr.spectra} & USINE & 1.8-2.5 & 0.8-2 & 0.4-0.6 & 5$^{\ast}$ & Injection fitted by species \\
\citet{derome:2019.usine.bc.mock.data.cr.comparisons} & USINE & 2.3$^{\ast}$ & 1.8-2.3 & 0.45-0.65 & 10$^{\ast}$ & Varying cross-sections, production \\
\hline
\citet{recchia:2024.cr.spectral.modeling.features.bumps.wiggles} & Custom 1D & 2.35$^{\ast}$ & 2.3$^{\ast}$ & 0.7$^{\ast}$ & 4$^{\ast}$ & Inhomogeneous $\kappa$ in disk \\
\citet{jacobs:2023.isotope.ratios.large.halo.size.required.constrain.volume.of.inhomogenous.slower.diffusion.zones.in.model} & Custom 1D & 2.3$^{\ast}$  & 2-4  & 0.4-0.7 & 7.5 (6.4-8.5) & Non-uniform $\kappa$ important \\ 
\citet{evoli:2020.weighted.slab.model.cr.fits.halo.size} & Weighted Slab & 2.3 & 2.2-4.5 & 0.54 & 9 (6-12) & Simplified analytic model, Be fit \\
\hline
}
\end{footnotesize}

\begin{figure}
	\centering
	\includegraphics[width=\columnwidth]{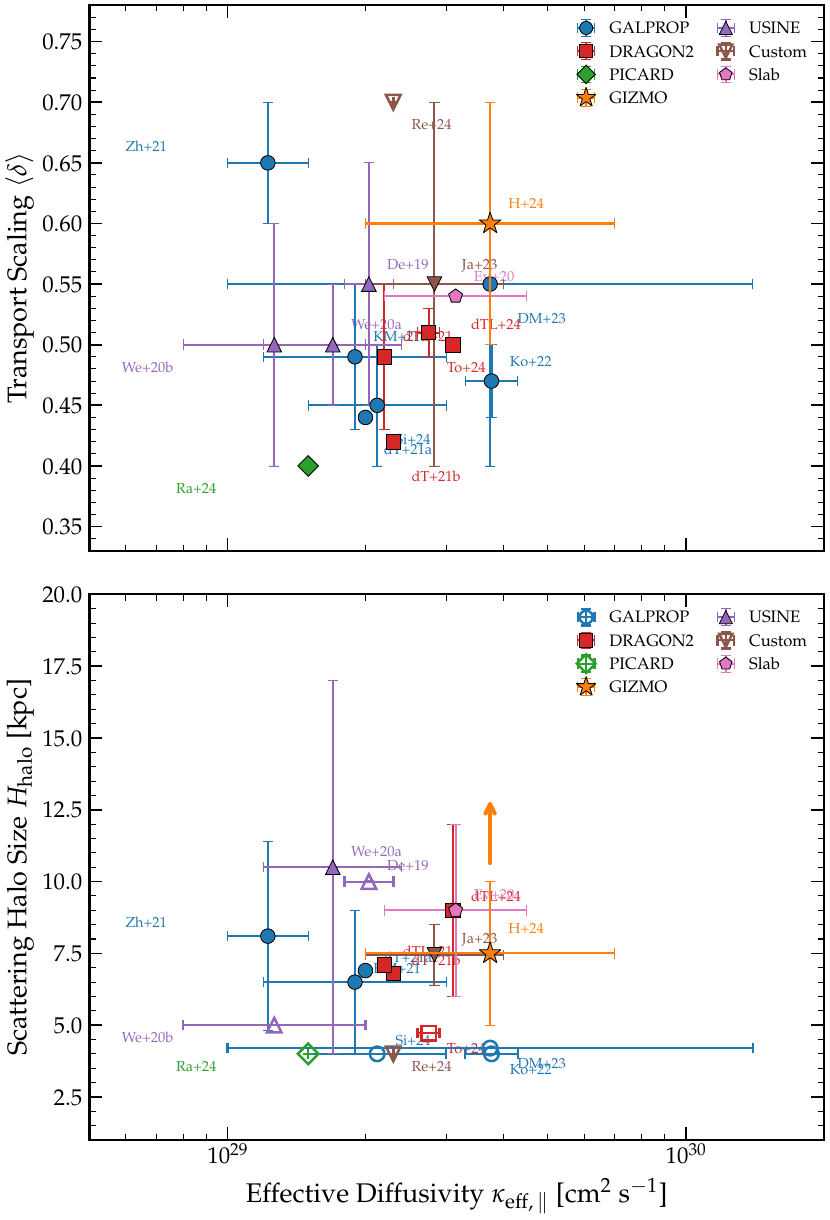}
	\caption{Visual summary of the spectral fit parameters from Table~\ref{tbl:spectral.fits}, in the $\kappa_{\rm eff}$--$\delta$ plane ({\em top}) and $\kappa_{\rm eff}$--$H_{\rm halo}$ plane ({\em bottom}). Points are color-coded by code/method. Hollow symbols indicate values that were held fixed (not fit) in that study. The upward arrow on the GIZMO point denotes the effectively infinite halo in the CR-MHD cosmological simulation. Despite a range of codes and methods, the models broadly converge on $\kappa_{\rm eff,\,\|} \sim 1$--$5 \times 10^{29}\,{\rm cm^{2}\,s^{-1}}$, $\delta \sim 0.4$--$0.7$, and $H_{\rm halo} \sim 4$--$12\,$kpc.
	\label{fig:spectral.fits.kappa.delta}}
\end{figure}

\subsection{What Do We Measure?}
\label{sec:obs.spectra}

In Galactic CR transport models (illustrated in Fig.~\ref{fig:transport.old.vs.new}), one assumes some injection spectrum and energy at sources (e.g.\ SNe), then integrates Eq.~\ref{eqn:twomoment} or some simplification thereof to predict the CR spectra of different species observed in the Solar neighborhood (e.g.\ Voyager, AMS-02, PAMELA, etc., with the unique advantage that Voyager does not need to correct at the same level for modulation of low-energy CRs by the heliosphere). 

Fig.~\ref{fig:specdemo} shows examples of these CR spectra from MeV-TeV, and the exceptional data constraining most of this dynamic range for just a subset of highly-relevant species. It also heuristically shows which parts of the spectra of different species contribute to different physical processes and observables. Table~\ref{tbl:summary} gives a high-level summary of what we learn from different CR spectra, species, and ratios, which we expand upon briefly here and in Figs.~\ref{fig:bc.be.epos}, \ref{fig:transport.old.vs.new}, \ref{fig:be.halo}, \&\ \ref{fig:epos.losses}. (also Table~\ref{tbl:spectral.fits}). It is helpful to illustrate this qualitatively by imagining that the steady-state spectrum of some species $s$ given by $dN_{s}/d T^{\rm kin}_{s} \sim (dN_{s}/d T^{\rm kin}_{s})_{\rm inj} \, \Delta t_{\rm res}(s,\,T^{\rm kin}_{s},\,...)$, where $(dN_{s}/d T^{\rm kin}_{s})_{\rm inj}$ is some effective injection spectrum from sources (taken to e.g.\ follow the observed supernovae rate in the Galaxy, or supernovae plus massive/fast OB winds), and $\Delta t_{\rm res}$ is an effective residence time, which itself is the shorter of either the timescale for CRs to lose their energy after injection ($\Delta t_{\rm loss} \sim |\mathcal{R}|^{-1}$, in Eq.~\ref{eqn:twomoment}), or escape from the ISM/Galaxy ($\Delta t_{\rm escape} \sim \ell_{\rm galaxy}^{2} / \kappa_{\rm eff}$ or $\sim \ell_{\rm galaxy} / v_{\rm st,\,eff}$ in terms of some effective diffusivity or streaming/advective speed). This is most definitely an over-simplification, but is useful to inform our intuition.

The direct CR spectra (e.g.\ Fig.~\ref{fig:specdemo}) of primary species (those produced primarily in these initial acceleration regions, like H, He, C, N, O, Fe, $e^{-}$) constrain this product. At a given rigidity (same rigidity, hence same gyro-resonant scattering physics), CRs should scatter identically so these should trace the same injection spectrum, but the degeneracy between $(dN_{s}/d T^{\rm kin}_{s})_{\rm inj}$ and $\Delta t_{\rm res}$ is partially broken because of strong losses (so $\Delta t_{\rm res} \rightarrow \Delta t_{\rm loss}$ instead of $\Delta t_{\rm escape}$) at e.g.\ kinetic energies $T_{\rm kin}/{\rm nuc} \lesssim 0.1\,$GeV/nuc (Coulomb/ionization losses in ionized/neutral gas scale as $\Delta t_{\rm loss} \propto T_{\rm kin}^{3/2}$ for hadrons and $\propto T_{\rm kin}$ for leptons), or $\gtrsim 10\,$GeV for leptons (synchrotron/inverse Compton $\Delta t_{\rm loss} \propto T_{\rm kin}^{-1}$). Where not loss-dominated, the ratios of these species imply that injection spectra are very close to universal in their momentum-space shapes, and the abundance ratios strongly constrain where and when CRs are accelerated (including what types of SNe could produce them, given their different intrinsic yields/ratios, and how much ISM gas must be swept up compared to pure ejecta, when the CRs are first accelerated; see e.g.\ \citealt{lingenfelter:2019.cr.sne.where.in.shock.accel.review}). 

Secondary species like B, Be, Li, $e^{+}$, $\bar{p}$, are produced primarily by reactions from other CRs in flight. These are observed in abundances (relative to e.g.\ CR H or CNO or $e^{-}$) vastly larger than present in SNe ejecta or the ISM, but can be produced by spallation reactions from collisions of intermediate/heavy nuclei with ISM nuclei, and/or pionic reactions with CR H. The production of these, and in particular various secondary-to-primary ratios (a couple shown in Fig.~\ref{fig:bc.be.epos}), constrains the grammage $X_{\rm cr} = \int n_{\rm gas} d\ell_{\rm path} = \int n_{\rm gas} v_{\rm cr} dt \sim \langle n_{\rm gas}\,c\,\Delta t_{\rm res}\rangle$, i.e.\ the column of ISM traversed by the producing species since their acceleration. Most famous of these is B/C (the boron-to-carbon CR flux ratio, as a function of CR kinetic energy per nucleon; Fig.~\ref{fig:bc.be.epos}), which is useful as B is produced by primaries like C conserving kinetic energy per nucleon (so the injection spectral shapes and isotopic ratios have almost no effect on this ratio), and both suffer minimal losses at $\gtrsim 0.1\,$GeV/nuc. But also very useful is $e^{+}/e^{-}$ (Fig.~\ref{fig:bc.be.epos}), since this ratio is trivially independent of Solar modulation at all energies, and since $e^{+}$ comes from protons and high-energy leptons suffer losses (Fig.~\ref{fig:epos.losses}), this can break some degeneracies. Similarly $\bar{p}/p$ and other ratios (not shown in Fig.~\ref{fig:bc.be.epos}, but used in the detailed models cited in Table~\ref{tbl:spectral.fits}) provide additional information (e.g.\ $\bar{p}/p$ being more sensitive to injection slope). 

Isotopic ratios, in particular radioactive-to-stable isotopes like $^{10}$Be/$^{9}$Be (where $^{10}$Be has a half-life $\sim 1.4\,\gamma$\,Myr), are powerful complementary constraints on $\Delta t_{\rm res}$ {\em alone} (Fig.~\ref{fig:bc.be.epos} \&\ \ref{fig:be.halo}; though since all Be is secondary, these probe residence time {\em since production}), as compared to the degenerate grammage (which goes like density times residence time). They can thus probe CR traversal out of the disk into a the CGM/halo, where densities are lower (illustrated qualitatively in Fig.~\ref{fig:transport.old.vs.new}, and quantitatively in Fig.~\ref{fig:be.halo}). 

There are also indirect radiation signatures of CRs: ($\sim\,$GeV) $\gamma$-rays, ($\sim$MHz-GHz) synchrotron plus ($\sim 1-100$\,keV) inverse Compton, and molecular recombination lines excited by CR ionization (examples shown in Fig.~\ref{fig:xgal.obs} \&\ summarized in Table~\ref{tbl:nonlocal.obs}). The smooth, galaxy-scale constraints from these are reviewed below, but for the Milky Way, they are not as constraining on their own owing to various degeneracies, but follow from the LISM constraints. For example, since the same pionic interactions produce most of the $\gamma$-rays and $e^{+}$, $\bar{p}$, getting the CR spectra ``correct'' in the LISM almost ensures the $\gtrsim$\,kpc Galactic {\em diffuse} $\gamma$-ray emissivity comes out correctly (see references in Table~\ref{tbl:spectral.fits} or review in \citealt{kronecki:2022.cosmic.ray.gamma.ray.review.not.calorimeters}). Similar arguments apply to inverse Compton and ionization, and for synchrotron uncertainties in the small-scale structure of ${\bf B}$ dominate the predictions (while for ionization the dominant uncertainty in the most-dense gas is the role of local low-energy CR sources like protostellar jet shocks).

Important complementary information about injection (when and where CRs are accelerated as well as injection spectra) comes from detailed modeling of individual SNRs as sources, in e.g.\ $\gamma$-rays or synchrotron or inverse Compton, from more detailed elemental and isotopic ratios of primary CRs (and from geochemical isotopic abundances), from CR isotropy and variability and detailed spectral structure of individual species, and from modeling the global Galactic SNe and other candidate source spatial distributions (e.g.\ source distance distribution from Galactic center). Energy-dependent CR anisotropy measurements also provide complementary discriminating power for transport models, as the predicted anisotropy amplitude and phase depend sensitively on the assumed scattering rate and local source distribution \citep[see e.g.][]{2018AdSpR..62.2731A,evoli:2020.weighted.slab.model.cr.fits.halo.size}.

\subsection{Key Lessons}
\label{sec:lessons}

A few qualitative conclusions emerge from this robustly, across many dozens of studies by different groups, using different codes and qualitatively different methods. These include e.g.\ time-static Fokker-Planck codes like GALPROP, DRAGON, or PICARD, semi-analytic models like USINE, modeling $\gamma$-ray and synchrotron spectra with CR priors, and dynamical spectrally-resolved CR transport models with codes like GIZMO. An attempt at a summary of state-of-the-art low-energy CR full-spectrum modeling results from these methods in the last few years is provided in Table~\ref{tbl:spectral.fits} (shown graphically in Fig.~\ref{fig:spectral.fits.kappa.delta}) and some are illustrated in Fig.~\ref{fig:transport.old.vs.new}.

\begin{itemize}
\item Most of the LISM CRs are accelerated in SNe shocks, in the early stages of shock development (entrained mass $\sim$ ejecta mass), from the Galactic radii where most SNe reside (the effective radius of young stars or SNRs, $\sim 5$\,kpc from Galactic center), with a quasi-universal momentum-space acceleration spectrum slightly steeper than $dN/d^{3}{\bf p}_{\rm cr} \propto p^{-4}$.

\item The residence time for high-energy ($\gtrsim 10\,$GeV) leptons and low-energy ($\lesssim 0.1\,$GeV) leptons and hadrons is governed by losses, namely synchrotron+inverse Compton and Coulomb+ionization, respectively. For the latter, it is crucial to include models of the ionized and neutral gas distribution. For $\gtrsim 0.1\,$GeV hadrons, the residence time is limited by escape (see references in Table~\ref{tbl:spectral.fits}), and in this regime it must decrease with energy/rigidity as $\Delta t_{\rm res} \propto E_{\rm cr}^{-\delta}$ with $\delta \sim 0.5$. For intermediate-energy ($\sim 0.1-10$\,GeV) leptons, escape likewise dominates over losses --- this follows from the full spectral fits in Table~\ref{tbl:spectral.fits} (the $e^{+}/e^{-}$ ratio in particular requires escape-dominated transport at these energies) --- though the transition from escape- to loss-dominated is more gradual for leptons than hadrons, as synchrotron+IC losses are non-negligible even at a few GeV.

\item Fundamentally, $\langle n_{\rm gas} \Delta t_{\rm res} \rangle \ll \langle n_{\rm gas,\,ISM} \rangle\, \langle \Delta t_{\rm res} \rangle$, so CRs must spend significant time outside of the disk in an extended ``CR scattering halo'' of low density (not accumulating much grammage) and size $\gtrsim$\,a few kpc, in order to simultaneously fit the data above. Accounting for this, inferred energy-averaged CR scattering rates must be $\sim 10^{-9}\,{\rm s^{-1}}$ near the peak of the spectrum. 
\end{itemize}

\subsection{Why Halos and Global Models?}
\label{sec:halos}

Older models of CR transport (generally pre-2000s) often adopted the leaky-box or nested leaky-box formalism, illustrated in Fig.~\ref{fig:transport.old.vs.new} This assumes a plane-parallel ``disk'' (really a section of an infinite sheet), with CR sources in the midplane and a boundary at some height $\pm z$ in the vertical direction (for original leaky box models, the boundary is at the edge of the thin disk, while for nested leaky boxes, there can be vertical stratification of the gas density in levels before some cutoff at the edge of the box) and periodicity or shear-periodicity assumed in the $xy$ directions. This is, of course, the same setup as stratified periodic boxes or shearing box simulations of a portion of the ISM used in some CR-MHD simulations.

While useful for some intuition-building, it has been clearly demonstrated in the astro-particle community that such models \textit{cannot} correctly reproduce LISM CR observations at modern precision (see references in Table~\ref{tbl:spectral.fits} and reviews in \S~\ref{sec:intro}, as well as \citealt{jones:1990.cr.leaky.box.model.failures,maurin:2002.cr.modeling.usine.need.galaxy.model.with.global.features,taillet:2003.spatial.sources.cr.diffusion.galaxy,ptuskin:2006.cr.transport.galaxy.leaky.box.limitations,codino:2008.leaky.boxes.misleading,jaupart:2018.gal.center.crs.important,schroer:2025.nested.leaky.box.failures.cr.transport.models}). 
Dozens of reasons for this have been noted in the literature, but it is worth reviewing a few of the most important. 
(1) Per \S~\ref{sec:lessons}, most of the LISM CR energy comes from closer to the effective radius of the galaxy where most SNe reside. This is impossible in a model without a global Galactic structure: leaky/shearing boxes necessarily predict the LISM CRs come primarily from nearby SNe within a radius of order the thin disk height $H_{\rm thin} \sim 200\,$pc. That violates constraints on CR isotropy, $\gamma$-ray observations of individual SNRs/clusters and detailed CR spectral features, CR variability, historical (geochemical) CR isotopic constraints, CR spectral shapes, isotopic ratios, cross-correlations with Galactic source distribution and emission models, and residence times. Together these place robust upper limits showing $\ll 10\%$ (and at $\sim$\,GeV, $\lesssim 1\%$) of observed LISM CRs of almost any species below $<$\,TeV come from sources within $\sim 1\,$kpc of Sol \citep{genolini:2017.higher.order.cr.statistics.strong.constraints.against.local.sources.gives.source.number.whole.galaxy,mertsch:2018.local.cr.sources.introduce.breaks.potentially.interesting.above.tev.not.below,kachelries:2018.sne.isotopic.ratios.earth.constraints.local.cr.sources,evoli:2020.weighted.slab.model.cr.fits.halo.size,evoli:2021.constraints.local.individual.cr.sources.stochasticity.strong.upper.limits,phan:2021.voyager.variability.data.constrains.individual.sources.and.source.number.millions.sources.mixed.needed,bitter:2022.local.pulsar.constraints.positron.excess,stall:2025.cr.upper.limits.individual.sources.from.spectral.features}.
(2) The observed CR density ($\bar{f}_{0}$) scale-height is $\gtrsim$\,kpc; this means capturing the vertical structure/escape of CRs requires a vertical box of height $\gg$\,kpc. But that necessarily badly violates the assumptions of (shear)periodic boxes. The approximations required to expand the dynamical equations locally are only valid for structure on scales much smaller than the disk scale-height/length \citep[references in][]{regev:2008.shear.periodic.boundary.limitations}.
(3) Related to (2), it is well-known that one cannot accurately model wind/outflow dynamics in such boxes, because their geometry implies infinite escape velocities and prohibits sonic points and other critical phenomena \citep{martizzi:2016.stratified.boxes.fail.at.winds.because.of.geometry.and.sonic.points,fielding:sne.vs.galaxy.winds}, an effect made more extreme with a thicker CR gradient scale-height \citep{chan:2021.cosmic.ray.vertical.balance}. 
(4) Per \S~\ref{sec:obs.spectra}-\ref{sec:lessons}, any model must reconcile CR grammage and residence time, and simultaneously fit primary spectra, secondary-to-primary (e.g.\ B/C, $e^{+}/e^{-}$, $\bar{p}/p$) ratios, and radioactive-to-stable ($^{10}$Be/$^{9}$Be) along with other light-element ratios (Li, $^{3}$He, etc.), and fundamentally this requires the existence of a large CR scattering halo. The bare-minimum halo needed to fit these in modern codes has a height of $\gtrsim 4\,$kpc (references in Table~\ref{tbl:spectral.fits}), and the more recent fits with free halo sizes in Table~\ref{tbl:spectral.fits} favor values more like $\gtrsim 7$\,kpc. In either case, this is comparable to the radial scale-length of the Galactic disk and much larger than its scale-height, so global structure must be included to model any gradients on those scales and obtain the correct CR residence times for their grammage.
(5) Periodic (or shear-periodic) boundaries make the problem fundamentally one-dimensional in $z$ by definition, but the statistics of random walks, and therefore the statistics of CR residence time {\em distributions} (not just their mean) and diffusion or super-diffusion, relevant for the full spectra of different species, are fundamentally different in three versus one or two dimensions.\footnote{For example, in 3D as the ``allowed'' halo/box size $\rightarrow \infty$ (as in full-galaxy CR-MHD simulations), the ``effective'' CR scattering halo size always asymptotes to a constant of a few kpc (of order the disk size/distance from the Sun to Galactic center), because above this height, the probability of diffusing ``back into'' the disk drops rapidly. In 1D with an infinite domain, CRs revisit the origin (periodic plane) an infinite number of times, meaning their flight statistics become strongly dependent on the details of the vertical boundary condition adopted (references above and \citealt{liang:2025.leaky.boxes.levy.flights.modeling.crs.transport}).}

What happens if one tries to reproduce CR spectra with older phenomenological leaky-box-type models, with freely-fitted transport coefficients? Again, some observations are essentially impossible to reconcile (e.g.\ radioactive isotopes) without obviously unphysical assumptions or contradicting other CR spectra observed. If one exclusively compares to CR spectra and a single ratio like B/C (e.g.\ pre-Voyager/AMS/PAMELA datasets), then solutions are possible, but these are severely biased. Even in global models, if the halo is artificially truncated at some height $H_{\rm halo} \lesssim $\,a few kpc (i.e.\ CRs are simply escaped or removed, as is usually assumed in vertically-stratified boxes with outflow boundaries) , then there is a well-known degeneracy \citep[illustrated recently in e.g.][]{korsmeier:2021.light.element.requires.halo.but.upper.limit.unconfined} where one obtains $\nu \propto 1/H_{\rm halo}$, which arises because the smaller $H_{\rm halo}$ reduces the maximum travel-distance of individual CRs to Sol, so $\nu$ must be artificially increased (the effective diffusivity or streaming speed $\kappa$ or ${\bf v}_{e}$ decreased) in order for those CRs to accumulate enough grammage. The problem is even more severe if one assumes a plane-parallel model (with or without a halo), as the median effective travel distance becomes effectively $\sim H_{\rm thin}$ (see (1) above). It is also well known that this artificially biases the dependence of residence time on rigidity \citep[see above and][]{vladimirov:cr.highegy.diff,vecchi:2022.rigidity.dept.cr.transport.params}, suppressing $\delta$ to values as low as $\sim 0.2-0.3$. These artificial assumptions and limited data are why one sometimes sees effective diffusivities as low as $D_{xx} \sim 10^{27}-10^{28}\,{\rm cm^{2}\,s^{-1}}$ quoted for $\sim$\,GeV CRs in  older papers (and still some textbooks and reviews). However, all of the studies referenced above and others \citep[e.g.][]{zhao:2021.spatially.dependent.cr.propagation.disk.halo.models,korsmeier:2021.light.element.requires.halo.but.upper.limit.unconfined,delaTorre:2021.dragon2.methods.new.model.comparison,nozzoli:2021.be.residence.time.crs,delatorreluque:2021.proton.antiproton.cr.modeling,hopkins:cr.multibin.mw.comparison,delatorre.luque:2024.gas.models.of.galaxy.key.for.scale.height.but.need.halo.for.crs} show that as long as the allowed CR scattering volume is sufficiently large, and the calculation global, the effective scattering halo size converges to a few kpc and these numbers become independent of the cutoff size of the halo. This is simple geometry, as CRs random-walking in 3D from the inner galaxy only have a significant probability of returning to the disk midplane near Solar galactocentric radii $R_{\odot}$ at distances order-of-magnitude similar to $\sim R_{\odot} - R_{\rm source} \sim$\,few kpc.

We emphasize that the above arguments are specifically about using local/shearing boxes for precision calibration of CR transport parameters against LISM spectra. Local box simulations remain valuable tools for studying many CR-relevant processes --- e.g.\ CR-driven instabilities, interactions with the multiphase ISM, local turbulence effects on CR transport, and CR microphysics --- where the goal is not to directly calibrate against LISM grammage and residence time constraints.

\begin{table*}
\begin{center}
\begin{footnotesize}
\caption{\label{tbl:nonlocal.obs}Non-Local-ISM Observations (Constrains variable \checkmark; Does not constrain \xmark)}
\begin{tabular}{ lllllll }
\hline\hline
Data & CR $P_{\rm cr}$ & Gas $P_{\rm gas}$ & CR Energies & $\kappa_{\rm eff}$ & Phases Seen & Limitations \\
\hline\hline
UV/X-ray absorption & \xmark & \checkmark (thermal) & $\sim$\,GeV $p$ (indirect) & (indirect) & CGM & limited sightlines, non-CR physics \\ 
X-ray emission & \checkmark $P_{\rm cr}$ & \xmark & $0.1-1\,$GeV $e^{\pm}$ & \checkmark & CGM & sensitivity, thermal-CR separation \\
$\gamma$-rays & $\langle P_{\rm cr} n_{\rm gas} \rangle$ & $n_{\rm gas}$ & $\sim 10\,$GeV $p$ & (indirect) & warm ISM, inner halo & sensitivity, resolution, degeneracy  \\
synchrotron & $\langle P_{\rm cr} B^{2} \rangle$ & $B^{2}$ & $\gtrsim 30$\,GeV $e^{\pm}$ & \xmark & ISM, inner halo & degeneracy, clumping, energies probed \\
molecular lines & $\xi_{\rm ion}$ & $T_{\rm gas}$ & $<0.1$\,GeV $p$+$e$ & \xmark & GMCs & calibration, chemistry, local sources \\
 \hline 
\end{tabular}
\end{footnotesize}
\end{center}
\end{table*}

Somewhat ironically, the importance of global models and scattering halos for correct CR transport is widely appreciated in the astro-particle and Galactic CR literature, but is still not widely appreciated in the galaxy formation and extragalactic literature, despite the latter being the origin of our knowledge of Galactic structure as well as showing for decades that magnetized gas does indeed exist in halos around galaxies extending far beyond a few kpc (out to several hundred kpc, at least).
Indeed, it is important to note the different usage in the CR and galaxy fields of the word ``halo.'' In the astro-particle/CR literature, ``halo'' (or ``CR scattering halo,'' for specificity) refers to a structure closer to the thick disk or corona or disk-halo interface. Such CR scattering halos \textit{must} exist with sizes $\gtrsim$\,a few kpc, so long as gas exists within a few kpc of the disk with sufficient magnetization such that CR gyro-radii and either CR scattering mean-free-paths and/or magnetic field radii of curvature are smaller than $\lesssim 10-30\,$kpc (a few times the halo size). 
In the extragalactic literature, ``halo'' or ``CGM halo'' typically refers to the extended (and magnetized) CGM of gas+dark matter from the disk out to $>100\,$kpc. From the extragalactic point of view, given even the weakest conceivable (and observationally allowed) inner-``CGM halo'' magnetic field strengths, it is impossible to avoid a $\gtrsim$\,few kpc ``CR scattering halo.''

\subsection{Open Questions and The Role of CR-MHD Dynamical Models}
\label{sec:dyn.models}

While there are very robust conclusions highlighted above, there are still regimes  -- even in the LISM -- where models feature large degeneracies (e.g.\ at energies $\ll$\,GeV where losses are important and data more limited, such models often introduce breaks into the injection spectra and $\kappa(R_{\rm cr})$ and differ dramatically in how they fit these regimes), and many unsolved puzzles and apparent anomalies remain \citep[see reviews in][]{gabici:2019.cr.paradigm.challenges.mostly.well.explained.in.galaxy.sims,kachelries:2019.cr.microphysics.review.obs.broad.range.anomolies.local.sources}. These are where dynamical CR-MHD models can make important contributions to our understanding of Galactic CR transport. 

Most of these anomalies, for example, are defined by discrepancies between the best-fit achievable with models like GALPROP/DRAGON/PICARD/USINE and CR spectra, which are significant in a $\chi^{2}$ sense but often still small in an ``absolute'' sense.
To give a specific example: perhaps the most well-known CR anomaly in the galaxy formation community is the Galactic center ``excess.'' This is defined by a $\sim 1\%$ excess of $\gamma$-ray emission between $\sim 1-10\,$GeV, above template fitting/extrapolation from known sources with models like GALPROP, within the central single Fermi beam ($\sim 10^{\circ}$) of the Galactic center \citep{hooper:2011.galactic.center.excess,calore:interp.galactic.center.excess,ackermann:2017.fermi.galactic.center.excess,dimauro:2021.galactic.center.excess.fermi.modeling}. 
Similarly the ``positron excess'' is a $\lesssim 10\%$-level effect (though it can rise to factor $\sim2$, depending on how synchrotron/inverse-Compton losses are modeled at $\sim 100-1000\,$GV; compare \citealt{evoli:2021.spatially.variable.cr.losses.key.to.positron.problems,hopkins:cr.multibin.mw.comparison,rocamora:2024.spatial.variability.in.inverse.compton.important.for.positron.excess}). And ``cocoons'' and reduced acceleration zones near pulsars represent a $\sim 5\%$ grammage effect at energies $\ll$\,TeV (significant at $\mathcal{O}(1)$ only in small spatial regions around pulsars, for CR energies between $\sim 20-100\,$TeV; see \citealt{ambrosone:2025.cr.source.cocoons.grammage}). 
Traditional Galactic CR transport models include detailed Milky Way source distribution templates, cross-sections and reaction rates, and include hundreds of in-principle adjustable parameters to be marginalized over for these fits as data improve (e.g.\ the models in Table~\ref{tbl:spectral.fits} typically include $\sim10+$ parameters \textit{just} for $D_{xx}(R_{\rm cr})$, sometimes with separate fits for different species, with dozens of CR injection parameters allowing species-to-species variations and non-monotonic energy-dependence). But they also make strong simplifying assumptions, typically including: (1) predicting only steady-state CR properties ($D_{t} \bar{f}_{0} = D_{t} \bar{f}_{1} = 0$); in (2) a time-static simplified analytic fitting-function based source/Galaxy model; (3) reducing CR transport to the Fokker-Planck equation (assuming near-isotropy everywhere, $D_{t} \bar{f}_{1} = 0$ and $c\rightarrow \infty$, dropping all streaming terms $\bar{v}_{A}$); with (4) isotropic diffusion ($D_{xx} \sim \kappa_{\|}/3$); and (5) dropping all small-scale (time-and-space-dependent) structure in ${\bf u}$, ${\bf B}$, $\rho$, $v_{A}$, etc. 

This raises a crucial question: if many of these anomalies are small quantitative (and not qualitative) effects, then how can we justify neglecting time-and-space variation in terms like $\bhat$, $u_{\rm B}$, $u_{\rm rad}$, $n_{\rm gas}$ (which appear in loss rates $\mathcal{R}$) as well as ${\bf u}$ and $\nabla{\bf u}$, $v_{A}$ and $\bar{v}_{A}$ (which appear in transport), and almost certainly $\bar{\nu}$ (if it depends on any of these or other ISM properties), which are known to vary by {\em orders-of-magnitude} in the ISM on spatial and timescales small compared to the CR propagation distances and residence times? Indeed more and more literature has shown we cannot: preliminary studies have shown that including any model of self-confinement/streaming, or allowing for a small number of nearby ($<1\,$kpc) sources/SNe, or modeling Galactic spiral arm structure, or including low/high diffusion ``sub-zones'' within the disk, or modeling highly spatially-variable synchrotron/inverse Compton losses, could explain or qualitatively change the interpretation of different anomalies \citep{2016ApJ...824...16J,serpico:2018.need.variable.cr.diffusion.models,kachelries:2019.cr.microphysics.review.obs.broad.range.anomolies.local.sources,evoli:2021.spatially.variable.cr.losses.key.to.positron.problems,zhao:2021.spatially.dependent.cr.propagation.disk.halo.models,jacobs:2023.isotope.ratios.large.halo.size.required.constrain.volume.of.inhomogenous.slower.diffusion.zones.in.model,dimauro:2023.cr.diff.constraints.updated.galprop.very.similar.our.models.but.lots.of.interp.re.selfconfinement.that.doesnt.mathematically.work,tovar:2024.inhomogeneous.diffusion.cr.spectra,rocamora:2024.spatial.variability.in.inverse.compton.important.for.positron.excess,recchia:2024.cr.spectral.modeling.features.bumps.wiggles,delatorre.luque:2024.gas.models.of.galaxy.key.for.scale.height.but.need.halo.for.crs,john:2025.cocoons.around.pulsars.for.cr.modeling}. Consistent with those papers, in the only study so-far to use global non-equilibrium CR-MHD simulations to predict full spectra in LISM-like conditions, \citet{hopkins:cr.multibin.mw.comparison} argued that CR ``weather'' (local fluctuations from variations in all the above and non-equilibrium effects) led to variations in CR spectra in space (selecting different stars as the mock ``Sun'') and time (over $\sim$\,Myr timescales) comparable to or larger than these observed effects. And \citet{thomas:2022.self-confinement.non.eqm.dynamics} also recently argued that CR properties at any given location are almost always non-equilibrium at a level $\gtrsim 10\%$. 
This does not mean said anomalies are unimportant, or that the conclusions from previous modeling are incorrect -- only that understanding these features and the uniqueness of different interpretations of structure at factor $\lesssim 2$ level in CR spectra requires more work with more physically-comprehensive models . And, as emphasized in \citealt{2016ApJ...824...16J}, it makes it even more critical that models compare to the full set of CR spectra simultaneously, and not just one or two primary/secondary species.

\begin{figure*}
	\centering
	\includegraphics[width=0.98\textwidth]{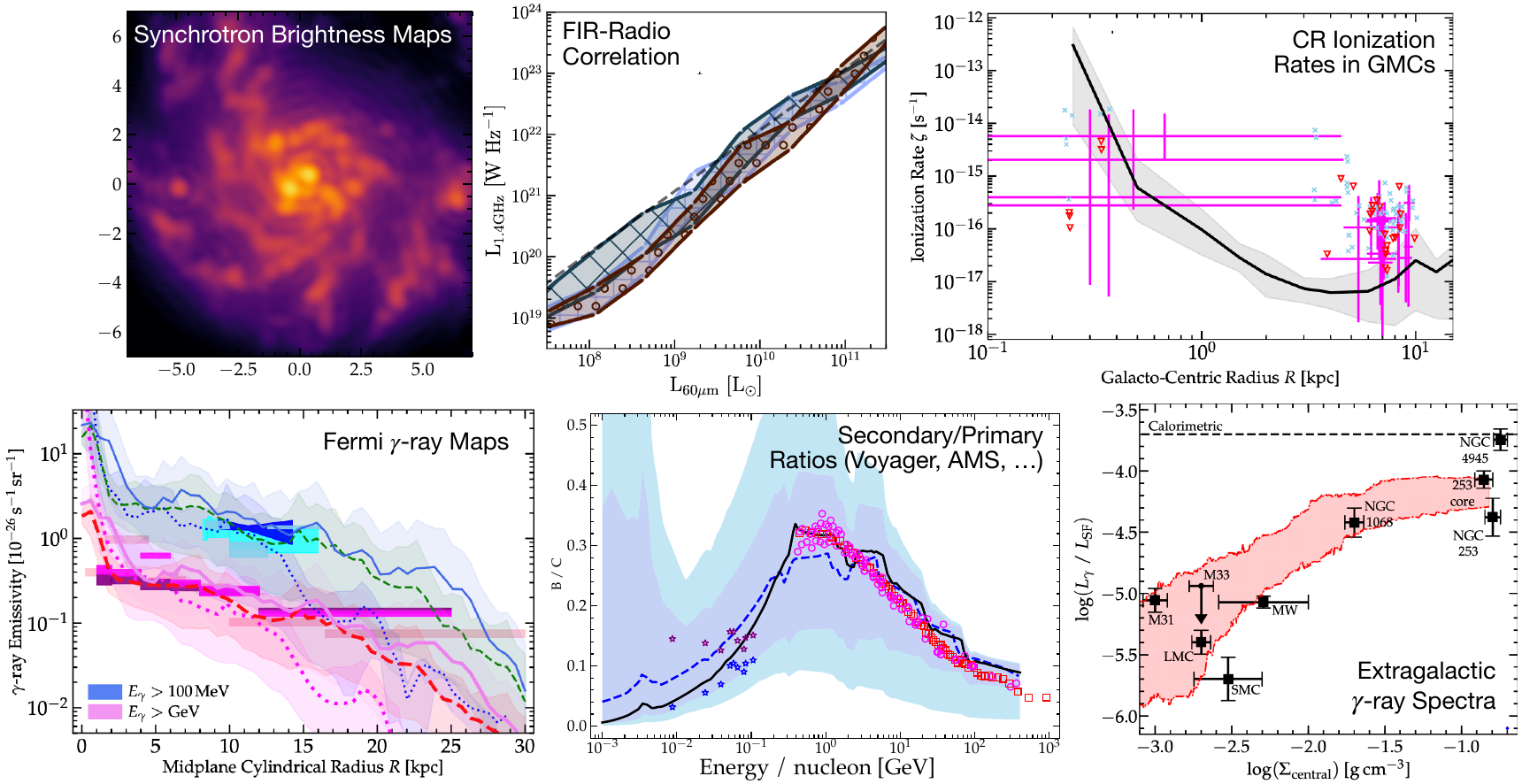}
	\caption{Examples of different observational CR constraints which can be reproduced using a universal CR scattering rate $\bar{\nu}$ calibrated to the LISM. Plots show example synchrotron brightness maps and the infrared-radio luminosity correlation of galaxies, CR ionization rates in individual Galactic molecular clouds, Fermi $\gamma$ ray emissivity profiles/maps of the galaxy (vs.\ frequency, $R$, and $z$), detailed CR spectra as Fig.~\ref{fig:specdemo}, and extragalactic $\gamma$ ray spectra. For each, points/bars show observations, while shaded ranges and lines show example predictions from CR-MHD galaxy formation simulations using a universal $\bar{\nu}(R)$ from the LISM. These demonstrate remarkable consistency, but in each case the observation is driven by the ISM of galaxies, and primarily by some volume-averaged or typical Milky Way-like ISM volumes.
	\label{fig:xgal.obs}}
\end{figure*}

\section{Extra-Galactic Observational Constraints}
\label{sec:extrasolar}

\subsection{Observations and Caveats}
\label{sec:extrasolar.obs}

Outside the Solar neighborhood, at the energies of interest we unfortunately have access at present only to secondary radiative tracers of CRs: e.g.\ $\gamma$-rays, X-rays, and radio emission, plus their indirect effects on other gas observable in e.g.\ optical/UV or line excitation. Each of these have different caveats.

Synchrotron measurements ($\sim 10\,$MHz to $>$\,GHz) reach the highest sensitivity and angular resolution, but: (1) are highly degenerate (depending on a convolution of ${\bf B}$ and the CR spectrum $\int B^{2}\,dN/dE^{\rm kin}_{\rm cr}(E_{\rm crit}[B])\,f(\nu,\,B,\,...)\,dz$); (2) are subject to strong clumping factor effects, as the emission in e.g.\ GHz bands can scale as $B^{4}$, and are correspondingly (3) biased to subvolumes with locally high-$B$; (4) trace primarily high-energy ($\gg 10\,$GeV) leptons, a small fraction of CR energy; and (5) struggle to be sensitive to low surface brightness features. As such their interpretation for constraints on CRs are highly non-unique and modeling papers have shown vastly different (allowed) $B$ and CR configurations can reproduce the same observations \citep{werhahn:2021.cr.calorimetry.simulated.galaxies,ponnada:2023.fire.synchrotron.profiles,ponnada:2023.synch.signatures.of.cr.transport.models.fire,ponnada:2024.fire.fir.radio.from.crs.constraints.on.outliers.and.transport,martin.alvarez:2023.mhd.cr.sims.synch.maps.similar.emission.regions.conclusions.to.fire.ponnada.papers.but.very.different.methods,dacunha:2024.synthetic.synchrotron.problems.with.obs.interp}. 

Fermi ($\sim$\,GeV) $\gamma$-ray emission can probe protons at $\gtrsim$\,a few GeV, but: (1) $\gamma$-rays have serious limitations in sensitivity, with only a handful of Local Group galaxies (e.g.\ the MW, SMC, LMC, M33, and M31) having interesting diffuse-gas emission detections or upper limits of their ISM \citep{lacki:2011.cosmic.ray.sub.calorimetric,tang:2014.ngc.2146.proton.calorimeter,griffin:2016.arp220.detection.gammarays,fu:2017.m33.revised.cr.upper.limit,wjac:2017.4945.gamma.rays,wang:2018.starbursts.are.proton.calorimeters,lopez:2018.smc.below.calorimetric.crs,chan:2018.cosmicray.fire.gammaray,hopkins:cr.transport.constraints.from.galaxies}; (2) low-surface-brightness emission from the CGM (with independent measurement of gas densities) is likely undetectable for individual galaxies other than the MW and M31 \citep{abdo:2010.outer.mw.gamma.ray.emission,ackermann.2011:diffuse.gamma.ray.cr.profile.constraints,tibaldo.2014:diffuse.gamma.ray.cr.profile.constraints,tibaldo.2015:diffuse.gamma.ray.cr.profile.constraints,tibaldo.2021:diffuse.gamma.ray.cr.profile.constraints,acero:2016.gamma.ray.constraints.cr.emissivity,yang.2016:diffuse.gamma.ray.cr.profile.constraints,pothast:2018.radial.mw.gamma.profile.modeling.fermi,karwin:2019.fermi.m31.outer.halo.detection,recchia:2021.gamma.ray.fermi.halos.around.m31.modeling,do:2021.cr.transport.m31.halo.modeling}, though \citet{pshirkov:2024.fermi.stacking.extended.gamma.lstar.galaxies} have recently claimed a tentative detection of extended $\gamma$-ray emission from stacking Fermi data for nearby ($\lesssim 15$\,Mpc) $L_{\ast}$ galaxies, which if confirmed would provide important new constraints (and imply very large CR pressures); and perhaps some  brightest cluster cores (but those are not of interest here);  (3) Fermi is limited in resolution to $\sim$\,degree scales, making it impossible to probe diffuse emission on meso-scales even in the MW, and making all extragalactic detections unresolved; and (4) the $\gamma$-ray emission is degenerate, depending on the product of CR spectra and gas densities. 

Certain molecular lines (radio-infrared) can probe low-energy ($\ll 0.1\,$GeV) leptons and protons, but 
(1) only if we {\em assume} CRs dominate the ionization of the species, which is not always obvious (there can be contributions from microturbulence, shocks, X-rays, and more, see \citealt{gabici:2022.low.energy.cr.ionization.review,ravikularaman:2025.cr.ionization.gal.center.cant.be.crs.alone.limits.other.nonthermal}). Even then they 
(2) require detailed molecular spectra of individual cloud regions, so have thus far been primarily used within the Milky Way. Moreover they (3) are subject to large systematic uncertainties in their calibration -- the recovered ionization rates $\xi_{\rm ion}$ are still debated at the order-of-magnitude level \citep{bovino:2020.cr.ionization.rate.estimators.new.model.and.calibration,bialy:2022.cr.ionization.rate.estimators.h2,redaelli:2024.model.calibration.cr.ionization.rate.estimators.some.off.orders.of.magnitude,obolentseva:2024.cr.ionization.rate.gmcs.much.lower.than.historical.estimates} -- owing to complex dependence on detailed chemical networks with assumptions about shielding, molecular species ratios, catalysis on dust grains, cross-sections, secondary CR ionizations, and more. They also (4) require assumptions about the CR spectral shape and hadron-to-lepton ratio (since they only constrain a total ionization rate); (5) are biased owing to different local emission in inhomogeneous clouds and heavily modulated by Coulomb+ionization losses (so it is not clear if emission comes more from skin layers, for example); and (6) could easily be dominated by local sources (e.g.\ low-energy CR acceleration in proto/pre-stellar jets/winds, rather than the diffuse ISM CR background; \citealt{padovani:2009.cr.ionization.gmc.rates.model.w.alt.sources,gaches:2018.protostellar.cr.acceleration,phan:2018.cr.ionization.models.need.local.sources,pineda:2024.gmc.map.of.cr.ionization.shows.local.sources.dominate.elevated.ionization,fitzaxen:2024.cr.transport.into.gmcs.suppressed.starforge,luo:2024.cr.ionization.rate.hot.cores.local.sources}). 

Soft X-ray ($\sim$\,keV) inverse Compton (IC) emission from CRs scattering CMB photons can probe CR leptons at the peak of the spectrum ($\sim 0.1-1$\,GeV), and uniquely constrain their full spectral shape in the CGM \citep[because the CMB spectrum is perfectly known, there are no degeneracies here][]{hopkins:2025.crs.inverse.compton.cgm.explain.erosita.soft.xray.halos}, but: (1) this emission is quite low surface brightness (well below the meta-galactic background), thus far being potentially detected and spatially-resolved only with sensitive soft X-ray surveys like eROSITA in massive stacks of thousands of MW and M31-mass galaxies \citep{zhang:2024.hot.cgm.around.lstar.galaxies.xray.surface.brightness.profiles}; and (2) requires separation of the IC from thermal (free-free plus line) emission of hot gas (very similar in spectral shape), which is in practice quite challenging. Within the ISM of galaxies, both of these challenges are more severe, plus the IC-scattered radiation spectrum is unknown, making this far less constraining.

Many even more indirect constraints have been proposed. For example, \citet{ji:fire.cr.cgm,ji:20.virial.shocks.suppressed.cr.dominated.halos,li:2021.low.z.fire.cgm.probes,lu:2025.cr.transport.models.vs.uv.xray.obs.w.cric} argued (as we show below) that CR pressure could strongly alter the global thermodynamic evolution of the CGM, which in turn leads to different gas thermal phase structure for the medium, and therefore different predictions for optical/UV/X-ray absorption-line studies sensitive to these. While important consistency checks for full galaxy formation CR-MHD simulations, these cannot meaningfully be used as CR constraints in isolation, because such phase structure is famously sensitive to a host of deeply-uncertain non-CR microphysics (e.g.\ un-resolved instabilities and plasma physics, details of stellar and AGN feedback and star formation). But there are more robust indirect constraints. For example, \citet{butsky:2022.cr.kappa.lower.limits.cgm} showed that measurements of gas column densities (from UV absorption, or radio dispersion measures) at different CGM distances around star-forming galaxies sets strong lower limits to the effective isotropic diffusivity $\kappa_{\rm eff}^{\rm iso}$ of CRs at the same radii, provided only two (much more well-supported) assumptions: (1) that most CR energy is not lost in the ISM, and (2) that the observed gas is not unbound/ejected on short timescales.

\subsection{The ISM in the MW and Other Galaxies}
\label{sec:xgal.ism}

Fig.~\ref{fig:xgal.obs} shows examples of CR modeling of observables from \S~\ref{sec:extrasolar.obs} in the ISM of different galaxies. While synchrotron, $\gamma$-rays, and molecular line diagnostics do allow us to probe CRs outside the LISM, the vast majority of these observations are still sensitive to ISM CRs, \textit{within those galaxies}, as opposed to more diffuse CGM emission at $\gg 10\,$kpc from galaxies.

Within the Milky Way ISM, CR ionization rates, synchrotron, and $\gamma$-ray emission maps are all consistent with a model where the CR pressure/energy density increases towards the Galactic center, vaguely as $\propto 1/r$ (more weakly than the observed turbulent, radiation or magnetic energy density; \citealt{ackermann.2011:diffuse.gamma.ray.cr.profile.constraints,tibaldo.2014:diffuse.gamma.ray.cr.profile.constraints,tibaldo.2015:diffuse.gamma.ray.cr.profile.constraints,acero:2016.gamma.ray.constraints.cr.emissivity,yang.2016:diffuse.gamma.ray.cr.profile.constraints,2018AdSpR..62.2731A,tibaldo.2021:diffuse.gamma.ray.cr.profile.constraints,hopkins:cr.multibin.mw.comparison}). This is qualitatively expected from global CR transport models in the LISM (\S~\ref{sec:halos}), as the source density is peaked in the center and for steady-state flux in a few-kpc or larger halo, $4\pi\,r^{2}\,F_{\rm cr} \sim \kappa_{\rm eff}\,e_{\rm cr}/r \sim $\,constant -- though it further strengthens the argument that transport within the Galaxy is ``effectively'' diffusive, rather than e.g.\ advective. There are subtle anomalies: for example the TeV excess (probably associated with ``cocoons'' around local pulsars,
and related to the LISM positron excess and ``bump'' in the proton spectrum at $\sim 10-100\,$TeV)\footnote{These observational hints of ``cocoons'' and positron excess are potentially very interesting from a CR source/acceleration and transport microphysics point of view. However we stress that they are not very significant from a galactic ISM-integrated and/or CR-energy-integrated point of view. A robust conclusion from many studies on these seems to be that CRs accumulate something like an ``extra,'' weakly-energy-dependent grammage of $\sim 0.4\,{\rm g\,cm^{-2}}$ within some region of size ${\rm pc} \lesssim R \lesssim 200\,{\rm pc}$ around sources \citep[probably millisecond pulsars; see][and references therein]{ambrosone:2025.cr.source.cocoons.grammage}. But this is only a $\sim 1-5\%$ correction to the total grammage of low-energy CRs $\sim 0.1-1\,$GeV. And for typical dense clouds/GMCs/superbubbles/etc., or pulsars within the central few kpc of the Galaxy where TeV excesses are most significantly detected \citep{albert:2024.tev.halos.around.pulsars.crs,john:2025.cocoons.around.pulsars.for.cr.modeling}, the gas surface densities are $\gtrsim 100\,\Sigma_{100}\,{\rm M_{\odot}\,pc^{-2}}$, implying CRs need to be ``slowed down'' to effective streaming speed in these regions on initial escape of $v_{\rm st,\,eff}^{\rm iso} \lesssim 8000\,{\rm km\,s^{-1}}\,\Sigma_{100}$ or parallel $v_{\rm st} \sim 0.08\,\Sigma_{100}\,c$, in order to accumulate said grammage \citep{evoli:2021.spatially.variable.cr.losses.key.to.positron.problems,krumholz:2024.beaming.argument.to.constrain.cr.transport.near.sources.would.apply.to.inhomogeneous.transport.models.like.irynas}. This is already much faster than the steady-state streaming speeds predicted for low-energy CRs, and only becomes significant for TeV CRs -- where the effect is actually measured and where the grammage correction is significant.}
or the ``galactic center excess'' (at $\sim 3-10\,$GeV in $\gamma$-rays). 
And the ionization data -- depending on which diagnostic and calibration is adopted (\S~\ref{sec:extrasolar.obs}) -- could imply a significant excess of very low-energy ($\sim$\,MeV) CRs near young star-forming regions, the most popular explanation for which is acceleration of low-energy CRs by local sources like (proto)stellar jets/winds \citep{phan:2018.cr.ionization.models.need.local.sources,gabici:2022.low.energy.cr.ionization.review,pineda:2024.gmc.map.of.cr.ionization.shows.local.sources.dominate.elevated.ionization,luo:2024.cr.ionization.rate.hot.cores.local.sources}. 
But per \S~\ref{sec:dyn.models} these are second-order effects involving as little as $\lesssim 0.1\%$ of the total CR pressure/energy density at those locations -- though they can be very interesting in their own right for constraints on the microphysics of pulsars, protostellar jets, and CR pevatrons.

Outside the Galaxy, a few bright, nearby ($\lesssim$\,few Mpc) star-forming ($\dot{M}_{\ast} \sim 1-20\,{M_{\odot}\,{\rm yr}^{-1}}$), Milky Way-mass galaxies have $\sim$\,kpc-scale spatially resolved synchrotron maps within a few to ten kpc off the disk \citep{beck:2015.b.field.review}, whose profiles and fluctuations within the ISM (e.g.\ in face-on galaxies) are similar to the Galaxy \citep{basu:2015.synchrotron.spectral.index.and.cr.properties.nearby.sf.galaxies,beck:2016.galactic.random.bfields.strong}. 
While not especially constraining to CRs alone given the degeneracy with magnetic field structure, these synchrotron maps are all at least consistent with other galaxies having radial and vertical ISM CR energy-density profiles, CR ISM diffusion coefficients/scattering rates, and ISM CR lepton spectral shapes broadly similar to the Milky Way \citep{ponnada:2023.fire.synchrotron.profiles,ponnada:2023.synch.signatures.of.cr.transport.models.fire,martin.alvarez:2023.mhd.cr.sims.synch.maps.similar.emission.regions.conclusions.to.fire.ponnada.papers.but.very.different.methods,dacunha:2024.synthetic.synchrotron.problems.with.obs.interp}, while the maps of edge-on systems imply that almost all galaxies almost certainly have ``CR scattering halos'' extending to $\gtrsim$\,a few kpc \citep{krause:2018.radio.synchrotron.halo.scale.heights,krause:2020.spiral.galaxy.halo.magnetic.geometries.and.coherence,heesen:2021.radio.cosmic.ray.escape.sub.calorimetric}. Attempts have been made to use resolved ISM synchrotron maps to constrain CR transport parameters \citep[e.g.][]{vanheesen:2023.cr.diffusion.coefficient.estimation.resolved.ISM.maps.smoothing.not.capturing.physics.}, but these depend sensitively on a number of strong assumptions (strict equipartition, calorimetry, steady-state, diffusion-only, a laminar ISM/no turbulence, etc.) and as such forward-modeling shows the same data can be reproduced with a huge range of transport models (Ponnada et al., in prep.). There could be significant non-linear effects of CR dynamics on the phase structure of the ISM \citep{ponnada:2023.synch.signatures.of.cr.transport.models.fire} or evolution of the radio-infrared correlations \citep{ponnada:2024.fire.fir.radio.from.crs.constraints.on.outliers.and.transport}, but these are akin to effects of CRs on galactic star formation or outflows in that they are degenerate with other uncertain feedback physics.  

Beyond these, almost all observations sensitive to CRs in the ISM of other galaxies are \textit{integrated} detections of the total ISM emission. This includes a few bright very-nearby galaxies detected by Fermi in $\gamma$-rays \citep{ackermann:2012.fermi.gamma.rays.ism.cr.emission.modeling,rojas.bravo:2016.new.fermi.upper.limits.lots.of.sub.calorimetric.galaxies,griffin:2016.arp220.detection.gammarays,fu:2017.m33.revised.cr.upper.limit,lopez:2018.smc.below.calorimetric.crs}, and many galaxies detected in their integrated $\sim$\,GHz radio emission (usually studied in the far infrared-radio correlation; \citealt{lacki:2010.fir.radio.highz.synchrotron,magnelli:2015.fir.radio,delhaize:2017.fir.radio,wang:2019.fir.radio.corr,werhahn:2021.cr.calorimetry.simulated.galaxies,roth:2023.fir.radio.modeling,roth:2024.fir.radio.crs.constraints,ponnada:2024.fire.fir.radio.from.crs.constraints.on.outliers.and.transport}). 
The interpretation of these is even less unique owing to the degeneracies in \S~\ref{sec:extrasolar.obs} plus the fact that they integrate over the entire ISM. Thus while CR-MHD simulations and other modeling efforts have shown that all of these observations can be reproduced with the same empirical CR scattering model fitted to the LISM spectra as described in \S~\ref{sec:obs.lism} \citep[references above and][]{lacki:2011.cosmic.ray.sub.calorimetric,fu:2017.m33.revised.cr.upper.limit,lopez:2018.smc.below.calorimetric.crs,chan:2018.cosmicray.fire.gammaray,su:turb.crs.quench,hopkins:cr.multibin.mw.comparison,ponnada:2024.fire.fir.radio.from.crs.constraints.on.outliers.and.transport}, studies like \citet{buck:2020.cosmic.ray.low.coeff.high.Egamma,hopkins:cr.transport.constraints.from.galaxies,ponnada:2023.synch.signatures.of.cr.transport.models.fire,ponnada:2024.fire.fir.radio.from.crs.constraints.on.outliers.and.transport} have shown that CR propagation models which make very different assumptions for how $\bar{\nu}$ varies with ISM properties like $B$, etc.\ can produce quite similar predictions.

\begin{figure*}
	\centering
	\includegraphics[width=0.48\textwidth]{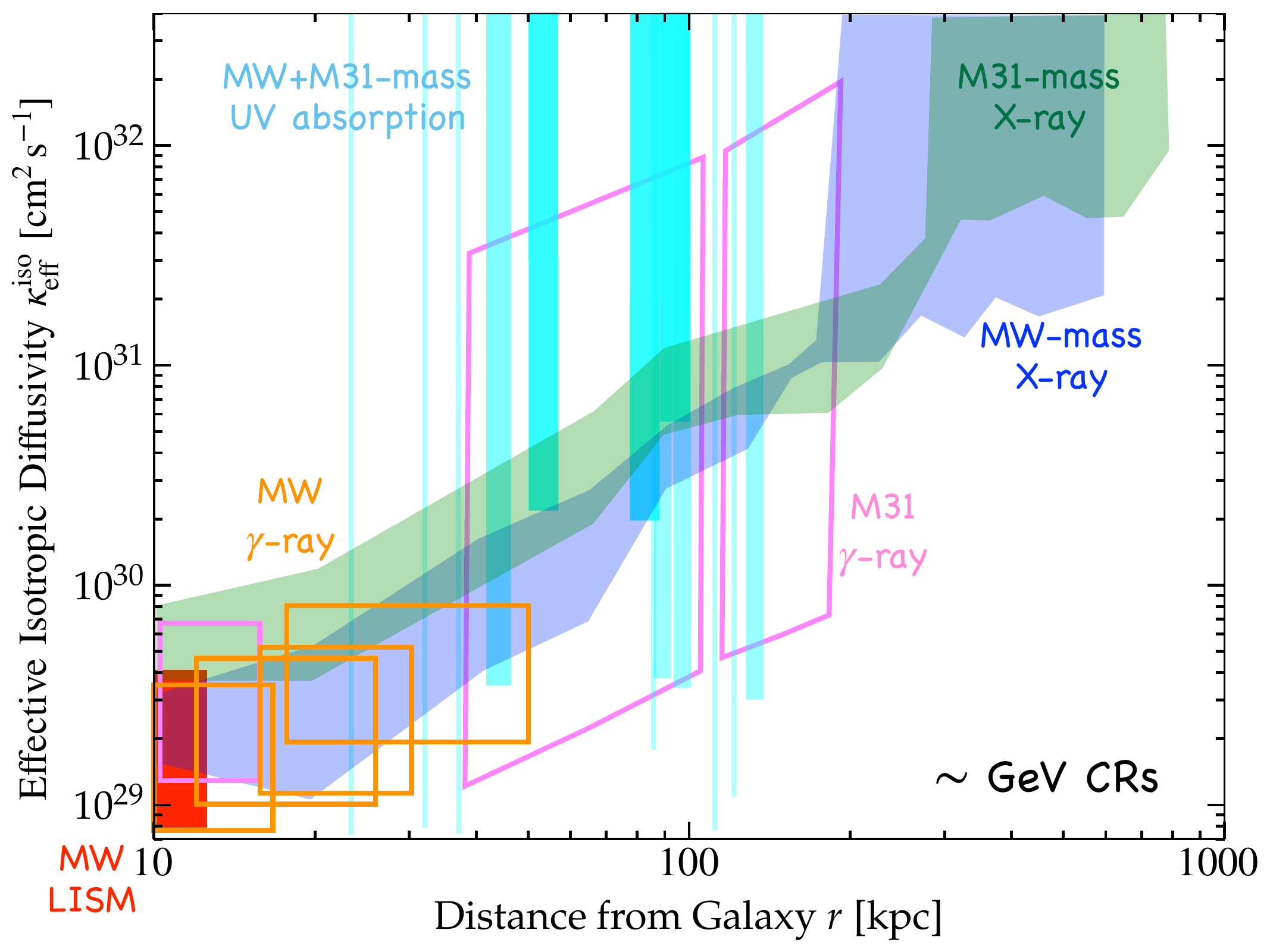}
	\includegraphics[width=0.48\textwidth]{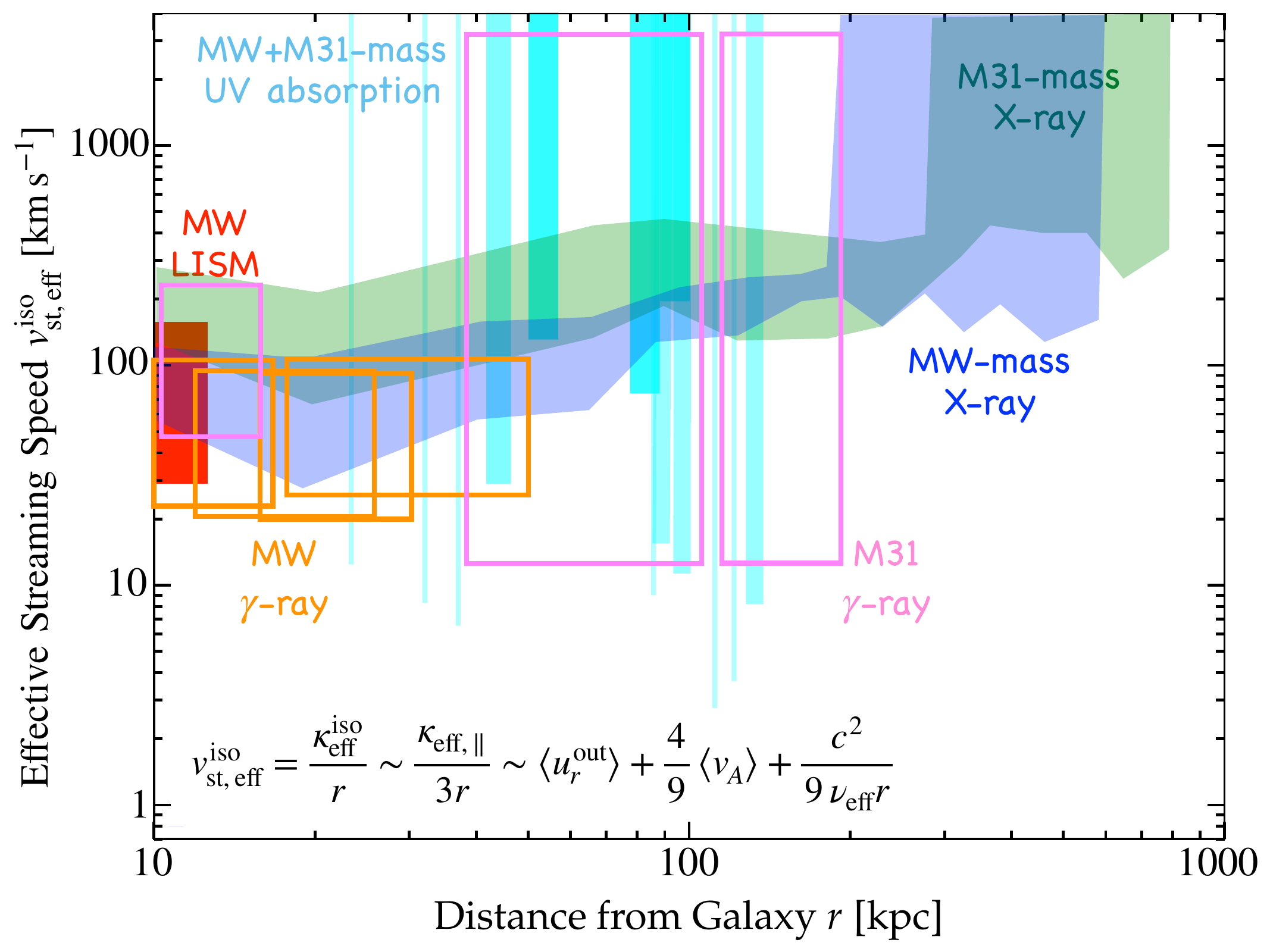}
	\caption{Compilation of constraints on CR transport around $\sim$\,GeV ($\sim 0.1-10\,$GeV) at different galacto-centric radii in the CGM. We plot both effective isotropic diffusivity $\kappa^{\rm iso}_{\rm eff}$ ({\em left}) and streaming speeds $v^{\rm iso}_{\rm st,\,\rm eff}$ ({\em right}). We compile constraints from the Milky Way (MW) LISM (reviewed above); MW and M31/Andromeda $\gamma$ rays (where both extended diffuse $\gamma$-ray emission with an LISM-like spectrum is detected by Fermi, and column density of gas at the same distance can be determined, to calculate an emissivity); stacked X-ray soft X-ray ($\sim$\,keV) emission from MW or M31-mass galaxies from eROSITA (assuming this comes from inverse Compton); and indirect constraints (shaded bars since the constraints are lower limits at each $r$) from UV absorption lines from virial pressure arguments (see \S~\ref{sec:obs.cgm}). While constraints in the CGM remain sparse, what is remarkable is that all of these data points are consistent with one another and can be fit with a roughly  typical $v_{\rm st,\,eff}^{\rm iso} \sim 100\,{\rm km\,s^{-1}}$ in the CGM (see \S~\ref{sec:effective}, Eq.~\ref{eqn:effective.nu.ism.cgm}).
	\label{fig:cgm.kappa.obs}}
\end{figure*}

\subsection{The CGM of the MW and Other Galaxies}
\label{sec:obs.cgm}

As noted above, we have far fewer CGM constraints, but there are some recent improvements. 
Fig.~\ref{fig:cgm.kappa.obs} compiles observational constraints on the effective, isotropically-averaged diffusivity and/or streaming speed (defined so the CR flux is $F_{\rm cr} \sim \kappa_{\rm eff}^{\rm iso} \nabla e_{\rm cr} \sim v_{\rm st,\,eff}^{\rm iso} e_{\rm cr}$) of CRs in the CGM at different galacto-centric radii from just outside the LISM ($\sim 10\,$kpc) to Mpc scales (well into the IGM). 
The observations shown are primarily sensitive to $\sim$\,GeV CRs (though different in detail per \S~\ref{sec:extrasolar.obs}, as e.g.\ $\gamma$-rays probe somewhat higher-energy protons, X-rays somewhat lower-energy leptons), and focus on MW+M31-mass halos, so we restrict to these limits. 
We show the MW LISM constraints (\S~\ref{sec:obs.lism}) for reference.

The MW+M31 represent the two systems where diffuse CGM $\gamma$-ray emission has been detected, together with independent gas density/column measurements \citep{abdo:2010.outer.mw.gamma.ray.emission,ackermann.2011:diffuse.gamma.ray.cr.profile.constraints,tibaldo.2014:diffuse.gamma.ray.cr.profile.constraints,tibaldo.2015:diffuse.gamma.ray.cr.profile.constraints,tibaldo.2021:diffuse.gamma.ray.cr.profile.constraints,acero:2016.gamma.ray.constraints.cr.emissivity,yang.2016:diffuse.gamma.ray.cr.profile.constraints,pothast:2018.radial.mw.gamma.profile.modeling.fermi,karwin:2019.fermi.m31.outer.halo.detection,recchia:2021.gamma.ray.fermi.halos.around.m31.modeling,do:2021.cr.transport.m31.halo.modeling}. The emissivities versus energy are consistent with the LISM proton spectrum of either galaxy simply escaping/propagating in steady-state, as expected since the hadronic losses should be weak outside $\gtrsim 10\,$kpc from the galaxy center given the measured densities; the implied fluxes are converted into equivalent $\kappa_{\rm eff}^{\rm iso}$ in \citet{hopkins:2025.crs.inverse.compton.cgm.explain.erosita.soft.xray.halos}. Note the constraints for M31 are still relatively poor, but give some information; the MW measurements of emissivity are quite accurate, but there the dominant uncertainty comes from not knowing exactly where along a line-of-sight the emission originates. Using the indirect method from \S~\ref{sec:extrasolar.obs}, \citet{butsky:2022.cr.kappa.lower.limits.cgm} used measurements of HI gas columns (from UV absorption measurements) around star-forming galaxies where the CR injection rate into the CGM can be estimated to place lower limits to $\kappa_{\rm eff}$, requiring only that the HI not be in rapid outflow. These lower limits are shown for their sample (primarily MW+M31-mass galaxies). And \citet{hopkins:2025.crs.inverse.compton.cgm.explain.erosita.soft.xray.halos} noted that if the eROSITA measurement of diffuse soft X-ray surface brightness is interpreted as inverse-Compton emission, the stacked soft X-ray profiles from \citet{zhang:2024.hot.cgm.around.lstar.galaxies.xray.surface.brightness.profiles} of thousands of MW+M31-mass galaxies can be turned into a corresponding constraint on some median CR profile and therefore diffusivity (of $\sim 0.1-$few\,GeV $e^{-}$).

Remarkably, even though these are indirect, and come from very different samples (MW+M31, stacks, or galaxies with background quasars) using qualitatively different techniques, all of these estimates are consistent with one another, and with extrapolation to MW LISM constraints at $R < 10\,$kpc. Very robustly, the effective $\kappa_{\rm eff}$ appears to increase with galacto-centric distance, as many models predict (though see \S~\ref{sec:uncertainty}). But  in terms of an effective, isotropically-averaged (radial) streaming speed $v_{\rm st,\,eff}^{\rm iso}$, all of these data are reasonably consistent with a more or less constant $v_{\rm st,\,eff}^{\rm iso} \sim 100\,{\rm km\,s^{-1}}$. Thus, at least \textit{on average}, it seems like CR streaming in the CGM of MW/M31-mass galaxies can be represented quite simply, like how the ISM observables of other galaxies above can be reasonably captured with a universal $\nu$ or $\kappa$.

This is an extremely useful constraint for models. However, given the crudeness of the constraints, it does not tell us how scattering rates $\nu$ vary with plasma properties directly. And the constraints in Fig.~\ref{fig:cgm.kappa.obs} are broad-binned in CR energy, meaning we do not know the rigidity-dependence of this streaming over the $\sim 0.1-10\,$GeV range where it applies.\footnote{The $\gamma$-ray points in Fig.~\ref{fig:cgm.kappa.obs} give some information in principle, but the constraints on spectral shape are sufficiently weak in M31 that these are not too constraining, and the dynamic range of distance $r$ in the MW data are such that one can similarly well fit the data at $>10\,$kpc with an rigidity-independent $v_{\rm st,\,eff}^{\rm iso}$ or with a rigidity dependence $\propto R_{\rm cr}^{1/2}$ similar to the LISM.} We also stress that while this behavior can be represented by a constant $v_{\rm st,\,eff}$ in space: (1) this could come from actual ``streaming,'' or advective/convective/outflow velocities (and is similar to \Alf\ or typical outflow speeds); but (2) could come from an effective diffusivity or scattering rate which simply scales, on average, as $\kappa \propto r$ or $\nu \propto r^{-1}$ (or not too different from this, since the constraints allow a range $\kappa \propto r^{0.8 - 1.7}$); or (3) a similar ``effective'' transport law can arise from a single scattering rate if the CR scattering is super-diffusive, as compared to diffusive, in pitch-angle space \citep{liang:2025.leaky.boxes.levy.flights.modeling.crs.transport}.

\subsection{An Effective Phenomenological Scattering Model}
\label{sec:effective}

Combining all of the above, to lowest order, it appears that all of the data from the MW ISM and extragalactic constraints reviewed above can be reasonably reproduced if we assume a phenomenological scattering rate of the form (the resulting $\kappa_{\rm eff}$ and $v_{\rm st,\,eff}$ as a function of galacto-centric radius are shown alongside the observational constraints in Fig.~\ref{fig:cgm.kappa.obs}):
\begin{align}
\label{eqn:effective.nu.ism.cgm} {\bar{\nu}} &\sim \bar{\nu}_{0}\,\left( \frac{R_{\rm cr}}{\rm GV} \right)^{-\delta}\,\left(1 + \frac{\ell_{\ast}}{\ell_{0}} \right)^{-1}  \ , 
\end{align}
with $\bar{\nu}_{0} \sim \beta\,10^{-9}\,{\rm s^{-1}}$, $\delta \sim 0.5-0.6$, $\ell_{0}\approx v_{\rm cr}^{2}/(9\,v_{\rm st,\,\infty}\,\bar{\nu}_{0}\,[R_{\rm cr}/{\rm GV}]^{-\delta}]) \sim 4\,{\rm kpc}\,\beta\,R_{\rm GV}^{1/2}\,v_{\rm st,\,\infty,\,100}^{-1}$ in terms of some constant asymptotic streaming speed $v_{\rm st,\,\infty} \equiv v_{\rm st,\,\infty,\,100}\,100\,{\rm km\,s^{-1}}$, and $\ell_{\ast}$ some characteristic macro-scale which scales like $\mathcal{O}(r)$ in the CGM.\footnote{Briefly, it is worth noting that a qualitatively similar phenomenological model, of the form: $\bar{\nu} \sim \bar{\nu}_{0}\,(R_{\rm cr}/{\rm GV})^{-\delta}\,(1 + \ell_{1}/\ell_{\ast})$, with $\ell_{1} \sim \alpha\,\ell_{\rm mfp} \sim \alpha\,v_{\rm cr} / (\bar{\nu}_{0}\,(R_{\rm cr}/{\rm GV})^{-\delta})$ with $\alpha \sim 4$ or so, and $\ell_{\ast} \sim \ell_{\nabla}$, could reproduce the scattering ``cocoons'' around certain sources (\S~\ref{sec:dyn.models}), since in those are seen precisely at CR energies and scales around sources where the size of the system (e.g.\ the pulsar wind nebulae) and corresponding CR gradient length scales become small compared to the nominal LISM CR scattering mean free path $\ell_{\rm mfp}$. However this model extrapolates poorly to high energies $\gg$\,TeV where $\ell_{1}$ would become very large.}

For example, motivated by extrinsic turbulence-type theories, we could take $\ell_{\ast} \sim \ell_{A}$, the \Alf\ scale of turbulence, which is $\sim 10\,$pc in the LISM (so the $\ell_{\ast}/\ell_{0}$ term is negligible and we recover the usual LISM power-law $\bar{\nu}(R)$), but this becomes $\gtrsim r$ in the CGM, so $\ell_{\ast}$ becomes larger than $\ell_{0}$ and $\bar{\nu} \rightarrow v_{\rm cr}^{2}/(9\,\ell_{\ast}\,v_{\rm st,\,\infty})$ which gives the desired asymptotic streaming speed.\footnote{This can also happen in the LISM at extremely low non-relativistic hadron energies, but then streaming is plausible and allowed physical behavior in those limits as well.}
Alternatively, motivated by self-confinement-type theories, we could take $\ell_{\ast} \sim \ell_{\nabla}$ (the CR gradient length scale at each $R_{\rm cr}$), which similarly ensures the desired behavior. In either case this gives $\kappa_{\|} \equiv v_{\rm cr}^{2}/(3\,\bar{\nu}) \sim 3\beta\,R_{\rm GV}^{1/2} (1+\ell_{\ast}/\ell_{0})^{-1}\times10^{29}\,{\rm cm\,s^{-1}}$, $D_{xx} \sim \kappa_{\|}/3$. 

For the spectrally-integrated equations, one can simply take a weighted-average of Eq.~\ref{eqn:effective.nu.ism.cgm}, i.e.\ $\hat{\nu} \sim 10^{-9}\,{\rm s^{-1}}\,(1+\hat{\ell}_{\ast}/\hat{\ell}_{0})^{-1}$ with $\hat{\ell}_{0} \sim 4\,{\rm kpc}\,v_{\rm st,\,\infty,\,100}^{-1}$.

\begin{table*}
\begin{center}
\begin{footnotesize}
\caption{Importance of low-energy CRs.\label{tbl:dyn.fx}}
\begin{tabular}{ lclll }
\hline\hline
Phase & Scattering Rate & Dynamical/Pressure Effects & Thermochemical Effects & Dominant Loss \\
\hline\hline
Warm ISM & $\sim 10^{-9}\,{\rm s^{-1}}$ & weak: $P_{\rm cr} \sim P_{\rm th},\,P_{\rm B},\,P_{\rm rad},\,P_{\rm turb}$, but  & weak & hadrons: Coulomb+pionic \\
 & & $|\nabla P_{\rm cr} | \ll |\nabla P_{\rm th}|,\,|\nabla P_{\rm rad}|,\,|\nabla P_{\rm turb}|$,  $t_{\rm diff} \ll t_{\rm dyn}$ & & leptons: synchrotron+IC \\
 \hline
 Neutral ISM/ & very low? & weak: $P_{\rm cr} \ll P_{\rm B},\,P_{\rm turb}$ & ionization important & hadrons: ionization+pionic \\
 GMCs & & & for chemistry & leptons: ionization+synchrotron \\
 \hline
 CGM/ & Eq.~\ref{eqn:effective.nu.ism.cgm}? & unknown, but could be strong & indirect: $P_{\rm cr}$ modifies TI & hadrons: pionic (weak) \\
 IGM & (unconstrained) & $|\nabla P_{\rm cr}| \gg |\nabla P_{\rm th}|,\,\rho\,|\nabla \Phi|$ allowed & if $t_{\rm diff} \lesssim t_{\rm condense}$ & leptons: CMB-IC \\
 \hline
 AGN/ & ? & weak in disk/torus; could be & weak except in molecular & extremely rapid if  \\
 jets/bubbles & (unconstrained) & strong in bubbles/jets/CGM & torus (like GMCs) & close to disk/horizon/AGN \\
 \hline
\end{tabular}
\end{footnotesize}
\end{center}
\end{table*}

\section{Dynamical Importance of CRs}
\label{sec:fx}

Now we briefly review the potential \textit{dynamical} importance of CRs in different Galactic/extra-galactic environments, summarized in Table~\ref{tbl:dyn.fx}.

\subsection{AGN}
\label{sec:fx:agn}

CR \textit{acceleration} in AGN remains an incredibly active and interesting area of theory, and this has important implications potentially for modeling ultra-high-energy CRs, outflows like the Fermi bubbles, the structure of blazar jets, Comptonization and hard X-ray/$\gamma$-ray emission from AGN, synchrotron radiation from compact radio cores, and more. However, the direct \textit{dynamical} effects of CRs on the structure of most of the gas near AGN is almost certainly small. One can immediately rule out CRs as a significant source of pressure in AGN accretion disks or dusty torii, for example, as the densities are so high that (1) the loss timescales are vastly shorter than dynamical times of the system, and (2) the implied luminosities in $\gamma$-rays (for hadronic CRs) or synchrotron+IC (for leptons) would be many orders-of-magnitude larger than observed \citep{hopkins:superzoom.disk}. Likewise, CR ionization could play a role in the outermost, molecular accretion disk, but this is only insofar as it is similar to the neutral ISM (discussed below) -- the inner region is strongly thermally and photo-ionized at a level which makes CR ionization and heating irrelevant \citep{hopkins:superzoom.agn.disks.to.isco.with.gizmo.rad.thermochemical.properties.nlte.multiphase.resolution.studies,koutsoumpou:2025.cr.ionization.agn.starburst.galaxies.very.plausible.high.ion.conditions}. 

However, CRs injected by AGN (e.g.\ in blazar jets) could be extremely important for AGN feedback: quenching galaxies, solving the cooling flow problem in massive clusters, pushing baryons out of halos to explain Sunyaev-Zeldovich and weak-lensing observations \citep[see][]{pinzke.pfrommer:2010.cluster.gamma.ray.emission.simple.scalings.for.specific.advection.acceleration.models,ensslin:2011.cr.transport.clusters,su:turb.crs.quench,su:2021.agn.jet.params.vs.quenching,su:2023.jet.quenching.criteria.vs.halo.mass.fire,su:2024.fire.jet.sim.using.acc.jet.prescriptions.from.cho.multiscale.experiments,su:2025.crs.at.shock.fronts.from.jets.injection,wellons:2022.smbh.growth,mercedes.feliz:2023.agn.feedback.positive.negative,cochrane:2023.agn.winds.galaxy.size.effects,byrne:2023.fire.elliptical.galaxies.with.agn.feedback,ruszkowski.pfrommer:cr.review.broad.cr.physics,ponnada:2024.fire.fir.radio.from.crs.constraints.on.outliers.and.transport,quataert.hopkins:2025.crs.massive.halos.blowout.rvir.cosmology.constraints,koutsoumpou:2025.cr.ionization.agn.starburst.galaxies.very.plausible.high.ion.conditions}. But in these cases the action is on scales of the CGM, and the real question for CR dynamics/transport is the dynamics of CRs (from any source) in the CGM, discussed below. Though of course understanding where, when, and how CRs are accelerated and injected is crucial for this as well -- the point is again that the AGN are very interesting as a CR source, but CRs cannot dominate the gas dynamics in the AGN accretion disk or related environments \citep[see more discussion in][]{ruszkowski.pfrommer:cr.review.broad.cr.physics}.

These conclusions are not very sensitive to CR transport physics: if transport is ``fast'' near-horizon, CRs escape, while if CR transport is relatively slow, they lose their energy (but this amounts to a small fraction of the luminosity across wavelengths). Of course if CRs are accelerated near-horizon in jets and bulk-transported outwards by the jet, they will be carried ``rapidly'' with said jet, but it remains unclear whether the acceleration zone would actually be near-horizon (or further out via conversion of Poynting flux inside the jet or at a jet deceleration shock).

\begin{figure*}
	\centering
	\includegraphics[width=0.9\textwidth]{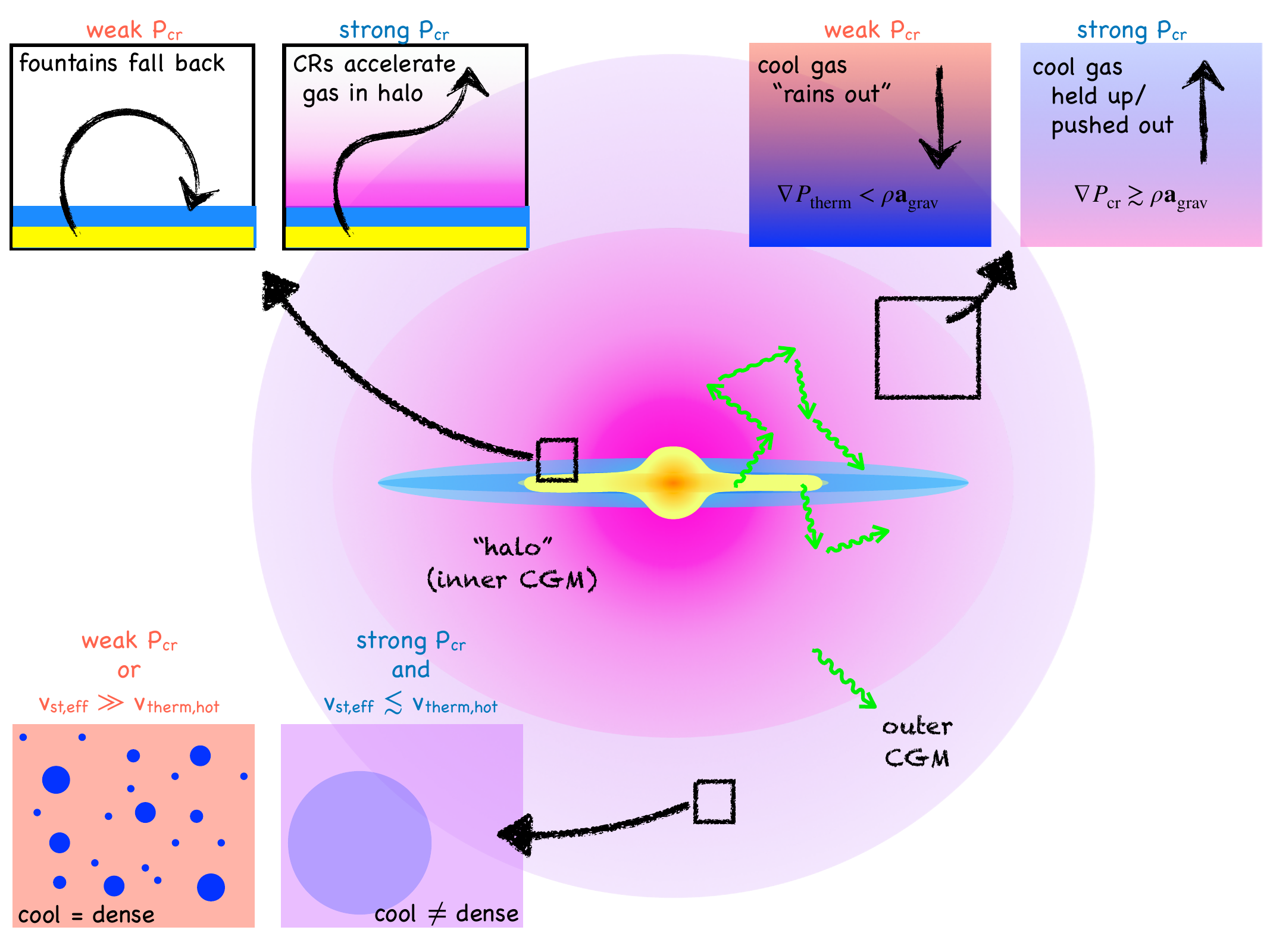}
	\caption{Illustration of the dynamical effects of CRs in the CGM. 
	At the disk-halo interface, ``failed outflows'' or ``fountain flows'' launched by SNe in the midplane would decelerate at $\sim$\,kpc above the disk as they adiabatically cool and are pressure-confined by the hot gas pressure of a thermal-pressure-supported CGM, but can be re-accelerated by CRs whose pressure gradient scale-length above the disk is $\sim$\,kpc. At larger radii, if the CR pressure is comparable to or larger than the thermal pressure in the outer CGM, cool gas can be stably supported or held up against gravity or accelerated into slow (outflow speed much lower than virial velocities) but highly mass-loaded outflows, whereas in a thermal-pressure supported halo gas which cools necessarily loses pressure support against gravity and so rains or precipitates out onto the central galaxy. On smaller scales, CRs can qualitatively alter the thermal instability, if the CR pressure is significant and the effective CR transport time is not much larger than the effective condensation time of large cooling structures ($v_{\rm st,\,eff} \lesssim$ thermal sound speeds in the hot, cooling phase). This leads to a more homogeneous medium with gas at different temperatures co-existing at low densities, while in the thermal-pressure-dominated case cold clouds are necessarily compressed to much higher densities and smaller sizes.
	\label{fig:cgm.cr.fx}}
\end{figure*}

\subsection{Neutral ISM, GMCs, \&\ Starbursts} 
\label{sec:fx.gmc}

In the dense, neutral ISM and GMCs, CR pressure is almost certainly subdominant to turbulent/magnetic/gravitational pressures. Even taking the most optimistic upper estimates of CR pressures inferred in cold clouds (scaling from ionization rates in \citealt{indriolo:2009.high.cr.ionization.rate.clouds.alt.source.models,indriolo:2015.cr.ionization.rate.vs.galactic.radius}, assuming a universal proton spectrum from MeV-GeV), the CR pressure $P_{\rm cr}$ would be much smaller than typical magnetic ($P_{\rm B} = B^{2}/8\pi$), or turbulent ($P_{\rm turb} \approx \rho\,v_{\rm turb}^{2}/2$), or virial ($P_{\rm vir} \approx G\,M^{2}/R^{4}$) pressures. Moreover CR pressures even that high (let alone higher) are already strongly ruled out by $\gamma$-ray observations \citep{krumholz:2023.cosmic.ray.ionization.gamma.ray.loss.budgets,ravikularaman:2025.cr.ionization.gal.center.cant.be.crs.alone.limits.other.nonthermal}. Likewise CR heating is almost certainly negligible compared to very fast cooling rates in GMCs \citep{wolfire:1995.neutral.ism.phases,kim:tigress.ism.model.sims,grudic:starforge.methods,fitzaxen:2024.cr.transport.into.gmcs.suppressed.starforge}. And CR effects diminish further at larger column densities \citep{indriolo:2012.cr.ionization.rate.vs.cloud.column,sabatini:2020.cr.ionization.rate.clumps,sabatini:2023.cr.ionization.rate.cores,socci:2024.orion.cr.ionization.vs.column}. They can be even further diminished by streaming losses and other boundary effects \citep{bustard:2020.crs.multiphase.ism.accel.confinement,fitzaxen:2024.cr.transport.into.gmcs.suppressed.starforge}

However, it is widely-known that CRs {\em do} have important effects on these environments, via their effects on chemistry. With ionization fractions $\lesssim 10^{-8}-10^{-6}$ in GMCs (which are strongly self-shielded to UV radiation), CRs can become the dominant ionization mechanism, providing free electrons that  modify the cooling physics and catalyze important chemical reactions, charge dust grains, and regulate the coupling and resistivity of magnetic fields to the gas. There are multiple recent, comprehensive reviews on the subject (e.g.\ \citealt{padovani:2020.cr.low.energy.ionization.fx.on.sf,gabici:2022.low.energy.cr.ionization.review,owen:2023.cr.review.galaxies.feedback}), so we will not discuss it further except to say that while the CR transport physics in the highly-neutral regime (which may strongly damp gyro-resonant modes via ion-neutral collisions) remains highly uncertain, the dominant uncertainties in modeling and observing these regimes may lie less with the CR transport physics and more with (1) the chemistry itself (highly uncertain as it depends on the size distribution of grains, catalysis of different reactions, micro-scale clumping and relative species velocities, etc.) and (2) presence or absence of ``local sources'' (see \S~\ref{sec:dyn.models}). 

The nuclei of starburst galaxies fall in-between this regime and AGN in most parameters, and observations similarly indicate that CR pressure cannot be dominant, with indeed most of the CR energy in hadrons lost to pionic/catastrophic processes (e.g.\ $\gamma$-ray luminosities being calorimetric; see \citealt{lacki:2010.fir.radio.conspiracy,lacki:2011.cosmic.ray.sub.calorimetric,krumholz:2020.cr.transport.starbursts.upper.limit.kappa.gamma.rays,crocker:2021.cr.pressure.starbursts.upper.limit.kappa.cr.pressure.weak,crocker:2021.starburst.no.cr.winds.only.lower.density.galaxies}). Here, the empirical result and upper limit to CR pressure is robust, and this does set a non-trivial upper limit to CR transport speeds: if CRs escape too quickly, then starbursts would no longer be approximate proton calorimeters (they would more closely resemble MW-like and smaller galaxies, which are observed to be factors of $\sim 100$ below the calorimetric limit -- i.e.\ $99\%$ of CRs escape to the CGM). Following \citet{krumholz:2020.cr.transport.starbursts.upper.limit.kappa.gamma.rays,crocker:2021.cr.pressure.starbursts.upper.limit.kappa.cr.pressure.weak}, if we allow for a simple toy model of a CR halo (with low densities) of size $\sim L_{\rm halo,\,kpc}\,{\rm kpc}$ plus a dense starburst of surface density $\Sigma \sim  \Sigma_{\rm g\,cm^{-2}}\,{\rm g\,cm^{-2}}$ and scale-height/half-thickness $H \sim  H_{\rm dense,\,100\,pc}\,100\,{\rm pc}$ (typical $\Sigma_{\rm g\,cm^{-2}}$, $H_{\rm dense,\,100\,pc} \sim 1$ in starburst nuclei), then the CR escape time being longer than or comparable to their hadronic loss timescale (so they can be calorimeters as observed) requires an effective diffusivity $\kappa \lesssim 3\times 10^{31} L_{\rm halo,\,kpc}^{2} \Sigma_{\rm g\,cm^{-3}} / H_{\rm dense,\,100\,pc}$. This is completely consistent with the standard LISM diffusion coefficient, but tells us, as noted in those papers, that ion-neutral damping cannot boost the effective diffusivity or streaming speed by a huge factor in these starburst environments (despite the gas being overwhelmingly neutral).

\begin{figure}
	\centering
	\includegraphics[width=0.98\columnwidth]{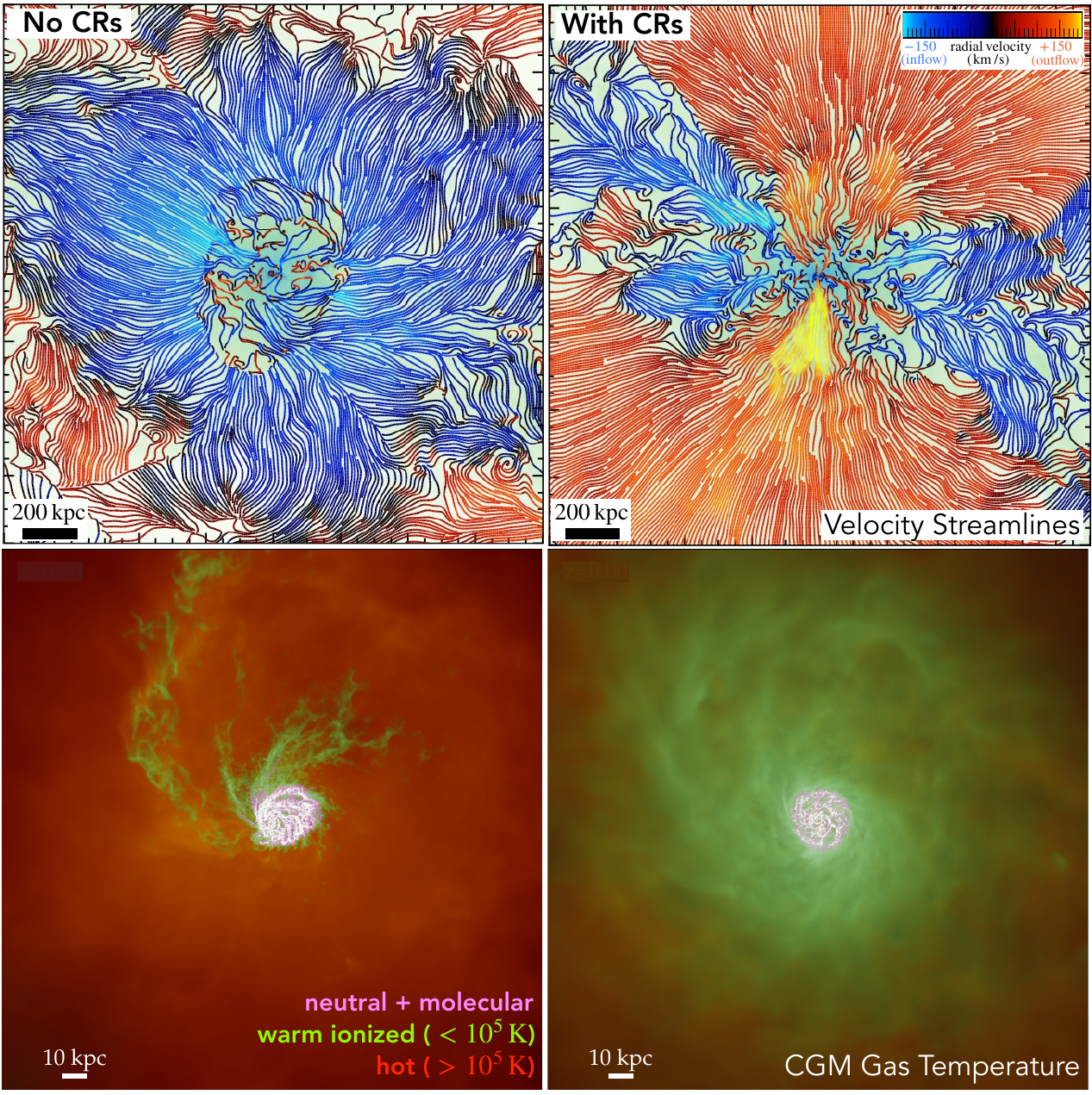}
	\caption{Illustration of the effects of CRs on the CGM in a CR pressure dominated halo ({\em right}) compared to a simulation without CRs ({\em left}), in cosmological CR-MHD galaxy formation simulations that explicitly evolve Eqs.~\ref{eqn:energy} for CR transport with an assumed universal scattering rate set to ISM-inferred values. The effects from Fig.~\ref{fig:cgm.cr.fx} are evident. (1) CR pressure continuously accelerates fountain flows into large-scale outflows reaching $\gtrsim$\,Mpc at sub-virial velocities, while outflows are strongly trapped and inflow dominates down to the disk in the no-CRs case. (2) CR pressure supports extended, cooler (warm ionized) gas in the halo, while without CRs the volume-filling halo phase can only be hot gas at the virial temperature, with small, dense clouds/filaments of cool gas raining onto the galaxy.
	\label{fig:cgm.cr.fx.sims}}
\end{figure} 

\subsection{Warm (\&\ Hot) ISM}
\label{sec:fx.ism}

In the warm (let alone hot) ISM, direct CR effects on thermochemistry become much weaker, as heating and ionization are dominated by photo-ionization and radiative recombination, shocks, turbulent dissipation, reconnection, and radiative+mechanical stellar feedback \citep{tielens:2005.book,draine:ism.book}. 
CR pressure effects are also weak. In the warm ISM this may be surprising as CR energy densities are similar to thermal, turbulent, radiation, and magnetic. But while $e_{\rm cr} \sim 3\,P_{\rm cr} \sim {\rm eV\,cm^{-3}}$ is similar to $e_{\rm thermal} \sim (3/2)\,P_{\rm therm}$, the {\em gradient} in CR pressure $\nabla P_{\rm cr} \sim P_{\rm cr} / \ell_{\nabla,\,\rm cr}$ is much weaker than the thermal/magnetic/radiation/turbulent pressure gradients (e.g.\ $P_{\rm therm}/\ell_{\nabla,\,\rm therm}$) by at least an order of magnitude, as CRs are distributed much more smoothly ($\ell_{\nabla,\,\rm cr} \gtrsim$\,kpc, while $\ell_{\nabla,\,\rm therm} \lesssim 100\,$pc). Related, if gas is compressed or shocked within the disk (let alone in smaller structures that will become GMCs, etc.), the CR diffusion time $\lesssim 3\,\bar{\nu}\,(100\,{\rm pc})^{2}/c^{2} \sim 10^{4}\,{\rm yr}$ is much smaller than salient dynamical times, so CRs escape rather than behaving adiabatically and doing PdV work on the gas. More subtle effects of CRs on the structure of turbulence and various mixing instabilities can arise, but these are generally small \citep[e.g.][]{bustard:2020.crs.multiphase.ism.accel.confinement,bustard:2023.cr.damping.turbulence}.

In the hot ISM, the imbalance is even more extreme in heating/ionization. For pressure, the diffuse or volume-filling hot phases have similar thermal pressure to warm ISM, so the same imbalance with the diffuse/volume-filling CR field persists, while for dense hot phases (e.g. SNRs and SNe bubbles), these are by definition powered by SNe which have not cooled, so even if CRs are perfectly-confined their pressure is lower than thermal by the ratio of CR energy to mechanical energy of SNe (a factor $\sim 10$). Moreover constraints on grammage accumulated near sources ($\lesssim 0.4\,{\rm g\,cm^{-2}}$; \citealt{krumholz:2024.beaming.argument.to.constrain.cr.transport.near.sources.would.apply.to.inhomogeneous.transport.models.like.irynas,ambrosone:2025.cr.source.cocoons.grammage}) indicate that CRs cannot be too strongly-confined in those regions. 

In principle, if one artificially forced much slower CR transport, in turn forcing the CR pressure gradient to $\sim 100$ times smaller scales than observed, one could generate large CR pressure effects in these environments. But this would strongly violate all of our direct observational constraints on CRs in the LISM and MW galaxy more broadly. So these conclusions are insensitive to CR transport models in detail, so long as those models match the wealth of observational constraints (the warm LISM being the regime where CR transport is best-constrained).

\begin{figure}
	\centering
	\includegraphics[width=0.98\columnwidth]{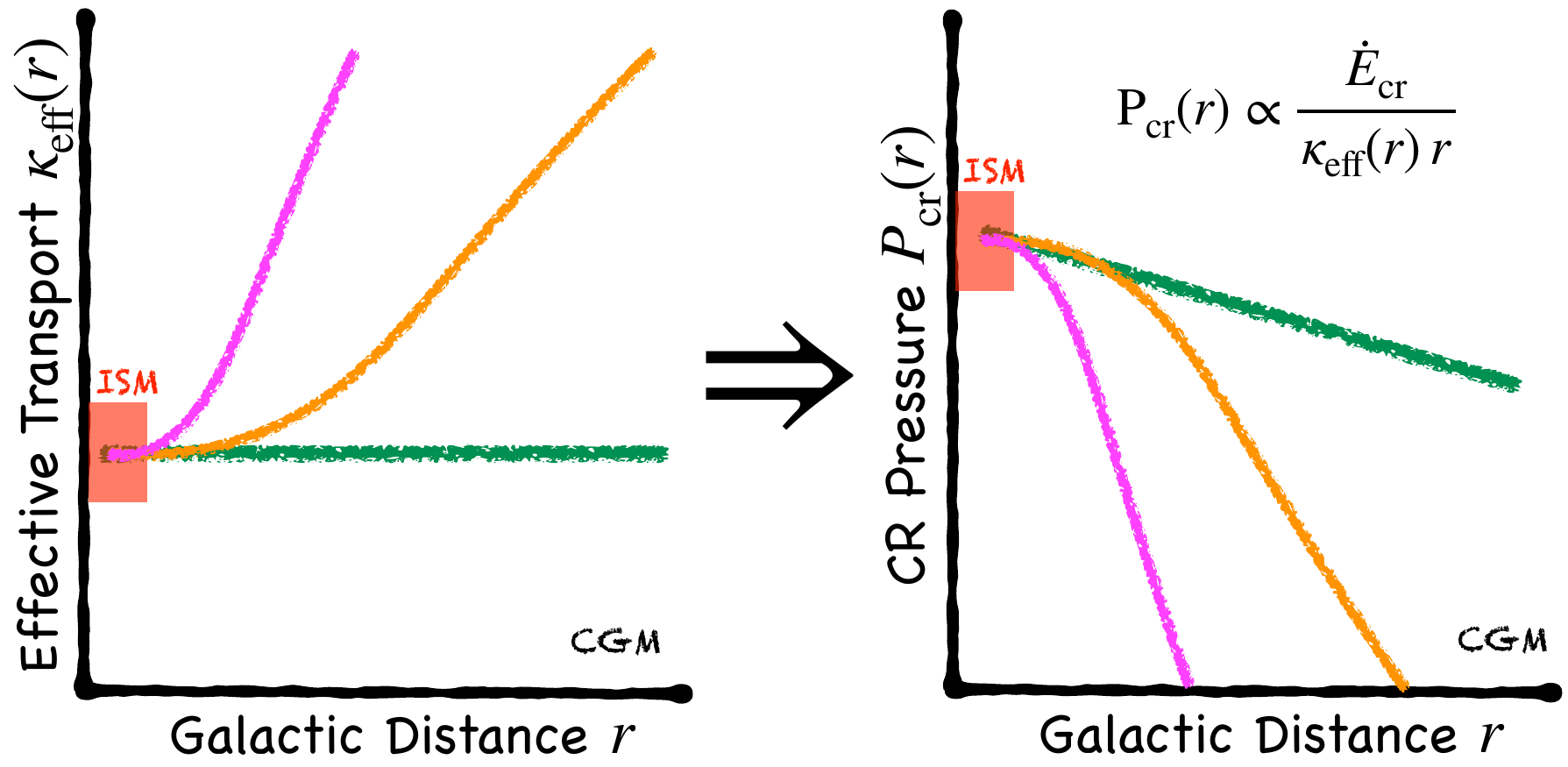}
	\caption{Cartoon illustrating how the importance of CRs in the CGM (Figs.~\ref{fig:cgm.cr.fx}-\ref{fig:cgm.cr.fx.sims}) depends sensitively on CR transport physics. 
	{\em Left:} Different illustrative models for how the effective diffusivity $\kappa_{\rm eff}$ (which includes diffusive/super-\Alf{ic}, streaming, and advective transport), averaged within spherical shells, depends on galactocentric radius $r$ (see Table~\ref{tbl:scattering.models} for physical examples). All are anchored to the same ISM constraints by construction, but depending on the physics, one could have constant-$\kappa_{\rm eff}$, $\kappa_{\rm eff} \propto r$ (constant-$v_{\rm stream}$), or $\kappa_{\rm eff} \propto r^{\alpha}$ rising more steeply or even falling (rapidly falling or rising CR scattering rates $\langle \bar{\nu}\rangle \propto r^{-\alpha}$). In steady-state in the CGM, assuming some smooth injection rate (proportional to the SNe and therefore star formation rate of the galaxy) $\dot{E}_{\rm cr}$, energy conservation implies the CR pressure $P_{\rm cr} \sim e_{\rm cr}/3 \sim \dot{E}_{\rm cr} / (12\pi\,\kappa_{\rm eff} r)$. Since $P_{\rm cr}$ is comparable to gas thermal pressure $P_{\rm therm}$ in the ISM, whether the effects in Figs.~\ref{fig:cgm.cr.fx}-\ref{fig:cgm.cr.fx.sims} occur depend on whether $P_{\rm cr}(r)$ falls more or less rapidly than $P_{\rm therm}(r)$, which is sensitive to the behavior $\kappa_{\rm eff}(r)$ in the CGM, far away from the observational constraints in Figs.~\ref{fig:specdemo}-\ref{fig:epos.losses}.
	\label{fig:cgm.cr.cartoon}}
\end{figure} 
\begin{figure*}
	\centering
	\includegraphics[width=0.8\textwidth]{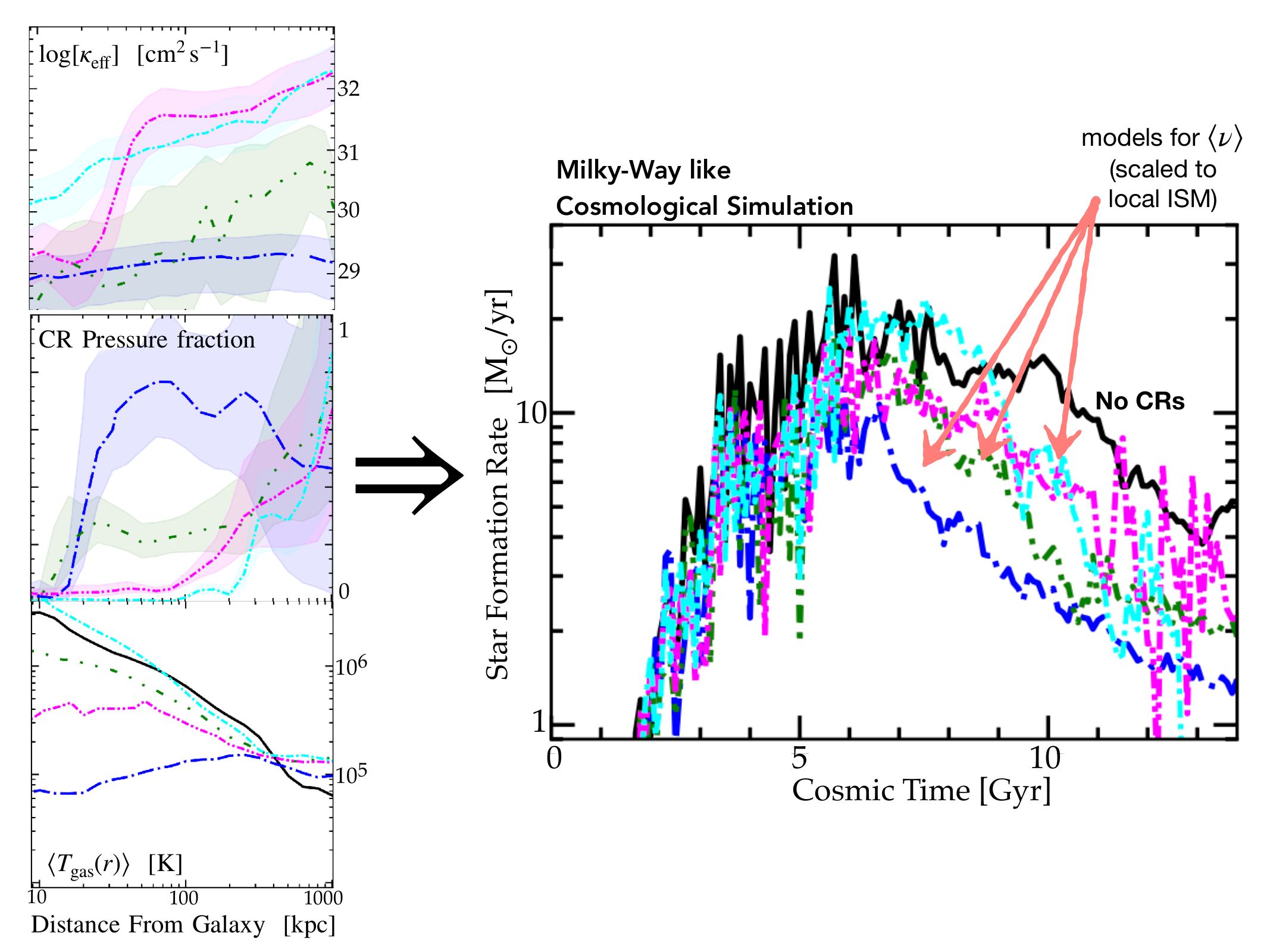}
	\caption{Quantitative demonstrations of the effects of different $\kappa_{\rm eff}(r)$ models on the CGM (per Fig.~\ref{fig:cgm.cr.cartoon}) in cosmological CR-MHD simulations (akin to those in Fig.~\ref{fig:cgm.cr.fx.sims}).
	Different models for $\kappa_{\rm eff}$ from Table~\ref{tbl:scattering.models}, all calibrated to give similar normalizations in the LISM, are evolved, giving ({\em left}) different $\kappa_{\rm eff}(r)$ in the CGM and correspondingly different CR pressure fractions at different radii (Fig.~\ref{fig:cgm.cr.cartoon}). Per Fig.~\ref{fig:cgm.cr.fx}, this in turn alters the CGM phases/temperatures, and (through inflow/outflows) modify galaxy star formation rates/masses.
	\label{fig:cgm.cr.sims}}
\end{figure*}

\subsection{CGM/IGM}
\label{sec:fx:cgm}

In the CGM, especially around dwarf and MW/M31-mass galaxies at low redshifts, it is rather easy in principle for CRs to dominate the pressure/energy density, owing to the  low thermal pressure, $P_{\rm therm} \sim 0.001\,{\rm eV\,cm^{-3}}\,(n/0.001\,{\rm cm^{-3}})\,(T/10^{4}\,{\rm K})$ and large scale-lengths $\gg$\,kpc. Indeed, most studies of CRs in galaxy formation have found this is the regime where global/cosmological effects of CR are maximized \citep{jubelgas:2008.cosmic.ray.outflows,booth:2013.cr.wind.early.idealized.disk.sims,Chen16,Rusz17,Buts18,chan:2018.cosmicray.fire.gammaray,ji:fire.cr.cgm,ji:20.virial.shocks.suppressed.cr.dominated.halos,su:turb.crs.quench,su:2021.agn.jet.params.vs.quenching,su:2024.fire.jet.sim.using.acc.jet.prescriptions.from.cho.multiscale.experiments,hopkins:2020.cr.transport.model.fx.galform,hopkins:cr.transport.constraints.from.galaxies,chan:2021.cosmic.ray.vertical.balance,hopkins:2021.sc.et.models.incompatible.obs,ponnada:2023.synch.signatures.of.cr.transport.models.fire,ponnada:2024.fire.fir.radio.from.crs.constraints.on.outliers.and.transport,ruszkowski.pfrommer:cr.review.broad.cr.physics,ramesh:2024.tng.plus.our.subgrid.crs.very.strong.fb.fx,martin.alvarez:radiation.crs.galsim.similar.conclusions.fire,martin.alvarez:2025.pandora.cr.bfield.highz.galaxy.burstiness.sf.winds,lu:2025.cr.transport.models.vs.uv.xray.obs.w.cric,dome:2025.no.burstiness.change.w.bfields.crs.other.physics.if.feedback.present.in.dwarfs.highz.galaxies,farcy:2025.cr.feedback.eor.galaxies.crs.increased.with.sne.decreased.lower.escape.fraction.but.similar.stellar.masses}. The most important effects are illustrated in heuristic form in Fig.~\ref{fig:cgm.cr.fx}, and in simulation examples in Fig.~\ref{fig:cgm.cr.fx.sims}. 

First, at the disk-halo interface $\sim$\,kpc above the disk (the CR scattering halo), most outflows driven by mechanical feedback have velocities below escape and will fall back or fountain, in the absence of CRs. But the known $\sim$\,kpc-scale CR pressure gradient -- while too shallow to be important within the star-forming/thin disk -- can now pick up on this material once it reaches $\gtrsim$\,kpc elevation, and re-accelerate it further out into the CGM. 

In the more extended CGM, the CR pressure gradient scale length should be of order the galacto-centric distance $r$, like other gradients (turbulent and thermal and magnetic and radiation), so if CRs dominate the pressure this can further accelerate material. Most notably since the acceleration from CRs $\sim \rho^{-1}\partial_{r} P_{\rm cr}$ is independent of gas temperature, this can carry away material which is otherwise quite cold or warm ($T \lesssim 10^{5}\,$K, i.e.\ below the virial temperature), and can carry it slowly (highly sub-sonic relative to the virial temperature, hence velocities well below the escape velocity) owing to the continuous acceleration \citep{Ipav75,Brei91,Ever08,Dorf12,uhlig:2012.cosmic.ray.streaming.winds,hanasz:2013.cosmic.ray.winds,booth:2013.cr.wind.early.idealized.disk.sims,Pakm16,lazarian:2016.cr.wave.damping,Rusz17,wiener:2017.cr.streaming.winds,zweibel:cr.feedback.review,Mao18:cr.winds.uplift.from.pressure,Jaco18,farber:decoupled.crs.in.neutral.gas,holguin:2019.cr.streaming.turb.damping.cr.galactic.winds,hopkins:2020.cr.outflows.to.mpc.scales,crocker:2021.starburst.no.cr.winds.only.lower.density.galaxies,girichidis:2021.cr.transport.w.spectral.reconnection.hack,quataert:2021.cr.outflows.diffusion.staircase,quataert:2022.isothermal.streaming.wind.analytic.cr.wind.models,thomas:2022.self-confinement.non.eqm.dynamics,sike:2024.cr.winds.pfrommer.model.launch.warm.gas}. 

At the same time, in the outer CGM, if CR pressure is important {\em and} CR diffusion/streaming is slow enough that at the largest (initial/driving) scales of the thermal instability, condensing/cooling modes in the CGM can carry CRs as they compress, then different thermal phases (cold/warm/hot gas) do not have to be in thermal pressure equilibrium (which would require $n\,k\,T=$\,constant, so $\rho \propto T^{-1}$, i.e.\ cold gas being denser). This allows for co-spatial gas at different temperatures, and diffuse cool gas, rather than shattering of the CGM into tiny cloudlets \citep{butsky:2020.cr.fx.thermal.instab.cgm,huang:2022.cr.mediated.thermal.instability,weber:2025.cr.thermal.instab.cgm.fx.dept.transport.like.butsky.study}. 

In turn, these processes have many indirect/non-linear effects on galaxy formation, most notably carrying outflows to the outer CGM/IGM, suppressing inflows, and therefore suppressing baryonic masses of galaxies. These have all been seen in many studies (extensive references above). AGN are just as capable of sourcing the CRs, in principle (see \S~\ref{sec:fx:agn}), as SNe, as once the CGM is reached the galaxy is effectively a point-source so the gas is agnostic to the source of the CR acceleration.

Many of the studies above, and especially comparisons like those in \citet{hopkins:2020.cr.transport.model.fx.galform}, have shown that the CGM/IGM is also where the effects of CRs are most sensitive to the CR transport physics, and those physics are most poorly constrained. Although there are some recent constraints reviewed in \S~\ref{sec:obs.cgm}, these are very recent and are only valid around MW/M31-mass systems on-average at $z\sim 0$, in a gross ensemble sense, and still have factor of several uncertainties. Extrapolation to dwarf galaxies, high redshifts, clusters, extreme plasma environments, beyond the virial radius (the IGM), and more remains deeply un-constrained. And this matters for the simple reason illustrated in Fig.~\ref{fig:cgm.cr.cartoon}. If galaxies are small compared to the CGM and the CGM is, to lowest-order, quasi-spherically symmetric, with low densities (so CR losses are relatively small), then for a given CR ``injection rate'' $\dot{E}_{\rm cr}$ (really the rate of CR \textit{escape} from the galaxy), the steady-state CR pressure at a given radius scales as $\sim \dot{E}_{\rm cr}/4\pi\,v_{\rm st,\,eff}^{\rm iso}\,r^{2} \sim \dot{E}_{\rm cr}/4\pi\,\kappa_{\rm eff}^{\rm iso}\,r$ -- i.e.\ inversely with the effective streaming speed or diffusivity. A number of detailed simulations (which allow for time-dependent CR-MHD, anisotropy, multi-phase structure, CR losses, etc.) have shown this is in practice a surprisingly accurate approximation to the average CR pressure gradients in the CGM \citep[see][]{hopkins:cr.mhd.fire2,ji:fire.cr.cgm,hopkins:2020.cr.transport.model.fx.galform,ji:20.virial.shocks.suppressed.cr.dominated.halos,butsky:2022.cr.kappa.lower.limits.cgm,hopkins:2022.cr.subgrid.model,ponnada:2023.synch.signatures.of.cr.transport.models.fire,lu:2025.cr.transport.models.vs.uv.xray.obs.w.cric}. But that means that if CR scattering rates decrease sufficiently rapidly in the CGM (plausible, given the lower densities, weaker magnetic fields, weaker turbulence, etc.), CRs will diffuse/stream ever-faster and their pressure will fall more rapidly, meaning they can never dominate the pressure far from the galaxy (to drive winds, suppress inflows, etc). Likewise modifying the thermal stability of the medium requires both a relatively large CR pressure and not-too-fast CR diffusion, since if CRs had an infinite diffusivity they would form a perfectly-uniform background pressure which could not enforce local gradients (Fig.~\ref{fig:cgm.cr.fx}).  It is worth noting that almost all of the referenced studies above simulating CRs in the CGM assume a single universal-in-time-and-space diffusion/streaming prescription. 

The practical effect of changing the local scaling of $\bar{\nu}$ (as a function of different plasma properties), and therefore effective $\kappa(r)$ or $v_{\rm st}(r)$, is shown in Fig.~\ref{fig:cgm.cr.sims}, with simulations from \citet{hopkins:2020.cr.transport.model.fx.galform}. As expected, models with higher $\kappa_{\rm eff}$ in the CGM produce lower steady-state CR pressure, more closely resembling simulations which neglect CRs entirely in their predictions for gross CGM properties like temperature, or effects of CR winds and pressure on suppressing star formation. Interestingly, of the models the authors surveyed therein which were calibrated to be consistent with LISM observations inside the ISM at $z=0$, almost all produced effects ``in-between'' two extremes: a model neglecting CRs entirely, and a model which assumed a universal constant diffusivity (the most common simulation choice) set to the LISM value.

\section{How Uncertain Is CR Transport, Really?}
\label{sec:uncertainty}

\begin{table*}
\begin{center}
\begin{footnotesize}
\caption{Modern CR Transport/Scattering Rate Models (Grouped by Category)\label{tbl:scattering.models}}
\begin{tabular}{| llll |}
\hline
Assumption/ & Effective $\kappa$ at $\sim$\,GeV & $\kappa \propto R_{\rm cr}^{\delta}$ (rigidity scaling: 1-100\,GV) & Notes \\
Model & (dimensional scaling with plasma properties) & (observed $\delta \sim 0.4-0.7$) &  \\
\hline\hline
Advective & $\sim v_{\rm wind}\,\ell_{\nabla} \sim v_{\rm esc}(M_{\rm halo},\,\ell_{\nabla})\,\ell_{\nabla}$ & $0$ & Winds, etc (\S~\ref{sec:advective}) \\
\hline\hline
\multicolumn{4}{|c|}{Self-Confinement ($\delta^{\prime}$ assumes model fine-tuned to observed CR spectral shapes, but any $\delta > 0$ rapidly decays to $\delta \le 0$; \S~\ref{sec:super.alfvenic})} \\
\hline
Saturated & $\sim v_{A}\,\ell_{\nabla} \propto {B\,\ell_{\nabla}}{f_{\rm ion}^{-1/2} \rho^{-1/2}}$ & $0$ & pure/\Alf{ic} streaming \\ 
Damped: Dust & $\sim \frac{\ell_{g} \Gamma_{\rm dust} e_{\rm B}}{v_{A,\,i}\,\nabla_{\|} P_{\rm cr}} \propto {B^{1.8}\,f_{\rm dg}\,\ell_{\nabla}}{e_{\rm cr}^{-1}}$ & $\delta^{\prime} \sim\xi_{d}+\alpha \sim 1.1$, decays to $\le 0$ & weak-RDI limit $\Gamma$ \\
Damped: Ion-Neutral & $\propto \rho^{3/2}\,f_{\rm ion}^{1/2} f_{\rm neutral} T_{\rm gas}^{1/2}\,\ell_{\nabla}\,e_{\rm cr}^{-1}$ & $\delta^{\prime} \sim1+\alpha \sim 1.7$, decays to $\le 0$ & 
$\Gamma \propto f_{\rm neutral} T_{\rm gas}^{1/2} \rho$, if $f_{\rm ion} \ll 1$ \\ 
Damped: Ion Landau & $\propto f_{\rm ion}\,\rho\,B^{-1/2} v_{\rm turb} \ell_{\nabla}^{1/2} T_{\rm gas}^{1/2}\,e_{\rm cr}^{-1}$ & $\delta^{\prime} \sim1/2+\alpha \sim 1.2$, decays to $\le 0$ &  $\Gamma \sim 0.4\,c_{s}\,(k_{\|}/\ell_{A})^{1/2}$ \\
Damped: Turbulent & $\propto f_{\rm ion} \rho^{1/2} B^{1/2} v_{\rm turb} \ell_{\nabla}^{1/2} e_{\rm cr}^{-1}$ & $\delta^{\prime} \sim1/2+\alpha \sim 1.2$, decays to $\le 0$ & $\Gamma \sim v_{A,\,\rm ideal}\,(k_{\|}/\ell_{A})^{1/2}$\\ 
Damped: Nonlinear Landau & $\propto f_{\rm ion}^{1/4} B T_{\rm gas}^{1/4} \ell_{\nabla}^{1/2} e_{\rm cr}^{-1/2}$ & $\delta^{\prime} \sim (1+\alpha)/2 \sim 0.9$, decays to $\le 0$ & $\Gamma \propto c_{s} k\, \mathcal{E}_{\pm}$  \\
\hline\hline
\multicolumn{4}{|c|}{Extrinsic Turbulence (accounting for anisotropy when $\ell_{g} \ll \ell_{A}$ and damping when $\ell_{g} \ll \ell_{\rm visc}$; \S~\ref{sec:et.alfven}-\ref{sec:et.fast})} \\
\hline
\Alf\ or Slow & $\propto B\,\ell_{\nabla}\,v_{\rm turb}^{-2}\,\rho^{1/2}$ & $0$ (balanced) or $<0$ (unbalanced) & cannot assume isotropy at $\ell_{g} \ll \ell_{A}$ \\ 
Fast & $\propto \ell_{\nabla}^{1/3} v_{\rm turb}^{-2/3} B^{2/3} f_{\rm ion}^{-1/3} \rho^{-1/3}$   & $0$ (TTV) or $<0$ (resonant) & viscous damping, $\mathcal{E}\propto k^{-2}$ \\ 
 & $\propto T_{\rm gas} f_{\rm ion}^{3/10} \rho^{3/10} B^{1/5} \ell_{\nabla}^{1/10} v_{\rm turb}^{-1/5}$  & $\le 0$ & viscous damping, $\mathcal{E}\propto k^{-3/2}$ \\
 & $\propto \ell_{\nabla} B T_{\rm gas}^{1/2} \rho_{i}^{-1/2} v_{\rm turb}^{-2}$  & $\le 0$ & Landau damping, $\mathcal{E}\propto k^{-3/2}$ \\
 Nonresonant & $\propto \ell_{0} B^{2} \delta B[\ell_{0}]^{-2}$ or $\ell_{0}^{-1}\delta B[\ell_{0}]^{-2}$  & $0$ ($\ell_{\rm modes} \gg \ell_{g}$) or $2$  ($\ell_{\rm modes} \ll \ell_{g}$) & \S~\ref{sec:no.resonant} \\ 
 \hline\hline
\multicolumn{4}{|c|}{Intermittent/Patchy (\S~\ref{sec:intermittency})} \\
\hline
Micro-Mirrors & $\propto T_{\rm gas}^{-1/2} B^{-1}$ & $2$ (if volume-filling) & requires small collisionality \\
Folds/field reversals & depends on driver of folds & traces fold size-spectrum & requires intermittent $\sim$\,au folds \\
Nonlinear ``patches'' & $\propto \ell_{g}\,f_{\rm vol}^{-1}[\ell=\ell_{g}]$ & patch size spectrum $f_{\rm vol}\propto \ell^{1-\delta}$ & transverse patch size $\sim \ell_{g}$\\
\hline
\end{tabular}
\end{footnotesize}
\end{center}
\end{table*}

Given the questions raised related to the effects of different Galactic CR transport models, and our inability to extrapolate present observational constraints to different extragalactic environments, it is important to ask -- just how uncertain, really, is CR transport from a (plasma physics) theoretical point of view? 
From \S~\ref{sec:transport}, it is important to note that the dominant unknown today in CR transport is not the formalism or closures or transport equations themselves: we have a reasonably good handle on the correction equations to integrate (even if some older implementations still use ad-hoc equations not valid in various limits). The overwhelming uncertainty is in the micro-physical scattering rates $\bar{\nu}$, and ensuing effective transport coefficients (e.g.\ ``effective'' $\kappa$ or $v_{\rm st}$).

\subsection{Different Models for CR Transport: Dependence on Plasma Properties}
\label{sec:uncertainty:plasma.props}

Table~\ref{tbl:scattering.models} lists a number of distinct microphysical models for CR scattering rates, including models with simple advection, pure streaming, un-saturated/damped streaming, extrinsic turbulence, and some recently-proposed nonlinear and/or intermittent mechanisms. For each, we note how the effective diffusivity $\kappa_{\rm eff} \equiv \langle |F_{\rm cr} |\rangle / \langle | \nabla_{\|} e_{\rm cr} | \rangle$ at energies around $\sim$\,GeV is predicted to scale with plasma properties ($\kappa_{0}$ or $D_{0}$). We also then look at their dependence on CR energy or rigidity in LISM like conditions ($\delta$). 
We stress these are all state-of-the-art models with serious calculations in the last decade or so looking at their potential leading role in CR scattering \citep[the scalings shown are compiled and used in a wide range of studies including][]{yan.lazarian.04:cr.scattering.fast.modes,yan.lazarian.2008:cr.propagation.with.streaming,lazarian:2016.cr.wave.damping,zweibel:cr.feedback.review,Rusz17,farber:decoupled.crs.in.neutral.gas,thomas.pfrommer.18:alfven.reg.cr.transport,thomas:2021.compare.cr.closures.from.prev.papers,krumholz:2020.cr.transport.starbursts.upper.limit.kappa.gamma.rays,kempski:2020.cr.soundwave.instabilities.highbeta.plasmas.resemble.perseus.density.fluctuation.power.spectra,kempski:2023.large.amplitude.fluctuations.and.cr.scattering,kempski.li.2024:unified.cr.scattering.plasma.scattering.from.strong.field.curvature.intermittent.ism.structure.explained.together,kempski:2021.reconciling.sc.et.models.obs,shalaby:2021.cr.acoustic.type.instability,lazarian:2021.cr.scattering.mirrors,hopkins:cr.transport.constraints.from.galaxies,hopkins:2021.sc.et.models.incompatible.obs,squire:2021.dust.cr.confinement.damping,ji:2021.mhd.pic.rsol,xu:2022.cr.streaming.turb.ism.mirrors.classical.damping,schekochihin:2020.mhd.turb.review,thomas:2022.self-confinement.non.eqm.dynamics,lazarian:2023.turb.dcf.method.agrees.skalidis.scaling.is.the.right.one,butsky:2023.cosmic.ray.scattering.patchy.ism.structures,reichherzer:2023.micromirror.cr.confinement.hot.galaxy.cluster.icm,fitzaxen:2024.cr.transport.into.gmcs.suppressed.starforge,ponnada:2024.fire.fir.radio.from.crs.constraints.on.outliers.and.transport,sike:2024.cr.winds.pfrommer.model.launch.warm.gas,barretomota:2024.mirror.scattering.ism.crs,habegger:2024.cr.propagation.feedback.in.multiphase.ism.turbulence,ewart:2024.cr.confinement.micromirror.cgm.of.massive.clusters,lu:2025.cr.transport.models.vs.uv.xray.obs.w.cric}.
 
Physically, these differences come from different assumptions and priors regarding (1) which terms in e.g.\ Eq.~\ref{eqn:twomoment} are most important; (2) what micro-scale \textit{types} of structures in the magnetic fields are most important for scattering; and (3) what determines the power spectra, amplitudes, and/or volume-filling factors of those structures (e.g.\ what physics sources or drives the structures, and what physics suppresses or damps them). Each permutation of these produces different scalings for $\kappa_{\rm eff}$. 
 
First consider their systematic dependence on plasma properties around a fixed energy $\sim$\,GeV or rigidity $\sim$\,GV 
(what is most relevant for e.g.\ spectrally-integrated CR transport). 
We will consider their energy/rigidity dependence (given otherwise fixed plasma properties) below. 
If we take the inclusive set of {\em just} the models in Table~\ref{tbl:scattering.models} (by no means exhaustive), and wished to parameterize the resulting scaling, we require $\kappa_{\rm eff}$ of the form:
\begin{align}
\label{eqn:kappa.eff.compare} \kappa_{\rm eff} &\propto v_{\rm esc}^{\alpha_{v}} \ell_{\rm cr,\nabla}^{\alpha_{\nabla}} B^{\alpha_{B}} f_{\rm ion}^{\alpha_{i}} \rho^{\alpha_{\rho}} e_{\rm cr}^{\alpha_{\rm cr}} f_{\rm dg}^{\alpha_{\rm dg}} T_{\rm gas}^{\alpha_{T}} f_{\rm neutral}^{\alpha_{n}} v_{\rm turb}^{\alpha_{t}} \ , 
\end{align}
where $v_{\rm esc}$, $\ell_{\rm cr,\,\nabla}$, $B$, $f_{\rm ion}$, $\rho$, $e_{\rm cr}$, $f_{\rm dg}$, $T_{\rm gas}$, $f_{\rm neutral}$, $v_{\rm turb}$ represent an escape/virial velocity, CR gradient scale-length, magnetic field strength, gas ionization fraction, gas density, CR energy density, dust-to-gas mass ratio, gas temperature, neutral fraction, and turbulent velocity (at $\sim \ell_{\rm cr,\,\nabla}$), each scaling with a power law which could lie in the range:
\begin{align}
\nonumber 0 &\le \alpha_{v} \le 1 \ ,& \ 0 \le \alpha_{\nabla} \le 1 \\ 
\nonumber -1 &\le \alpha_{B} \le 2 \ ,& \ -1/2 \le \alpha_{i} \le 1 \\ 
\nonumber  -1/2 &\le \alpha_{\rho} \le 3/2 \ ,& \ 0 \le \alpha_{\rm dg} \le 1 \\ 
\nonumber  -1/2 &\le \alpha_{T} \le 1/2 \ ,& \  0 \le \alpha_{n} \le 1 \\
\nonumber  -2 &\le \alpha_{t} \le 1 \ ,& \ -1 \le \alpha_{\rm cr} \le 0 
\end{align}
And for completeness note that the energy dependencies range from $-2 \le \delta \le 2$. 

Clearly, there is a huge range of possible scalings! Even for an obvious parameter like $B$, these models range from predicting $\kappa \propto 1/B$ to $\kappa \propto B^{2}$ -- and others surveyed in the literature \citep[see][]{hopkins:cr.transport.constraints.from.galaxies} can give dependencies as extreme as $\propto B^{\pm 4}$. 
As a result, even if we were to re-normalize each of the models in Table~\ref{tbl:scattering.models} to give the same $\kappa_{\rm eff}$ in LISM conditions, if we extrapolate them all to CGM conditions at $r \gtrsim 100\,$kpc, we obtain predicted $\kappa_{\rm eff}$ ranging over $>7$ orders of magnitude!

\subsection{Classical CR Scattering Models and their Failure to Reproduce Proper Rigidity Dependence}
\label{sec:uncertainty:rigidity}

Before proceeding, we stress that the arguments below apply to each classical model considered {\em in isolation} as the dominant, volume-filling source of CR scattering across a broad rigidity range in the ISM. They do not rule out hybrid or environment-dependent scenarios in which different physical mechanisms dominate in different energy/rigidity regimes or environments (e.g.\ self-confinement at low energies in the disk, extrinsic turbulence at high energies in the halo). Indeed, such mixed scenarios are part of the motivation for the newer models discussed in \S~\ref{sec:intermittency}. With that caveat, it remains the case that {\em none} of the classical CR scattering models reviewed in Table~\ref{tbl:scattering.models} predicts anything close to the correct dependence of CR scattering on rigidity or CR energy in the LISM, if it is taken to be the sole mechanism! 
Indeed, the classic theoretical models for the source of CR scattering widely considered in the literature for the last $\sim 60$ years face many serious problems if one attempts to use them to predict scattering rates $\bar{\nu}$ and subsequent CR spectra and products in the MeV-TeV range in the ISM. Some of these  are reviewed in \citet{hopkins:2021.sc.et.models.incompatible.obs} and \citet{kempski:2021.reconciling.sc.et.models.obs}, and some were recognized even in the first papers on the subject \citep{skilling:1971.cr.diffusion,1971BAAS....3..480C}, others in many subsequent papers \citep[see e.g.][]{chandran00,yan.lazarian.02,yan.lazarian.04:cr.scattering.fast.modes,cho.lazarian:2003.mhd.turb.sims,lazarian:2016.cr.wave.damping,fornieri:2021.comparing.et.models.data.et.only.few.hundred.gv}. These are the subject of many detailed studies, but we pedagogically summarize here one of the most serious issues, which robustly indicates that one of the core assumptions of these models must be incorrect.

One central assumption of these models -- often unstated -- is that the CR scattering in the ISM (at least as relevant for the LISM spectra) is dominated by volume-filling fluctuations, which are smooth on meso-scales and larger. If this is true, the fluctuations that scatter $\lesssim\,$TeV CRs must be low-amplitude, $|\delta {\bf B}| / |{\bf B}| \sim 10^{-4} - 10^{-3}$, and quasi-linear theory used to derive the CR scattering coefficients is an excellent approximation. 

In this case, we can approximate the scattering rate $\bar{\nu}$, contributed by fluctuations with wavelength $\lambda_{\|}$ and wavenumber $k_{\|} \sim 1/\lambda_{\|}$, as:
\begin{align}
\label{eqn:nu} \bar{\nu} \sim \frac{3\pi}{16} \frac{k_{\|} v_{\rm cr}}{(1 + k_{\|}^{2} \ell_{g}^{2})} \frac{k_{\|} \mathcal{E}(k_{\|})}{e_{\rm B}}
\end{align}
where $\mathcal{E}$ is the power spectrum (so $k_{\|} \mathcal{E}(k_{\|}) \sim |\delta B(\lambda_{\|})|^{2}/8\pi$), and we are careful to denote the \textit{parallel} wavenumber is what enters here.
Note Eq.~\ref{eqn:nu} is a simple approximation to the behaviors $\bar{\nu} \propto k_{\|}^{2}\mathcal{E}(k_{\|})$ for $k_{\|} \ell_{g} \ll 1$ (long wavelengths) and $\bar{\nu} \propto  \ell_{g}^{-2} \mathcal{E}(k_{\|})$ for $k_{\|} \ell_{g} \gg 1$ (short wavelengths); full expressions for $\nu$ are derived in classic texts (and are quite complicated!) but Eq.~\ref{eqn:nu} is sufficient to capture all of the key behaviors from these more general expressions. 

Now consider different regimes from Table~\ref{tbl:scattering.models} in turn, bearing in mind that {\em whatever} physics controls $\kappa_{\rm eff}$ (regardless of whether or not it is actually diffusive), the residence time in the escape-limited regime must decrease with CR energy/rigidity, roughly as $\Delta t_{\rm res} \propto 1/\kappa_{\rm eff} \propto R_{\rm cr}^{-\delta} \propto E_{\rm cr}^{-\delta}$ with $\delta \sim 0.5$ or so, from $\sim 0.1-1$\,GeV to $\sim$\,TeV energies. This is one of the most robustly and significantly-observed facts we know about Galactic CR transport.

\subsubsection{Advective or Pure-Streaming (Saturated Self-Confinement) Models}
\label{sec:advective}

If $\bar{\nu}$ is sufficient large -- e.g.\ {\em something}, be it self-confinement, extrinsic turbulence, or any other physics, produces enough scattering waves $\mathcal{E}$, then $\kappa_{\|} \sim c^{2}/3\bar{\nu}$ becomes very small, and the CR transport becomes purely advective+streaming, i.e.\ 
\begin{align}
D_{t} \bar{f}_{1d} &\rightarrow - \nabla \cdot ({\bf v}_{e} \bar{f}_{1d}) + ... \\ 
\Delta t_{\rm res} &\rightarrow \frac{H_{\rm disk,\,halo}}{|{\bf v}_{e} \cdot \hat{z}|} 
\end{align}
with ${\bf v}_{e} \equiv {\bf u} + \bar{v}_{A} \bhat$. Trivially, this means CRs of different rigidity are moving {\em at the same speed}, which means the \textit{residence time is independent of rigidity} ($\delta = 0$), i.e.\ ${\bf v}_{e}$ has no dependence on CR momentum! Thus this is immediately ruled out at the dominant CR transport case over any appreciable dynamic range in $E_{\rm cr}$.

\begin{figure*}
	\centering
	\includegraphics[width=0.9\textwidth]{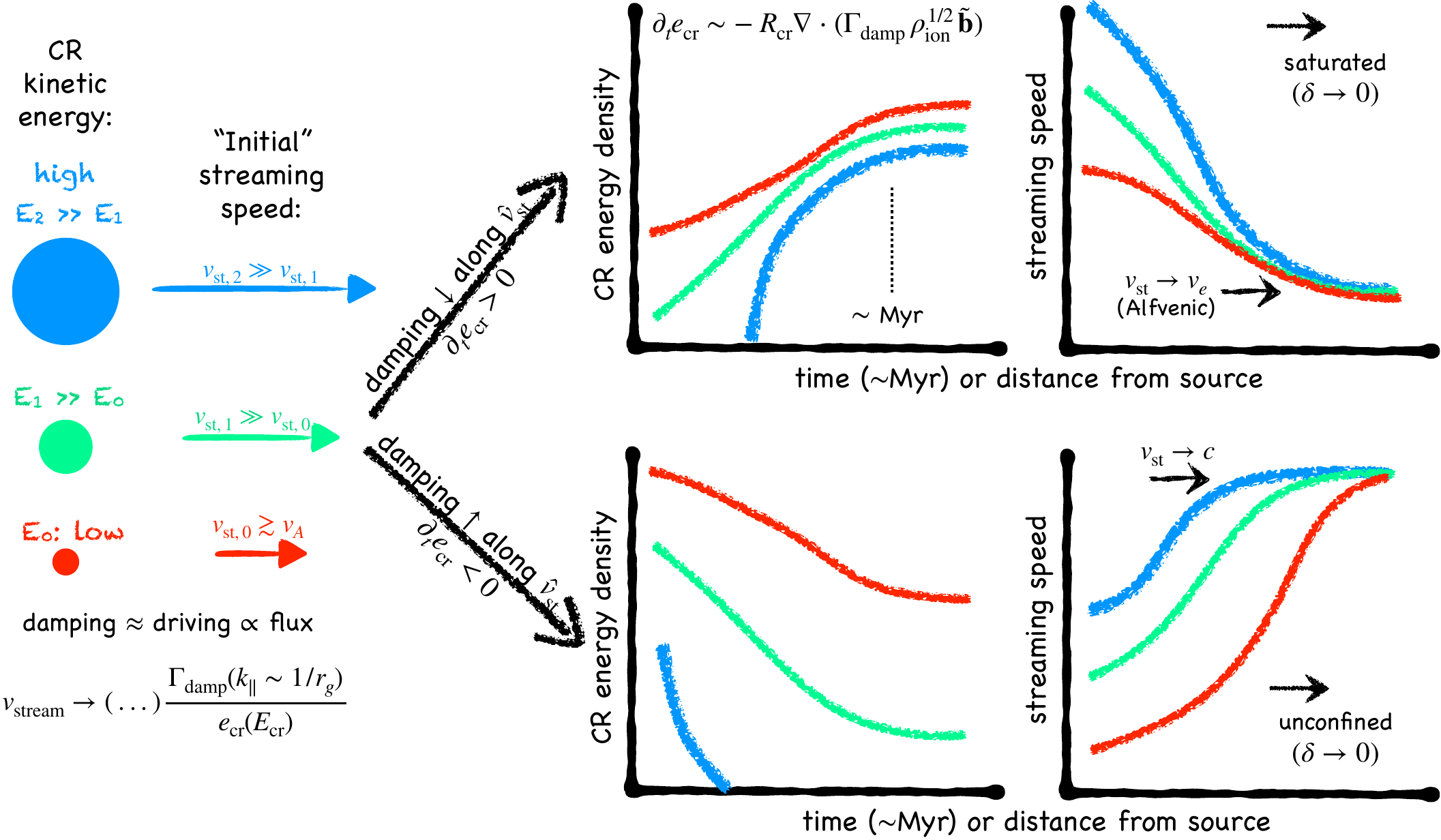}
	\caption{Illustration of the ``solution collapse'' problem in classical self-confinement transport models. 
	High-energy CRs must have shorter residence times, hence faster effective streaming speeds $v_{\rm st,\,eff}$. To prevent self-confinement from collapsing/trapping all CRs to stream at ${\bf v}_{e} = \bar{v}_{A}\bhat + {\bf u}$ (which would imply a clearly-ruled-out $\delta = 0$), this requires the growth of self-confinement-excited scattering modes (growth rate proportional to the super-\Alf{ic} CR flux $e_{\rm cr}\,v_{\rm stream}$ at each $E_{\rm cr}$) be balanced by damping $\Gamma_{\rm damp}$. These are equilibrium solutions to the CR flux ($\partial_{t} \bar{f}_{1}$ or $\partial_{t} F_{\rm cr}$) and scattering-mode ($\partial_{t} \bar{\nu}$ or $\partial_{t} \mathcal{E}$) equations, but {\em not} the CR number/energy/distribution function equation $\partial_{t} e_{\rm cr}$ (or $\partial_{t} \bar{f}_{0}$ or $\partial_{t} n_{\rm cr}$), and the normal CR flux $\nabla \cdot (v_{\rm cr} \bar{f}_{1} \bhat)$ becomes a source/sink term as shown. If $\Gamma_{\rm damp}$ decreases along the streaming direction, CRs pile up and their energy increases, more rapidly for higher-energy CRs, increasing $\bar{\nu}$ and $e_{\rm cr}$ until either radiative/catastrophic losses dominate at all energies (ruled-out) or all CRs reach the saturated self-confinement limit and lock to the same streaming speed ${\bf v}_{e}$ ($\delta \rightarrow 0$).
	Alternatively if $\Gamma_{\rm damp}$ increases along streaming, the opposite occurs, CRs de-confine rapidly until self-confinement becomes negligible (either free-streaming out at $\sim c$, or some other, non self-confinement scattering dominates).
	\label{fig:probs.sc}}
\end{figure*}

\subsubsection{Damped Streaming (Un-Saturated Self-Confinement) Models}
\label{sec:super.alfvenic}

The next case is illustrated in Fig.~\ref{fig:probs.sc}. 
In self-confinement models,\footnote{Here ``self-confinement'' refers to the broad category of models in which the scattering modes producing $\nu$ are directly driven by some positive power of the CR energy density or flux.} if the self-confinement is incomplete or unsaturated ($\bar{\nu}$ remains finite) because the modes $\mathcal{E}$ are damped, then one can have diffusive-like or ``super-\Alf{ic} streaming'' (i.e.\ CRs are not trapped together moving at ${\bf v}_{e}$). 
To see the behaviors in this limit, return to the gyro-averaged CR transport equations in \S~\ref{sec:transport.meso} (for more details, see e.g.\ Appendices in \citealt{hopkins:2021.sc.et.models.incompatible.obs}). 
Everything here can be derived from the general Vlasov equation for CRs \citep[e.g.][]{1975MNRAS.173..255S}, but it is easier to see if we start from the multi-moment equations for $f$. 
In self-confinement, there are three coupled equations: CR number/energy ($\bar{f}_{0}$), flux/anisotropy ($\bar{f}_{1}$ or $\langle \mu\rangle \equiv \bar{f}_{1}/\bar{f}_{0}$), and amplitude of scattering waves $\mathcal{E}$, specifically gyro-resonant parallel \Alf\ waves ($k_{\|} \approx 1/\ell_{g}$), since these are directly driven by the CR streaming itself. 
\begin{align}
\nonumber D_{t}& \bar{f}_{0}  + \nabla \cdot  (v_{\rm cr} \bhat\,\bar{f}_{1})  
=  j_{0} + 
\\
\nonumber &  
\frac{1}{p_{\rm cr}^{2}}\frac{\partial}{\partial p_{\rm cr}}\left[ p^{3}_{\rm cr}\,\left\{  \left( \mathcal{R} + \mathbb{D}:\nabla{\bf u} + \bar{\nu}  \frac{v_{A}^{2}}{v_{\rm cr}^{2}} \chi \psi_{p} \right) \bar{f}_{0}  
+ \bar{\nu} \frac{\bar{v}_{A}}{v_{\rm cr}}\,\bar{f}_{1}
\right\} \right] \ ,  \\
\nonumber 
D_{t} & \bar{f}_{1} +  
v_{\rm cr} \bhat \cdot \nabla \cdot (\mathbb{D} \bar{f}_{0}) = -\bar{\nu} \left[ \bar{f}_{1} +   \frac{\bar{v}_{A}}{v_{\rm cr}} \chi \psi_{p} \bar{f}_{0}  \right] + j_{1}  \ , \\
\nonumber D_{t} & \hat{\mathcal{E}}_{\pm} 
\pm \nabla \cdot \left( v_{A}\, \hat{\mathcal{E}}_{\pm}  \bhat\right) = - \frac{\hat{\mathcal{E}}_{\pm}}{2}\,\nabla \cdot {\bf u} + S_{\pm} - \Gamma\,\hat{\mathcal{E}}_{\pm} \ ,
\end{align}
where $\hat{\mathcal{E}}_{\pm} \equiv k_{\|}\mathcal{E}_{\pm}(k_{\|} = 1/\ell_{g}[R_{\rm cr}])$ is the energy in gyro-resonant \Alf\ waves propagating in the forward/backward ($\pm$) direction, $\Gamma$ some damping,\footnote{Here $\Gamma$ represents any damping processes operating on parallel \Alf\ waves at $k_{\|} \sim \ell_{g}$. Our derivation is agnostic to the form of $\Gamma$, but if it depends on $\hat{\mathcal{E}}_{\pm}$ or $\bar{f}_{1}$, then in our final expression we should take $\Gamma$ evaluated for their local-equilibrium values derived below. Known damping sources include turbulent shearing and induced ion-Landau damping 
$\Gamma \sim  [v_{A,\,{\rm ideal}}+0.4\,c_{s}]\,(k_{\|}/\ell_{A})^{1/2}$, 
dust damping 
$\sim 0.02\,k v_{A} f_{\rm dg}\,(k/k_{d})^{-\xi_{d}}$ (expressions in \citealt{squire:2021.dust.cr.confinement.damping}), 
ion-neutral collisional damping 
$\sim (\alpha_{i {\rm H}} + \alpha_{i {\rm He}})/2\rho_{i}$, 
and nonlinear Landau damping 
$\sim \sqrt{\pi} c_{s} k_{\|} \hat{\mathcal{E}}_{\pm}/8 e_{\rm B}$.
Turbulent, dust, and ion-neutral damping are straightforward and their dimensional scalings have been explicitly verified in PIC/MHD-PIC simulations \citep{bai:2019.cr.pic.streaming,hopkins:2019.mhd.rdi.periodic.box.sims,ji:2021.cr.mhd.pic.dust.sims,plotnikov:2021.cr.mhd.pic.sims.streaming.ion.neutral.strong.damping.deconfines,bambic:2021.mhd.pic.transport.inhomogeneous.ionization.effective.confinement.very.different.from.length.corr.of.turb.much.similar.to.cr.fluid.dynamics.models}. There may be other, undiscovered damping processes, and it has been suggested from kinetic theory and simulations that nonlinear Landau may not be able to operate  \citep{volk.cesarsky:1982.nonlinear.landau.damping,squire:2017.max.braginskii.scalings,holcolmb.spitkovsky:saturation.gri.sims,schroer:2025.nll.damping.sims.pic.but.weak.and.non.expected.saturation,lemmerz:2025.theory.extending.volk.cesarsky.resonant.trapping.crs.nll.minor.correction}, although for observed ISM/CGM parameters the nonlinear damping is already $\sim 4$ orders-of-magnitude smaller in the warm/hot volume-filling phases than turbulent+dust damping (at all $R_{\rm cr}$).} and $S_{\pm}$ is the source/driving term, which comes (by definition in self-confinement theories) from the CRs themselves. $S_{\pm}$ can be derived from simple energy conservation arguments (it is the energy being lost in the streaming loss term; see \citealt{thomas.pfrommer.18:alfven.reg.cr.transport}), or from 
kinetic theory  \citep{wentzel:1968.mhd.wave.cr.coupling,kulsrud.1969:streaming.instability,1975MNRAS.172..557S,1975MNRAS.173..245S,1975MNRAS.173..255S,holman:1979.cr.streaming.speed}, as\footnote{$S_{\pm}$ can also be written in terms of the streaming energy loss from the CR energy equation, per differential unit momentum (gyro-resonant $k_{\|}$)
\begin{align}
S_{\pm} &\sim -p\frac{\partial}{\partial p} \left\{\pm\frac{{v}_{A}}{v_{\rm cr}^{2}} {\nu}_{\pm} [F_{\rm cr} \mp 3\chi {v}_{A}(e_{\rm cr} + P_{\rm cr})  \right\}
\end{align}
or as
\begin{align}
S_{\pm} &\sim \gamma^{\pm}_{\rm growth} \mathcal{E}_{\pm}
\end{align}
where $\gamma_{\rm growth}^{\pm} \sim \pm (3\pi/16)\,\Omega\,({v}_{A}\,p/v_{\rm cr} e_{\rm B})\,(4\pi p^{3})\,[v_{\rm cr} \bar{f}_{1} \pm {v}_{A} \chi \psi_{p} \bar{f}_{0}]$ is the linear growth rate of the CR resonant streaming instability, often written (in the streaming direction which matters here) as $\gamma_{\rm growth} \sim \gamma\,\Omega (n_{\rm cr}/n_{\rm ion})\,(v_{\rm cr}/\bar{v}_{A})\,[\langle \mu\rangle - \bar{v}_{A}\,\chi\,\psi_{p}/v_{\rm cr}] = \gamma\,\Omega (n_{\rm cr}/n_{\rm ion})\,[v_{\rm stream}/v_{A} - \chi\,\psi_{p}]$ \citep{kulsrud.1969:streaming.instability,1975MNRAS.173..245S}.} 
\begin{align}
 S_{\pm} & \equiv \nu_{\pm} \frac{(\pm v_{A})}{v_{\rm cr}} 4\pi p^{3} E_{\rm cr} \left[ \bar{f}_{1} + \frac{(\pm v_{A})}{v_{\rm cr}} \chi\psi_{p} \bar{f}_{0} \right] \ ,
\end{align}
with the scattering rate
\begin{align}
\nu_{\pm}& \approx \frac{3\pi}{16}\,\Omega_{\rm cr}\,\frac{\hat{\mathcal{E}}_{\pm}}{e_{\rm B}}  =  \frac{3\pi}{16}\,\frac{v_{\rm cr}}{\ell_{g}}\,\frac{\hat{\mathcal{E}}_{\pm}}{e_{\rm B}} 
\end{align}
coming from the self-excited (gyro-resonant) waves \citep[e.g.][]{1975MNRAS.172..557S,thomas.pfrommer.18:alfven.reg.cr.transport}. 

In this scenario, we consider just transport (outside of sources so $j_{0}\rightarrow 0$, $j_{1}\rightarrow 0$), and the flux and wave amplitude equations evolve much faster than the $\bar{f}_{0}$ equation ($\bar{f}_{1}$ and $\mathcal{E}$ evolve on of order the CR streaming instability growth time $\sim 10^{7}\,\Omega_{\rm cr}^{-1}$ (since the linear growth rate $\gamma_{\rm growth} \sim \Omega_{\rm cr}\,(n_{\rm cr}/n_{\rm ion})\,(v_{\rm stream}/v_{A}) \sim 10^{-7}\,\Omega_{\rm cr}$ for typical ISM CR and ion densities) or scattering time $\sim \nu^{-1}$, both $\sim 10-100\,$yr, while $\bar{f}_{0}$ evolves on Myr), so we can treat them as in local steady-state ($D_{t} \bar{f}_{1}\rightarrow 0$, $D_{t} \mathcal{E}_{\pm} \rightarrow 0$), and neglect the advection/gradient terms in $D_{t} \hat{\mathcal{E}}_{\pm}$ (since those also evolve slowly). 
Note that the branch of $\hat{\mathcal{E}}_{\pm}$ and corresponding $\nu^{\pm}$ ($\hat{\mathcal{E}}^{\rm stream}$, $\nu^{\rm stream}$) in the streaming direction (sign $ - [\bhat \cdot \nabla (\mathbb{D} \bar{f}_{0} )] / |\bhat \cdot \nabla (\mathbb{D} \bar{f}_{0} )|$) will be amplified by $S$, while the counter-streaming $\hat{\mathcal{E}}^{\rm counter}$ is damped, so $\bar{\nu} \equiv \nu_{+} + \nu_{-} \rightarrow \nu_{\pm}^{\rm stream}$ and $\bar{v}_{A} \equiv v_{A}\,(\nu_{+} - \nu_{-})/(\nu_{+} + \nu_{-}) \rightarrow \pm v_{A}$, depending on whether the $+$ or $-$ direction is amplified, respectively. We therefore have: 
\begin{align}
\bar{f}_{1} &+ \frac{\bar{v}_{A}}{v_{\rm cr}} \chi \psi_{p} \bar{f}_{0} \rightarrow - \frac{v_{\rm cr}}{\bar{\nu}} \bhat \cdot \nabla \cdot (\mathbb{D} \bar{f}_{0}) \ , \\ 
\nonumber \hat{\mathcal{E}}_{\pm}^{\rm stream}& \rightarrow \frac{S_{\pm}}{\Gamma} \rightarrow \frac{{\nu_{\pm}}}{\Gamma} \frac{(\pm {v}_{A})}{v_{\rm cr}} 4\pi p^{3}  E_{\rm cr} \left[ \bar{f}_{1} + \frac{ (\pm {v}_{A} )}{v_{\rm cr}} \chi\psi_{p} \bar{f}_{0} \right]  \\ 
\nonumber &=
\frac{\bar{\nu}}{\Gamma} \frac{\bar{v}_{A}}{v_{\rm cr}} 4\pi p^{3}  E_{\rm cr} \left[ \bar{f}_{1} + \frac{ \bar{v}_{A} }{v_{\rm cr}} \chi\psi_{p} \bar{f}_{0} \right] \\ 
\nonumber &=  - \frac{\bar{v}_{A}}{\Gamma} 4\pi p^{3}  E_{\rm cr}  \bhat \cdot \nabla \cdot (\mathbb{D} \bar{f}_{0})
= \frac{{v}_{A}}{\Gamma} 4\pi p^{3}  E_{\rm cr}  | \bhat \cdot \nabla \cdot (\mathbb{D} \bar{f}_{0})|
\ , \\
\nonumber \hat{\mathcal{E}}_{\pm}^{\rm counter} &\rightarrow 0 \ , \\
\nonumber \bar{\nu} &\equiv \nu_{+} + \nu_{-} \rightarrow  \frac{3\pi}{16}\,\Omega_{\rm cr}\,\frac{\hat{\mathcal{E}}_{\pm}^{\rm stream}}{e_{\rm B}} \\
\nonumber &= \frac{3\pi^{2} v_{A} \Omega_{\rm cr} p^{3} E_{\rm cr}}{4 \Gamma e_{\rm B}}  | \bhat \cdot \nabla \cdot (\mathbb{D} \bar{f}_{0})| \ .
\end{align}

If we simply stopped here, and then inserted by fiat the observed CR spectral shapes (assuming a universal $\bhat \cdot \nabla \cdot (\mathbb{D} \bar{f}_{0}) \propto \bar{f}_{0} \propto p_{\rm cr}^{-4-\alpha}$ with $\alpha \sim 0.7$ above $\gtrsim$\,GV), this would appear to give $\kappa \propto v_{\rm cr}^{2}/\bar{\nu} \propto \Gamma / (\Omega_{\rm cr} p^{3} E_{\rm cr} p_{\rm cr}^{-4-\alpha}) \propto \Gamma \,p_{\rm cr}^{\alpha+1} \propto p_{\rm cr}^{\delta}$, which for different physical damping mechanisms (giving different scalings of $\Gamma(k_{\|} \sim 1/\ell_{g})$) might naively {\em appear} to give some $\delta > 0$ (though the values they give, noted in Table~\ref{tbl:scattering.models}, are still ruled-out).

But this is not self-consistent, because those spectral shapes/gradients are \textit{not} valid solutions of the CR number/energy/$\bar{f}_{0}$ equation!
To be self-consistent, insert our expression for $\bar{\nu}$ back into our expression for $\bar{f}_{1}$ to obtain:
\begin{align}
\bar{f}_{1} &+ \frac{\bar{v}_{A}}{v_{\rm cr}} \chi \psi_{p} \bar{f}_{0} \rightarrow \frac{4 v_{\rm cr} e_{\rm B} \Gamma}{3\pi^{2} p^{3} E_{\rm cr} \Omega_{\rm cr} \bar{v}_{A}} \ . 
\end{align}
The right-hand side here (the term $\propto \Gamma$) is the ``super-\Alf{ic} streaming'' (or ``diffusive like'') transport term that arises because of finite $\nu$ which itself (in this scenario) owes to finite damping $\Gamma$.
Now we can insert this expression into the CR number/energy/Fokker-Planck equation, and we have:
\begin{align}
\label{eqn:f0.damped.sc} D_{t}& \bar{f}_{0}  \rightarrow - \nabla \cdot  \left[ \frac{4 v^{2}_{\rm cr} e_{\rm B} \Gamma}{3\pi^{2} p^{3} E_{\rm cr} \Omega_{\rm cr} \bar{v}_{A}}\bhat
-\bar{v}_{A}\chi\psi_{p} \bhat \bar{f}_{0}  \right]  + 
\\
\nonumber &  
\frac{1}{p_{\rm cr}^{2}}\frac{\partial}{\partial p_{\rm cr}}\left[ p^{3}_{\rm cr}\,\left\{  \left( \mathcal{R} + \mathbb{D}:\nabla{\bf u} \right) \bar{f}_{0}  
-\bar{v}_{A} \bhat \cdot \nabla \cdot (\mathbb{D} \bar{f}_{0}) \right\} \right] \\ 
\nonumber 
&= -\frac{4 v_{\rm cr}}{3\pi^{2} p_{\rm cr}^{3} E_{\rm cr}} \nabla \cdot \left( \frac{\ell_{g} e_{B} \Gamma }{\bar{v}_{A}} \bhat \right)  - \nabla \cdot ({\bf v}^{\rm Alfvenic}_{\rm stream} \bar{f}_{0}) + ({\rm losses})\ .
\end{align}
In the last expression we note that this neatly divides into three terms: the ``super-\Alf{ic} streaming'' term in $\nabla \cdot (\bar{v}_{A}^{-1} \ell_{g} e_{B} \Gamma \bhat)$, the regular \Alf{ic} streaming term (streaming speed $-\bar{v}_{A} \chi \psi_{p} \bhat$), and losses which include the usual radiative/catastrophic ($\mathcal{R}$), adiabatic ($\mathbb{D}:\nabla {\bf u}$),  and ``streaming losses'' $-\bar{v}_{A} \bhat \cdot \nabla \cdot (\mathbb{D} \bar{f}_{0}) = v_{A} |\bhat \cdot \nabla \cdot \mathbb{D} \bar{f}_{0}|$. 

In this scenario (by definition) we are interested in cases where the ``super-\Alf{ic}'' term dominates the transport. Otherwise, either (1) if \Alf{ic} streaming dominates ($\Gamma$ small), we are back in the ``saturated/pure streaming'' limit of \S~\ref{sec:advective}; or (2) if losses dominate, all CR energy is lost near sources (discussed below). 
So to leading order:
\begin{align}
\label{eqn:source.f0.only} D_{t} \bar{f}_{0} & \rightarrow -\frac{4 v_{\rm cr}}{3\pi^{2} p_{\rm cr}^{3} E_{\rm cr}} \nabla \cdot \left( \frac{\ell_{g} e_{B} \Gamma }{\bar{v}_{A}} \bhat \right) =  -\frac{ v_{\rm cr}  \nabla \cdot \left( { \rho_{\rm ion}^{1/2}  \Gamma } \tilde{\bf b} \right) }{3 \pi^{5/2} p_{\rm cr}^{2} Z e E_{\rm cr}} 
\end{align}
where $\tilde{\bf b} \equiv (v_{A}/\bar{v}_{A})\,\bhat = \pm \bhat$ points in the streaming direction.
Or more simply, using $d e_{\rm cr} / d\ln{ p_{\rm cr}} = 4\pi p_{\rm cr}^{3} E_{\rm cr} \bar{f}_{0}$, we can write:
\begin{align}
\label{eqn:cr.source.egy} \frac{\partial}{\partial t} \left( \frac{d e_{\rm cr}}{d \ln p_{\rm cr} }\right) & \sim - \frac{30\,{\rm eV}}{{\rm cm^{3}\,Myr}}\, {\rm pc}\, \nabla \cdot \left(\psi \tilde{\bf b} \right)
\end{align}
with
\begin{align}
\nonumber \psi &\sim 30\,f_{\rm neutral} f_{\rm ion}^{1/2} T_{4}^{1/2} \rho_{24}^{3/2} R_{\rm GV} 
+ 
f_{\rm dg,\odot} \rho_{24}^{3/16} B_{\rm \mu G}^{13/8} R_{\rm GV}^{3/8} \\
\nonumber & + 
f_{\rm ion}^{1/2}T_{4}^{1/2} \ell_{A,10}^{-1/2} \rho_{24}^{1/2} B_{\rm \mu G}^{1/2} R_{\rm GV}^{1/2}
+ 
0.3\,f_{\rm ion}^{1/2} \ell_{A,10}^{-1/2} B_{\rm \mu G}^{3/2} R_{\rm GV}^{1/2} \\ 
 & + 
0.1\,f_{\rm ion}^{1/4} T_{4}^{1/4} \tilde{e}_{\rm cr, eV}^{1/2} \ell_{\rm \nabla,kpc}^{-1/2} \rho_{24}^{1/4}  R_{\rm GV}^{1/2}
\end{align}
where the terms in $\psi$ correspond to the damping rate $\Gamma$ from ion-neutral collisions \citep{lee.volk:1973.damping.alfven.waves.smalscale,foote.kulsrud:1979.damping}, dust \citep{squire:2021.dust.cr.confinement.damping}, turbulent ion-Landau \citep{zweibel:cr.feedback.review}, turbulent shear \citep{yan.lazarian.02,farmer.goldreich.04}, and non-linear Landau \citep{kulsrud.1969:streaming.instability,volk.mckenzie.1981,cesarsky.kulsrud:1981.cr.confinement.damping.hot.gas,volk.cesarsky:1982.nonlinear.landau.damping} processes, respectively.\footnote{With definitions: $f_{\rm neutral} \equiv \rho_{\rm neutral}/\rho$, $f_{\rm ion} \equiv \rho_{\rm ion}/\rho$, $T_{4} \equiv T_{\rm gas}/10^{4}\,{\rm K}$, $\rho_{24} \equiv \rho / 10^{-24}\,{\rm g\,cm^{-3}}$, $R_{\rm GV} \equiv R_{\rm cr}/{\rm GV}$, $f_{\rm dg,\,\odot} = \rho_{\rm dust}/0.01\,\rho$, $B_{\rm \mu G} \equiv |{\bf B}|/{\rm \mu G}$, $\ell_{A,\,10} \equiv \ell_{A}/10\,{\rm pc}$, $\tilde{e}_{\rm cr,\,eV} \equiv |d e_{\rm cr} / d\ln R_{\rm cr}|/{\rm eV\,cm^{-3}}$, $\ell_{\nabla,\,{\rm kpc}} \equiv \ell_{\nabla}/{\rm kpc} = (\tilde{e}_{\rm cr} / |\nabla \tilde{e}_{\rm cr}|)/{\rm kpc}$.}

As noted in some of the very first papers on self-confinement referenced above, this leads to the bizarre behavior that the flux term actually mathematically appears as a source/sink term in the equation for $\partial_{t} f$, which depends on the divergence of plasma properties but not on $f$ itself! Realistically, $\nabla \cdot (\psi \tilde{\bf b})$ will not generically vanish in the ISM, and indeed $\psi$ depends on quantities like gas densities, ionization and neutral fractions, that have gradient scale-lengths as small as sub-pc scales. This means there is no steady-state value of $f$: it will grow or decay rapidly on $\sim$\,Myr timescales. Moreover because $\psi$ depends on $R_{\rm cr}$, there is not even a steady-state {\em shape} of the CR spectrum (i.e.\ $\alpha$ and therefore $\delta$) that can be defined. 

Physically, one of two things happens, illustrated in Fig.~\ref{fig:probs.sc}. 
(1) Suppose the damping $\Gamma$ (or more accurately the function $\psi$) decreases along the CR streaming direction. Since the growth rate of the scattering modes $S$ depends directly on the CR flux $\bar{f}_{1} + (\bar{v}_{A}/v_{\rm cr})\,\chi\psi_{p} \bar{f}_{0} \propto  v_{\rm st,\,eff}\,e_{\rm cr}$ (in some interval of CR momentum/rigidity), this means weaker damping allows scattering modes to grow more rapidly, so the flux decreases along this direction until $S$ balances $\Gamma$. This in turn leads to a $\nabla \cdot {\bf v}_{\rm st,\,eff} < 0$ or pileup of CRs -- i.e.\ the source term Eqs.~\ref{eqn:f0.damped.sc}-\ref{eqn:cr.source.egy} is positive $D_{t}\bar{f}_{0} > 0$ so $e_{\rm cr}$ increases rapidly (faster with smaller gradients or larger $R_{\rm cr}$, per Eq.~\ref{eqn:cr.source.egy}). As $\bar{f}_{0}$ or $e_{\rm cr}$ increases, so the CR scattering mode amplitudes and advective/streaming term ($\propto \nabla \cdot ({\bf v}_{e} e_{\rm cr})$) and loss ($\propto \mathcal{R} e_{\rm cr}$) terms in $\partial_{t} f$ also increase. Since $\psi$ depends on positive powers of $p_{\rm cr}$, this happens faster for higher-energy CRs, flattening the CR spectra (to $\alpha < 0$; mathematically this is just the statement $\partial_{p} (D_{t} \bar{f}_{0}) > 0$). 
Eventually, $\bar{f}_{0}$ must become so large that either (a) the advective/streaming terms dominate in Eq.~\ref{eqn:f0.damped.sc}, i.e.\ one reaches the saturated limit of \S~\ref{sec:advective}, with $\delta \rightarrow 0$, or (b) losses dominate at all $E_{\rm cr}$ in Eq.~\ref{eqn:f0.damped.sc}, in direct contradiction to what we know about CR spectra (and for hadrons at $>0.1$\,GeV, giving $\delta < 0$ owing to the dependence of loss rates on rigidity). 

Alternatively (2), if $\psi$ (i.e.\ $\Gamma$) increases along the streaming direction, $D_{t} \bar{f}_{0} < 0$, CRs become progressively less confined and stream faster along that direction, $\nabla \cdot {\bf v}_{\rm st,\, eff} > 0$ and so $e_{\rm cr}$ decreases rapidly. Again this occurs most rapidly at high energies so $\alpha \rightarrow \infty$ ($\partial_{p} (D_{t} \bar{f}_{0}) < 0$), and (because now the loss and \Alf{ic} streaming/advection terms $\propto \bar{f}_{0}$ in Eq.~\ref{eqn:f0.damped.sc} are getting \textit{smaller}) it will only ceases when either (a) the CR scattering mean-free-path becomes macroscopic (larger than $\ell_{\nabla}$) and so CRs effectively free-stream at $c$ and $\delta \rightarrow 0$, or (b) some {\em other} physics dominates the scattering rate, so this derivation does not apply at all.

Note that properly modeling this solution collapse (1) requires accounting for the dynamical dependence of $\bar{\nu}$, $\hat{\mathcal{E}}_{\pm}$, $\bar{f}_{1}$ and $\bar{f}_{0}$ on local properties and each other (rather than just imposing some ``effective'' streaming or diffusion); 
(2) requires that one include (in the volume-filling phases that matter for CR confinement in the Galaxy) the known linear damping mechanisms in diffuse gas, not just non-linear Landau damping; 
and 
(3) requires high resolution, multi-phase, turbulent ISM dynamics, since the growth timescale of the solution collapse is directly proportional to the gradient length scale of $\psi$ and therefore variables like $B$, $T$, $\rho$, $f_{\rm ion}$ in the ISM (i.e.\ solution collapse is artificially suppressed if small-scale ISM structure is suppressed either because of resolution or physics/subgrid model limitations).

\begin{figure}
	\centering
	\includegraphics[width=1.02\columnwidth]{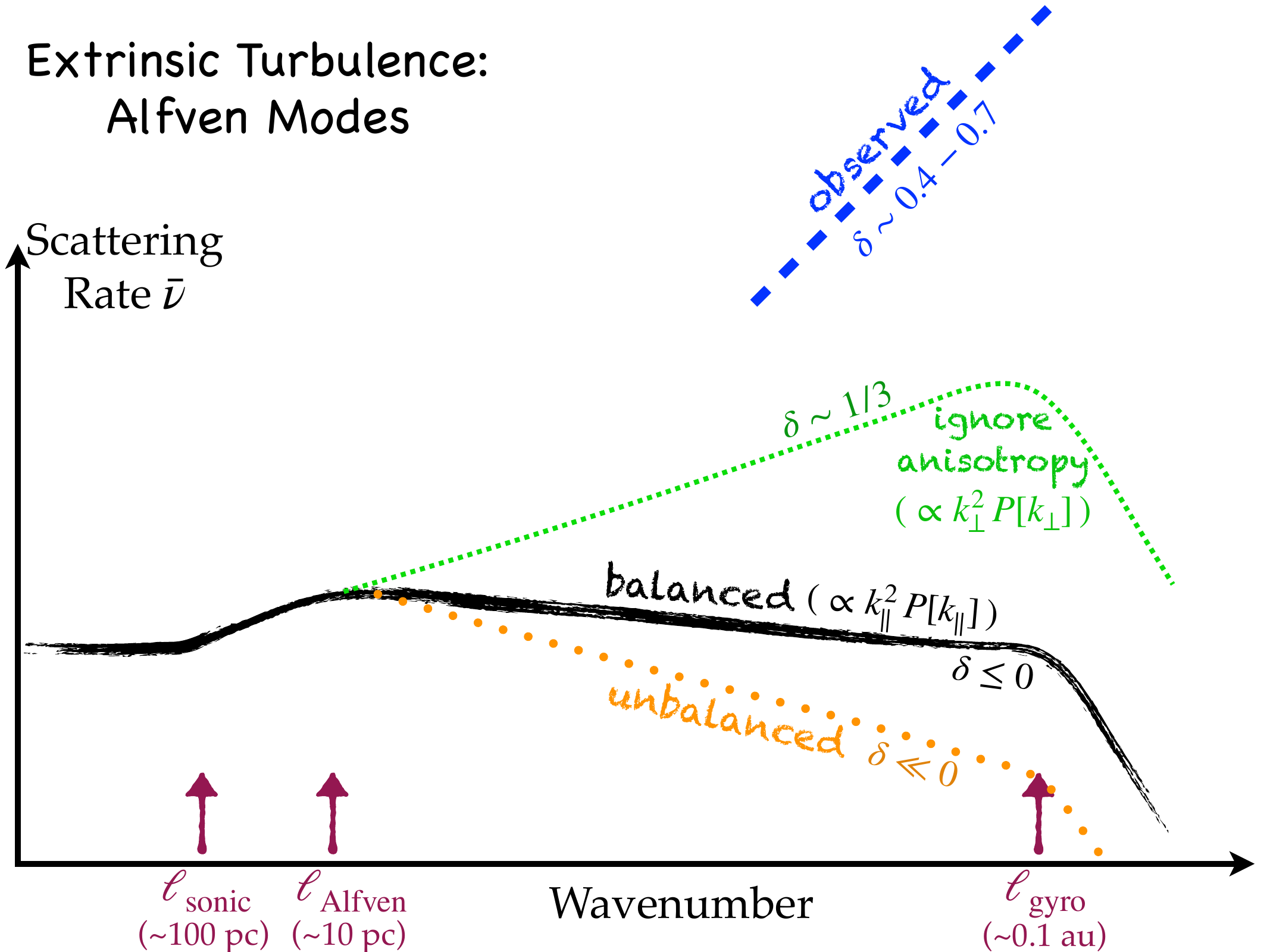}\\
	\vspace{0.5cm}
	\includegraphics[width=1.02\columnwidth]{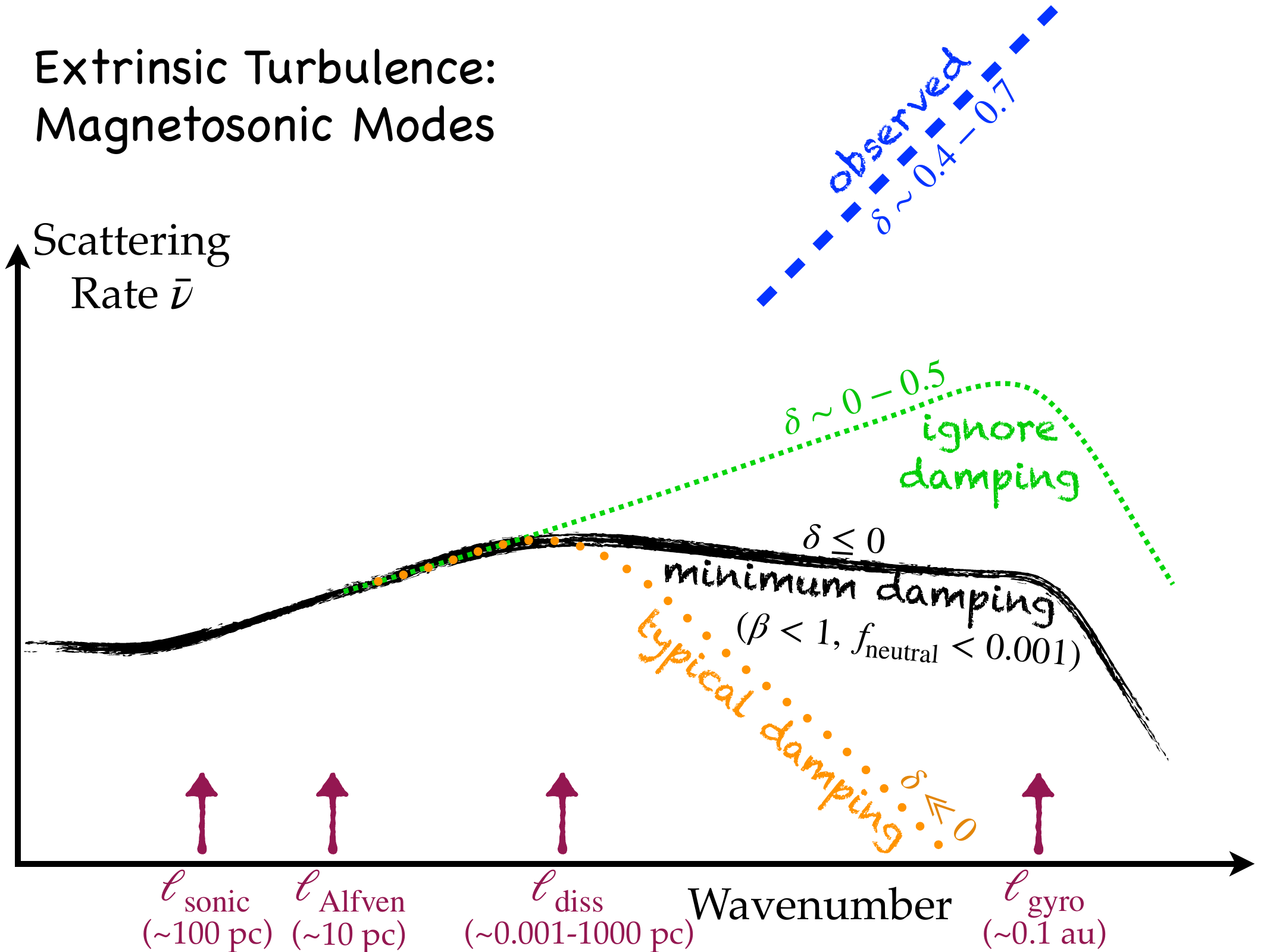}
	\caption{Problems in classical extrinsic turbulence models (CR scattering from a turbulent cascade from meso-scale). 
	We plot scattering rate $\bar{\nu}$ contributed by turbulent fluctuations on scale $\ell$, from driving to sub-gyro scales. To reproduce the observed dependence of $\bar{\nu}$ on rigidity for GeV-TeV CRs, this must match the ``observed'' line. 
	{\em Top:} \Alf{ic} (or slow) modes. If we ignore anisotropy -- i.e.\ take $\bar{\nu} \propto k^{2} P(k)$ or $\propto k_{\bot}^{2} P(k_{\bot})$ -- we predict a Kolmogorov spectrum which still under-predicts observed scattering rates but at least gives $\delta \sim 1/3>0$. But $\bar{\nu}$ depends on the {\em parallel} wavenumber $k_{\|} = {\bf k} \cdot \bhat$. On scales where $\delta B \ll B$ (below the \Alf\ scale $\sim 10\,$pc), it is mathematically impossible to have an isotropic \Alf{ic} cascade with $k_{\|} \sim k \sim k_{\bot}$: fluctuations perpendicular and parallel to mean field ${\bf B}$ are {\em not} equivalent, and $k_{\|}^{2} P(k_{\|})$ is suppressed. This suppresses the predicted scattering rates by factors $\gtrsim 10^{10}$. Critically-balanced cascades produce the maximum possible $\delta$ (slightly $<0$), while any un-balanced cascade produces more-negative $\delta \ll 0$.
	{\em Bottom:} Magnetosonic (fast) modes. These can be isotropic and have inertial-range spectra giving $\delta \sim 0-0.5$ (depending on model), but are strongly-damped below a dissipation scale $\ell_{\rm diss} \sim 0.001-1000\,$pc (warm ISM to CGM), including all gyro scales below TeV (warm ISM) to PeV (CGM). The minimum-possible damping (collisionless damping in a dust-free, fully-ionized plasma with $\beta_{\rm plasma} = P_{\rm thermal}/P_{\rm B} \ll 1$), truncates the spectrum to give $\delta$ slightly $<0$ {\em independent} of the inertial-range spectrum, while any stronger damping (e.g.\ non-zero dust or neutrals or $\beta_{\rm plasma} \gtrsim 1$) truncates spectra more sharply ($\delta \ll 0$).
	\label{fig:probs.et}}
\end{figure}

\subsubsection{Extrinsic Turbulence from \Alf{ic} Modes: Anisotropy}
\label{sec:et.alfven}

The primary alternative theory to self-confinement for decades has been ``extrinsic turbulence,'' in which one assumes $\mathcal{E}(k_{\|})$ in Eq.~\ref{eqn:nu} arise from a turbulent cascade from meso-scales or larger. In this case, the allowed range in $\delta$ translates directly to the required shape of the parallel power spectrum, $\mathcal{E}(k_{\|}) \propto k_{\|}^{-\xi}$ with $1.2 \lesssim \xi \lesssim 1.6$ ($\delta \approx 2-\xi$) -- but this is not actually possible at the scales of interest! Fig.~\ref{fig:probs.et} attempts to illustrate this qualitatively. 

Since we are (by definition) considering small-amplitude magnetic fluctuations, they can be decomposed into \Alf{ic}, slow and fast magnetosonic modes. First consider \Alf\ and slow modes (where especially \Alf\ modes can be weakly-damped down to the ion gyro scale). 
Here one can immediately show that any cascade model with $\delta > 0$ is mathematically impossible. 
The problem is anisotropy. If one simply replaced $k_{\|}^{2}\mathcal{E}(k_{\|})$ in Eq.~\ref{eqn:nu} with $k^{2}\mathcal{E}(k)$ or $k_{\bot}^{2} \mathcal{E}(k_{\bot})$, then it is possible to construct a cascade with power spectra giving $\delta \sim 0.3-0.5$ (depending on model assumptions; see \citealt{schekochihin:2020.mhd.turb.review}). 
However, CR scattering does not depend on $k_{\bot}$ or $k$, but on $k_{\|}$ \citep[see detailed discussion and derivations in e.g.][]{wentzel.1969.streaming.instability,kulsrud.1969:streaming.instability,1975MNRAS.172..557S,mckenzie.1982.streaming.instability.nonlinear,chandran00,yan.lazarian.02,cho.lazarian:2003.mhd.turb.sims,farmer.goldreich.04}, and one {\em cannot} assume isotropy ($k_{\bot} \sim k_{\|} \sim k$) for low amplitude ($|\delta {\bf B}|\ll |{\bf B}|$, or $k \gg 1/\ell_{A}$) fluctuations (precisely because $|\delta {\bf B}| \ll |{\bf B}|$, and so ${\bf B}$ imposes anisotropy). In that case, as shown in e.g.\ \citet{1994ApJ...432..612S,GS95.turbulence,boldyrev:2005.dynamic.alignment,Boldyrev2006} and many subsequent studies, the MHD equations themselves forbid any ``cascade'' (in the sense of any power being transferred from larger-to-smaller scales in $k_{\|}$ as opposed to $k_{\bot}$) which would have $\mathcal{E}(k_{\|})$ shallower than $k_{\|}^{-2}$, which gives $\bar{\nu} \propto k_{\|}^{0}$, i.e.\ $\delta=0$. This is independent of any model assumptions. Indeed, a perfectly critically-balanced cascade, which gives $\delta$ just slightly smaller than $0$ (owing to some logarithmic corrections that come from solving the full integral equations for CR scattering; see \citealt{chandran00}) turns out to be the maximum-possible value of $\delta$ one can achieve for a weak or strong cascade in \Alf/slow modes at scales below the \Alf\ scale ($|\delta {\bf B}|\ll {\bf B}$) -- any un-balanced cascade actually produces a {\em steeper} spectral truncation at scales below $\ell_{A}$, so a more negative $\delta$ \citep{lazarian:2016.cr.wave.damping,hopkins:2021.sc.et.models.incompatible.obs,kempski:2021.reconciling.sc.et.models.obs}. Likewise allowing for dissipative or kinetic effects only further truncates the spectra and pushes $\delta$ negative.

Even more serious than the effect on the {shape} of the spectrum, it is also important to note that this anisotropy \textit{enormously} suppresses the normalization of the CR scattering rates, reducing them by factors of $\gg 10^{10}$ \citep{chandran00,yan.lazarian.02} compared to the naive expectations of isotropic \citet{kolmogorov:turbulence} or \citet{kraichnan:1965.ik.aniso.turb} turbulence given the observed ISM power spectra (the ``great power law on the sky''; e.g.\ \citealt{lazio:2004.great.power.law.on.the.sky}).

\subsubsection{Extrinsic Turbulence from Magnetosonic Modes: Damping}
\label{sec:et.fast}

What about fast magnetosonic modes? These can be isotropic (avoiding the normalization and anisotropy problems of \Alf/slow cascades), and their {\em inertial range} spectra can again, in principle, give $\delta \sim 0-0.5$ depending on (very uncertain) theoretical assumptions. But in this case, damping truncates the spectrum and produces a similar problem, at any scale below the dissipation scale\footnote{Formally defined as the {\em maximum} scale, marginalizing over direction of ${\bf k}$, where the dissipation time for those modes becomes comparable to the cascade time.} of the fast modes. That dissipation scale depends on the detailed physical conditions, but for the weakest-possible damping conditions of interest in the warm ISM, scales as $\ell_{\rm diss}/0.01\,{\rm pc} \sim {\rm MAX}[ \ell_{\rm drive,100}\,\beta_{\rm plasma} \ , \ 0.1\,\ell_{\rm drive,100}^{1/3} T_{4}^{4/3} \beta_{\rm plasma}^{1/3} n_{1}^{-2/3} ]$ (where $\ell_{\rm drive,100} \equiv \ell_{\rm drive}/100\,{\rm pc}$, and $\beta_{\rm plasma} = P_{\rm thermal}/P_{\rm B} \sim c_{s}^{2}/v_{A}^{2}$), and increasing to as large as $\gg$\,kpc in the outer CGM \citep[][and references therein]{cho.lazarian:2003.mhd.turb.sims,lazarian:2016.cr.wave.damping,hopkins:cr.transport.constraints.from.galaxies,hopkins:2021.sc.et.models.incompatible.obs,kempski:2021.reconciling.sc.et.models.obs,schekochihin:2020.mhd.turb.review}. This means that damping cannot be ignored for any CRs at $\lesssim$\,TeV-PeV energies.\footnote{It is worth noting that the predicted damping scale in the warm LISM at $\sim$\,TeV gyro-resonant wavelengths very neatly explains the gradual ``ankle-like'' feature between $\sim 300-1000\,$GV observed in local CR spectra. One actually predicts that for a damping scale gyro-resonant with $\sim$\,TeV CRs such that extrinsic turbulence dominates CR scattering at $\gtrsim$\,TeV, the curvature as fit to the usual double-power-law features in the empirical models will begin to appear at a factor $\sim 2-3$ smaller energy (since this is sensitive to very small changes in the dependence of $\bar{\nu}$ on $R_{\rm cr}$). That potential association has been widely discussed (though alternative explanations exist; e.g.\ \citealt{qiao:2025.cr.knees.from.local.source.contributions}), however we caution that sometimes it is stated (incorrectly) that this indicates the transition from extrinsic turbulence to self-confinement. A more robust statement is that this could indicate a transition from extrinsic turbulence associated with a cascade from the \Alf\ scale in the LISM (\textit{if} LISM fast modes indeed obey the correct scalings and behaviors) to ``something else'' being the dominant driver of CR scattering at energies $\ll$\,TeV, as damping/dissipation suppresses scattering from the extrinsic cascade. In that interpretation, the ``first knee'' or ``proton knee'' at $\sim$\,PeV would correspond to CRs gyro-resonant with the \Alf\ scale, while the ``second knee'' or ``iron knee'' (where extragalactic CRs become dominant) at $\sim 10-100$ times larger energy to the turbulent driving scale ($\sim$\,kpc).}

If damping cannot be ignored, then for the {\em weakest possible} damping (collisionless damping in a dust-free, fully-ionized plasma with $\beta_{\rm plasma} \ll 1$), the spectrum is steepened to $\mathcal{E}(k_{\|}) \propto k_{\|}^{-2}$ or steeper, i.e.\ one obtains $\delta \le 0$. This is calculated explicitly for these limits for e.g.\ the minimum Landau and Braginskii damping in the references above. And if there is any stronger damping -- i.e.\ if dust grains are present (if $f_{\rm dg} \gtrsim 10^{-5}$), or if the neutral fraction is not extremely small ($f_{\rm neutral} \gtrsim 0.001$), or if $\beta_{\rm plasma} \gtrsim 1$ -- then damping cuts off the spectrum more sharply giving even more-negative $\delta$.

There are other important potential problems for fast-mode scattering on small scales, which have been discussed in e.g.\ \citet{kempski:2020.cr.soundwave.instabilities.highbeta.plasmas.resemble.perseus.density.fluctuation.power.spectra,kempski:2021.reconciling.sc.et.models.obs,kempski:2023.large.amplitude.fluctuations.and.cr.scattering,hopkins:2021.sc.et.models.incompatible.obs,schekochihin:2020.mhd.turb.review}. For example, many authors have argued on both analytic \citep{1973SPhD...18..115K,1976ZPhyB..23...89E,1992PhLA..166..243S,2000JPlPh..63..447G,2011PhyA..390.1534S,2008PhPl...15f2305K,2011PhRvL.107m4501G,2016MPLB...3050297S} and numerical \citep{1974PhLA...47..419E,1990ThCFD...2...73E,2006MNRAS.370..415M,2010PhRvE..81a6318L,2010ApJ...720..742K,2020PhRvX..10c1021M} grounds that fast modes on small scales would steepen into weak shocks (so there is no true cascade on small scales), or otherwise converge to Burgers-like power spectrum $\mathcal{E} \propto k_{\|}^{-2}$ such that $\delta \le 0$ even above the damping scale (with $\delta \ll 0$ below the damping scale). And there may still be a several order-of-magnitude normalization discrepancy for scattering rates of $\sim$\,GeV CRs.

\subsubsection{Small or Large-Scale Modes Only}
\label{sec:no.resonant}

Briefly, we can also immediately rule out any volume-filling model (obeying Eq.~\ref{eqn:nu}) where the dominant scattering modes (in the ISM) come from wavelengths either much smaller or much larger than the gyro radii of interest.  If all the scattering power comes from modes {\em below} the micro-scale ($\ell_{0} \sim 1/k_{\|,\,0} \ll \ell_{\rm g}$), then from Eq.~\ref{eqn:nu}, $\bar{\nu} \sim v_{\rm cr} \ell_{\rm g}^{-2} \mathcal{E}(\ell_{0})/e_{\rm B} \propto R_{\rm cr}^{-2}$, i.e.\ $\delta = 2$, far stronger than observationally allowed. This strongly constraints processes on the plasma kinetic scale (much smaller than the minimum $\ell_{\rm g}$ of interest), like micro-mirror scattering \citep{lazarian:2021.cr.scattering.mirrors,xu:2022.cr.streaming.turb.ism.mirrors.classical.damping,reichherzer:2023.micromirror.cr.confinement.hot.galaxy.cluster.icm,barretomota:2024.mirror.scattering.ism.crs,zhang:2024.cr.mirror.diffusion.in.nonlinear.turbulence,ewart:2024.cr.confinement.micromirror.cgm.of.massive.clusters}.
If all the power comes from modes {\em above} micro-scale ($\ell_{0} \sim 1/k_{\|,\,0} \gg \ell_{\rm g}$), then $\bar{\nu} \sim v_{\rm cr} \ell_{0}^{-2}\mathcal{E}(\ell_{0})/e_{\rm B} \propto R_{\rm cr}^{0}$, i.e.\ $\delta = 0$, akin to the truncated/suppressed turbulent spectra per \S~\ref{sec:et.alfven}-\ref{sec:et.fast}, and also immediately ruled out.

\begin{figure*}
	\centering
	\includegraphics[width=0.98\textwidth]{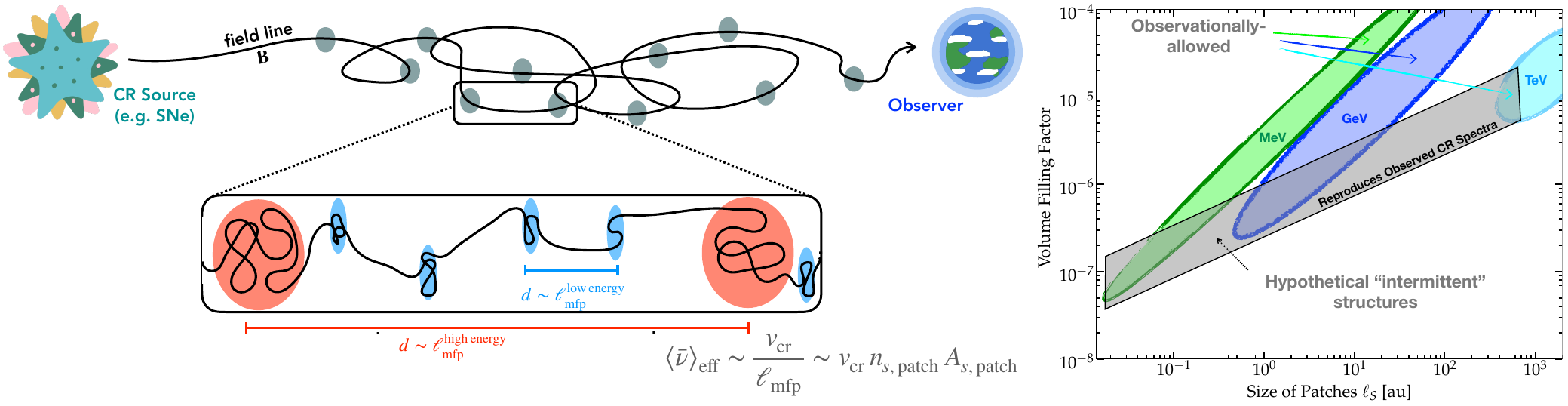}
	\caption{Illustration of intermittent/patchy CR scattering models. 
	{\em Left:} Instead of assuming CR scattering is volume-filling ($f_{\rm vol}\sim 1$) from low-amplitude fluctuations $|\delta B/B| \ll 1$ (which gives $\bar{\nu} \sim (v_{\rm cr}/\ell_{g})\,|\delta B(\ell_{g})/B|^{2}$), assume small ($f_{\rm vol} \ll 1$) structures/patches produce $\mathcal{O}(1)$ scattering (giving $\bar{\nu} \sim v_{\rm cr}\,n_{\rm patch} A_{\rm patch} \sim f_{\rm vol} v_{\rm cr}/\ell_{\rm s}$, for patches with some line crossing-size $\ell_{\rm s}$). 
	{\em Right:} Example of volume-filling factor $f_{\rm vol}$ of patches of a given $\ell_{\rm s}$ which would give the correct dependence of CR residence times on rigidity. These are strikingly similar to properties inferred for known radio scattering structures in the ISM, and could plausibly arise from many intermittent sources in turbulence.
	\label{fig:intermittency}}
\end{figure*} 

\subsection{Motivation for Alternative Drivers or Intermittent Scattering Models}
\label{sec:intermittency}

There are two basic categories of solutions to the fundamental challenges in \S~\ref{sec:uncertainty:rigidity}.

{\bf (1)} If scattering is indeed volume-filling and low-amplitude, then there must be some other driver of scattering modes (besides large-scale turbulence or CR streaming instabilities alone), which satisfies the following. (a) It drives parallel modes $\delta{\bf B}(k_{\|})$ explicitly on scales $k_{\|}^{-1} \sim 0.01-100\,$au (not cascading from much larger/smaller scales), with (b) a spectrum $\mathcal{E}(k_{\|}) \propto k_{\|}^{-\xi}$ where $1.2\lesssim \xi \lesssim 1.6$ and with appropriate normalization ($\delta B(k_{\|})/B \sim 0.0008\,(k_{\|}\,{\rm au}/B_{\mu G})^{1/4}$), and (c) the driving rate scales weakly or not at all with the CR flux. There are many proposed candidates which might satisfy this and operate on the salient scales, for example: charged dust resonant drag instabilities \citep{squire.hopkins:RDI}, plasmoid instabilities \citep{fielding:2022.plasmoid.instability.interface.ism.possible.cr.effects}, or Kelvin-Helmholtz and other mixing/boundary layer instabilities \citep{ji:2019.radiative.turbulent.mixing.layers,yang:2023.radiative.turbulent.mixing.layers}. But at present, it is not known whether any of these can satisfy the three criteria above under typical ISM conditions.

{\bf (2)} One can abandon the volume-filling assumption, and instead assume that CR scattering is dominated by intermittent or patchy structures that have a small volume-filling factor $f_{\rm vol} \ll 1$, as illustrated in Fig.~\ref{fig:intermittency} \citep{butsky:2023.cosmic.ray.scattering.patchy.ism.structures}.\footnote{Importantly, the volume filling factor distinguishes intermittent/patchy models of CR scattering from classical ``inhomogeneous'' scattering models which have been more widely-studied (see \S~\ref{sec:dyn.models}), in which the CR scattering is still volume-filling and low-amplitude but varies relatively smoothly on meso or larger scales.} The local microphysical scattering rate $\bar{\nu}$ can be large (even non-linear) within patches, but in this limit the {\em effective} scattering rate on meso-scales becomes 
\begin{align}
\nu_{\rm eff} \sim \frac{v_{\rm cr}}{\ell_{\rm mfp,\,\|}} \sim v_{\rm cr} n_{\rm patch} A_{\rm patch} \sim f_{\rm vol}\,\frac{v_{\rm cr}}{\ell_{\rm patch}}
\end{align}
where $n_{\rm patch}$ and $A_{\rm patch}$ represent the number density and effective cross section/area of patches with linear crossing size  $\ell_{\rm patch}$. While there are many classes of models that fit in this category \citep[reviewed in][]{butsky:2023.cosmic.ray.scattering.patchy.ism.structures}, a natural possibility is to assume a spectrum of structure sizes $\ell_{\rm patch}$ overlapping the $\sim 0.01-100\,$au range of interest, with high-energy CRs being weakly-scattered by those with $\ell_{\rm patch} \ll \ell_{\rm g}$. Then $\bar{\nu} \rightarrow (v_{\rm cr}/\ell_{g})\,f_{\rm vol}(\ell_{\rm patch} \sim \ell_{g})$, and the observed $\bar{\nu}$ as a function of rigidity is reproduced if the volume-filling factor scales something like what is shown in Fig.~\ref{fig:intermittency}.\footnote{In number density, $\delta=1/2$ becomes $n_{\rm patch} \propto \ell_{g}^{-(N_{\rm coll} - 1/2)}$ where $N_{\rm coll}$ is the number of collapsed dimensions of the scattering structures/patches ($N_{\rm coll}=1,\,2,\,3$ for sheets, filaments, or clumps, respectively), or (independent of the dimensionality) $f_{\rm vol} \propto \ell_{g}^{1/2}$.} Note $f_{\rm vol}$ can be very small: $\sim 10^{-7}-10^{-6}$ for au-scale structures, so these are indeed highly non-volume filling. While this model simply replaces our ignorance of $\mathcal{E}(k_{\|} \sim 1/\ell_{\rm g})$ with $f_{\rm vol}(\ell_{\rm patch} \sim \ell_{\rm g})$, there is no obvious theoretical challenge producing a spectrum of patch sizes in the correct range, and there are again many candidate scattering structures \citep[e.g.][]{dong:2018.plasmoid.instability.mhd.turbulence,schekochihin:2020.mhd.turb.review,lazarian:2021.cr.scattering.mirrors,dong:2022.reconnection.driven.intermittency.mhd.turb.sims,fielding:2022.plasmoid.instability.interface.ism.possible.cr.effects,lemoine:2023.cr.scattering.magnetic.fields.bends,kempski:2023.large.amplitude.fluctuations.and.cr.scattering,tharakkal:2023.cr.test.particle.sims.mirror.trapping,ntormousi:2024.strong.ism.intermittency.sims.from.rotation.and.sne.feedback.driving,beattie:2025.mhd.turb.boxes.spectra.intermittent.in.boldyrev.sense,kriel:2025.mhd.turb.sims.shock.steepening,hu:2025.cr.superdiffusion.mirror.diffusion.partially.ionized.plasmas.turbulence,reichherzer:2025.micromirror.confinement.crs.clusters}, including intermittent structures in ISM turbulence, plasma sheets, turbulent boundary/mixing layers, gyro-resonant magnetic mirrors/traps or non-linear plasmoid instabilities, weak shocks, regions with strong CR-dust coupling, or regions where self-confinement has locally ``run away'' per \S~\ref{sec:super.alfvenic} (as well as larger-scale structures which already can be ruled out, including stellar magnetospheres, planetary nebulae, HII regions, molecular clouds, spiral arms, and SNe remnants, on the basis of the arguments in \citealt{butsky:2023.cosmic.ray.scattering.patchy.ism.structures}). One particularly tantalizing possibility is that the spectrum of $f_{\rm vol}$ versus size in Fig.~\ref{fig:intermittency} is quite similar to what is required to explain radio scattering/scintillation long observed in the ISM \citep[e.g.][]{cordes:2001.ese.scattering.structures} -- a connection made quantitatively concrete in \citet{lemoine:2023.cr.scattering.magnetic.fields.bends,kempski.li.2024:unified.cr.scattering.plasma.scattering.from.strong.field.curvature.intermittent.ism.structure.explained.together}. But again, no specific physical model has been conclusively shown to produce the required $f_{\rm vol}(\ell_{\rm patch})$ (but see \citealt{wang:2024.obs.synchrotron.probes.mhd.turb.intermittency,ocker:2025.ns.bow.shocks.ism.cr.accel.and.intermittency,zhao:2025.obs.intermittent.shock.cr.scattering.fx}, for discussion of potential observables).

\section{Conclusions}
\label{sec:conclusions}

There has been tremendous progress in our understanding of CRs on galactic scales in the past decade. We now have well-posed, rigorously-derived CR-MHD methods for solving the non-equilibrium full-spectrum coupled CR-MHD equations on micro, meso, and macro scales, which do not require assumptions about local isotropy or scattering mean free paths. New observations have highlighted the crucial role of modeling LISM constraints in \textit{global} Galactic models, including the actual scale of the Milky Way and source distribution and $\gg$\,kpc extended scattering volumes (i.e.\ the inner CGM), in a globally 3D geometry. It has become clear that CRs can play a crucial role in the dynamics of the CGM, including its thermal phase structure, outflows, and inflows, and therefore could play a critical role in galaxy formation more broadly. Theories of CR scattering from the 1960's have finally been put to the test, and the simplest theoretical microphysical CR scattering models have been conclusively shown to be missing some key physical ingredients. 

We conclude with a highly biased list of the most important and wide-open questions which can be addressed in the next decade, from CR-MHD simulations on micro, meso, and macro scales.

\subsection{Micro-Scale}
\label{sec:conclusions:micro}

\begin{itemize}

\item What is the effective CR scattering rate given by highly \textit{nonlinear} structures in the ISM like folds, mirrors, or weak shocks? And how does this depend on CR rigidity or gyro radius relative to the size of those structures?

\item What are the volume filling factors ($f_{\rm vol}$) and statistics (e.g.\ effective dimensionality, $f_{\rm vol}$, cross-section $\ell^{2}$ and longitudinal size $\ell_{g}$) of intermittent structures in ISM turbulence which have been proposed for intermittent CR scattering? Are these compatible with the required statistics (of both $f_{\rm vol}$ and size spectrum) to reproduce CR scattering (as a function of rigidity) at $\sim$\,MeV-TeV energies?

\item What is the behavior on gyro-resonant scales of alternative ``small-scale extrinsic turbulence drivers'' in the ISM (e.g.\ turbulent mixing layers, boundary layers, resonant drag instabilities, plasmoid or tearing or other instabilities)? Can these produce a ``local'' turbulent cascade over the range of scales needed to explain CR scattering? And would they be volume-filling or act as more intermittent CR scattering sites?

\item What is the dependence of any of these mechanisms of ``meso-scale'' properties like the magnetic field strength $|{\bf B}|$, plasma $\beta$, CR-to-gas pressure ratio $\beta_{\rm cr}$, etc.? 

\end{itemize}

With progress on these fronts, these insights can be parameterized into scalings for effective $\bar{\nu}$ that can be utilized in meso-scale simulations.

\subsection{Meso-Scale}
\label{sec:conclusions:meso}

\begin{itemize}

\item What is the ``effective transport'' if there are large variations in the pitch-angle scattering rate on scales small compared to the scattering mean free path, as might be expected for intermittent scattering? While this must be representable on meso-scales with standard two-moment methods, can this be parameterized for macro-scale calculations with the usual two-moment equations with a single/constant ``effective scattering rate'' or does it require an additional term to represent some super-diffusive or hyper-diffusive behavior?

\item What are the observable signatures on meso scales of different CR scattering models, or inhomogeneity in the CR scattering rate on scales smaller than the CR scattering mean free path? Can they produce features like harps, $\gamma$-ray or synchrotron ``hot spots,'' beaming, cocoons, or other direct observables either in spatially-resolved indirect CR tracers or in CR spectra as observed in the LISM at the Solar location? 

\item What is the effective non-linear behavior of instabilities like the CR ``staircase'' instability, or ``solution collapse'' for super-\Alf{ic} streaming in self-confinement theories? What is the effective diffusivity that would emerge averaging over macro scales?

\item What does this tell us about the importance of Galactic ``weather'' for different CR anomalies and detailed behaviors, like the Galactic center $\gamma$-ray excess, or the LISM positron excess, or other related but small-amplitude features in CR diagnostics?

\item What is the dependence of any of these on ``macro-scale'' properties like the global (Galactic) CR energy density gradient, position in the CGM versus LISM versus galactic center, strength of turbulence and/or multi-phase ISM structure?

\end{itemize}

Progress on these fronts will inform scalings for effective $\bar{\nu}$ (and $\kappa_{\rm eff}$, $v_{\rm st,\,eff}$) in macro-scale simulations, but also enable direct observational tests on meso scales within the ISM and inner CGM of different microphysical CR scattering models.

\subsection{Macro-Scale}
\label{sec:conclusions:macro}

\begin{itemize}

\item What are the consequences for observations of different CR scattering models as parameterized in their ``effective'' dependence on macro-scale properties? How do they produce different galaxy-integrated $\gamma$-ray or synchrotron or soft-X-ray (inverse Compton) spectra/luminosities? How do they influence different, even-more-indirect observables like UV metal line absorption, or galaxy outflows, or baryon fractions in different halos? And which of these models can be conclusively ruled out (as compared to being degenerate with other modeling choices like the treatment of stellar/AGN feedback)?

\item What are the consequences for galaxy and star formation of these different models? Do some strongly suppress star formation in galaxies via enhanced outflows or suppressed inflows, or regulate/enhance ``bursty'' star formation in dwarf and/or high-redshift galaxies? Is CR feedback more ``preventive'' or ``expulsive,'' in different CR transport models? What does this imply for AGN feedback in the form of CRs and galaxy ``quenching''? \\

\item What new observations can be developed to distinguish between these models, especially in the CGM? Can future soft X-ray missions, for example, strongly constrain CR transport in the CGM of individual distant galaxies, and couple to new radio observatories to jointly model their synchrotron emission and simultaneously constrain the magnetic fields involved?

\end{itemize}

Progress on these fronts will inform our understanding of galaxy formation, star formation, supermassive black hole formation and growth, and cosmology as a whole, and may potentially solve many open problems in those fields, but will also let us understand how important the (currently poorly-understood) systematic uncertainties from e.g.\ our lack of understanding of CR transport microphysics really are. Moreover, they enable a host of new tests, as almost all observational constraints on CR transport in more extreme environments (starbursts, dwarf galaxies, clusters, AGN, quasars, high-redshift systems, etc.), which can probe extremes of parameter space, will be necessarily extragalactic and therefore depend on the macro-scale CR-MHD dynamics.

In conclusion, there is much to do, but it is an exciting time with rapid progress being made and the tools available now to address all of these questions.

\,
\\
\begin{acknowledgements}
We thank Jono Squire, Phillpp Kempski, Iryna Butsky, Eliot Quataert, Ellen Zweibel, Peng Oh, Yue Hu, Kung-Yi Su, Lucia Armillotta, Nadine Soliman, and Raphael Skalidis for a number of stimulating conversations.
Support for PFH was provided by a Simons Investigator grant. 
\end{acknowledgements}

\bibliographystyle{mn2e}
\bibliography{ms_extracted}

\begin{appendix}

\section{A Note on Ad-Hoc ``RHD-Like'' CR Transport Schemes in the Literature}
\label{sec:chang.jo.note}

As noted in the text, before more rigorous derivations, there were some ``ad-hoc'' multi-moment CR transport schemes proposed, most notably those in \citet{jiang.oh:2018.cr.transport.m1.scheme} (JO) and \citet{chan:2018.cosmicray.fire.gammaray} (C19). Importantly, both papers stated clearly that they were not attempting to derive the correct physical equations for CR transport, but merely proposing an ad-hoc CR flux equation (motivated by M1 radiation hydrodynamics [RHD]) to replace the common single-moment spectrally-integrated CR diffusion+advection equation (Eq.~\ref{eqn:onemoment}) with an alternative that would resolve certain numerical problems (while producing the desired one-moment behavior in special limits). These schemes have been widely propagated, so it is worth briefly discussing how they are inaccurate.

JO and C19 considered the spectrally-integrated CR equations for a universal spectral shape, non-relativistic fluid background and ultrarelativistic CRs, on ``macro'' scales. From Eq.~\ref{eqn:twomoment}, the correct transport equations are then:
\begin{align}
\nonumber D_{t} e_{\rm cr} +& \nabla \cdot ( F_{\rm cr} \bhat ) = S_{\rm net} - \mathbb{P}: \nabla {\bf u}  - \frac{\hat{\nu} }{c^{2}}\left[  \bar{v}_{A} F_{\rm cr} - 3\hat{\chi} v_{A}^{2}(e_{\rm cr}+P_{\rm cr}) \right] \ , \\ 
\label{eqn:correct} D_{t} F_{\rm cr} +& c^{2}\bhat\cdot \nabla \cdot \mathbb{P} = - \hat{\nu} \left[  F_{\rm cr} - 3\hat{\chi} \bar{v}_{A}(e_{\rm cr} +P_{\rm cr} ) \right] \ .
\end{align}
Instead, for the energy equation JO+C19 both adopted: 
\begin{align}
\label{eqn:JOC19.egy} D^{\bf JOC19}_{t} e_{\rm cr} & +  \nabla \cdot \left( F_{\rm cr} \bhat \right) \sim  S_{\rm net} - P_{\rm cr} \nabla \cdot {\bf u}  - v_{A} \left | \bhat \cdot \nabla P_{\rm cr} \right | 
\end{align}
while for the flux equation\footnote{Note that we have cast the JO+C19 formulations into consistent notation, which requires some tedious algebra. Both write their expressions in terms of derivatives of a vector flux (defined in different frames) which includes parallel and perpendicular components. The latter are either suppressed by powers of $\ell_{g} /\ell$ for physical scattering rates, or numerically exponentially damped on the scattering time and therefore were ignored, but care is needed treating derivatives of the flux directions that arise.} they assumed: 
\begin{align}
\nonumber D^{\bf JO}_{t} F_{\rm cr}  +  & \alpha^{2} c^{2} \bhat \cdot \nabla P_{\rm cr} \sim  \nabla \cdot \left(F_{\rm cr} {\bf u} \right) - \bhat \cdot \partial_{t}\left({\bf u}\,(e_{\rm cr} + P_{\rm cr}) \right) \\
\label{eqn:JO.flux} & -\alpha^{2} \hat{\nu} \left[ 1 + \frac{v_{A} \hat{\nu} }{c^{2}}\,\frac{(e_{\rm cr} + P_{\rm cr})}{|\bhat \cdot \nabla P_{\rm cr} |} \right]^{-1} F_{\rm cr} \ , \\ 
\nonumber  D^{\bf C19}_{t} F_{\rm cr} + &  \alpha^{2} c^{2} \bhat \cdot \nabla P_{\rm cr} \sim  - F_{\rm cr} \bhat \cdot \nabla \cdot \left({\bf u} \bhat \right)\\
\label{eqn:C19.flux} & -\alpha^{2} \hat{\nu} \left[ 1 + \frac{v_{A} \hat{\nu} }{c^{2}}\,\frac{(e_{\rm cr} + P_{\rm cr})}{|\bhat \cdot \nabla P_{\rm cr} |} \right]^{-1} F_{\rm cr} \ ,
\end{align}
where $\alpha \equiv \tilde{c}/c$ represented a numerical speed of light reduction factor (obviously the correct equations require $\alpha=1$).

There are three obvious differences from the correct equations. 
First, JO+C19 assumed an isotropic CR closure, so $\mathbb{P}\rightarrow P_{\rm cr} \mathbb{I}$ and $\mathbb{P}:\nabla {\bf u} \rightarrow P_{\rm cr} \nabla \cdot {\bf u}$, $3\hat{\chi}\rightarrow 1$, $\nabla \cdot \mathbb{P} \rightarrow \nabla P_{\rm cr}$. This means their equations can only be correct on macro, not meso scales (scales large in space/time relative to CR scattering, per \S~\ref{sec:transport.macro}) and neither correctly recovers CR free-streaming on any scale (in fact they can produce unphysical behaviors like negative CR energies in this limit; see \citealt{hopkins:m1.cr.closure,thomas:2021.compare.cr.closures.from.prev.papers}). 
Second, JO+C19 contain spurious source terms. In the flux equations this owes to the choice of how $\bhat$ was inserted as an ad-hoc ``projection operator'' from the RHD equations (rather than actually deriving the CR equations to leading order in $\ell_{g} / \ell$); in the energy equation to the fact that only the ``steady-state streaming loss'' term was imposed.\footnote{A modified version of JO in \citet{armillotta:2021.cr.streaming.vs.environment.multiphase} and subsequent work replaces $-v_{A} |\bhat \cdot \nabla P_{\rm cr}|$ in Eq.~\ref{eqn:JOC19.egy} specifically with 
\begin{align}
+\nonumber \frac{{\bf u} \cdot \partial_{t} \left[ F_{\rm cr}\bhat + {\bf u}\,(e_{\rm cr} + P_{\rm cr}) \right] }{\alpha^{2} c^{2}}
+ \frac{\hat{\nu}^{2}}{c^{2}} \left[ 1 + \frac{v_{A} \hat{\nu} }{c^{2}}\,\frac{(e_{\rm cr} + P_{\rm cr})}{|\bhat \cdot \nabla P_{\rm cr} |} \right]^{-1}  \frac{v_{A} \bhat \cdot \nabla P_{\rm cr}}{|\bhat \cdot \nabla P_{\rm cr}|} F_{\rm cr} \ , 
\end{align}
but these are also not the correct source terms.}
Third,  they both (incorrectly) attempt to capture streaming (which does not appear in RHD) by adding a multiplicative correction term to the scattering rate, rather than the (correct) flux difference $F_{\rm cr} - 3\hat{\chi}\bar{v}_{A}\,(e_{\rm cr} + P_{\rm cr})$ that should appear (reflecting the comoving \Alf\ frame which defines the CR anisotropy).
Note that the incorrect flux equations produce correspondingly incorrect CR forces on gas.\footnote{The correct force density, in the spectrally-integrated limit from Eq.~\ref{eqn:cr.gas.force} is ${\bf f}_{\rm cr} = -\nabla \cdot \mathbb{P} + \bhat \left[\bhat\cdot\nabla\cdot\mathbb{P} +  \hat{\nu}c^{-2} \left\{ F_{\rm cr} - 3\hat{\chi} \bar{v}_{A}\,(e_{\rm cr} + P_{\rm cr} ) \right\} \right] = -\nabla \cdot \mathbb{P} - \bhat\, c^{-2} D_{t} F_{\rm cr}$. C19 instead assumed ${\bf f}^{\bf C19}_{\rm cr} \sim -\nabla P_{\rm cr}$, while JO assumed ${\bf f}^{\bf JO}_{\rm cr} \sim - \nabla P_{\rm cr} - \bhat\,(\alpha c)^{-2} \partial_{t}[ F_{\rm cr} \bhat + {\bf u}(e_{\rm cr} + P_{\rm cr}) ]$.}

Formally, the equations in JO or C19 are \textit{only} correct if all the following limits are taken: 
$\alpha\,c\rightarrow \infty$ (Newtonian) \textit{and} $\hat{\nu}\rightarrow \infty$ (vanishing scattering mean-free path) \textit{and} $\hat{\chi}\rightarrow 1/3$ (isotropy) \textit{and} $D_{t} F_{\rm cr} \rightarrow 0$ (local flux equilibrium) \textit{and} $|\bar{v}_{A}| \rightarrow v_{A}$ (maximal streaming anisotropy) \textit{and} require $\bhat \cdot \nabla P_{\rm cr} \ne 0$. Then we trivially obtain
$F_{\rm cr} \rightarrow v_{\rm st}\,(e_{\rm cr} + P_{\rm cr}) - \kappa_{\|} \nabla e_{\rm cr}$ with $v_{\rm st} \rightarrow -v_{A} \bhat \cdot \nabla P_{\rm cr} /|\bhat \cdot \nabla P_{\rm cr} |$ and $\kappa_{\|} \rightarrow c^{2}/ 3\hat{\nu}$, and the spurious terms become vanishingly small, but we have gone back to original limits of the one-moment equations.
If $\alpha\,c$ is not much larger than all other terms, the spurious source terms do not order out, and the divisor term in $\hat{\nu}$ has a different behavior from the scattering term in the correct equations. If $\bar{\nu}$ is not extremely large, the flux equation (1) cannot come into steady state and (2) should represent free-streaming, for which the isotropy assumption is incorrect and thus the closure terms assumed are incorrect. If $D_{t} F_{\rm cr}$ is large then there is no guarantee that the spurious source terms order out, the divisor term again modifies the behavior, and there are incorrect source term in the CR energy equations. If $|\bar{v}_{A}| \ne v_{A}$ then the asymptotic streaming speed and streaming losses are incorrect. If $\bhat \cdot \nabla P_{\rm cr} =0$, then the flux equation locally reduces to $D_{t} F_{\rm cr}$ equal to only the spurious source terms. 

These constraints can be serious. For example, define $\cos{\theta_{bP}} \equiv (\bhat \cdot \nabla P_{\rm cr}) / (P_{\rm cr} / \ell_{\nabla})$ where $\ell_{\nabla} \sim \ell_{\nabla,\,{\rm kpc}}\,{\rm kpc}$ is some characteristic CR pressure gradient length. Then comparing the ordering of terms, supposing that ${\bf u}$ and/or $v_{A}$ can reach up to some characteristic maximum $v_{\rm sig,\,max} \sim v_{\rm max,\,300}\,300\,{\rm km\,s^{-1}}$, and that the CR flux must be allowed in generality to reach up to order $\mathcal{O}(\alpha\,c\,e_{\rm cr})$, while gradients in local fluid properties like ${\bf u}$, ${\bf B}$, $v_{A}$ can (and regularly do) have scale lengths reaching $\Delta x \sim \Delta x_{\rm pc} {\rm pc}$ in the ISM, then the requirement that the spurious source terms order out of the transport equations is equivalent to $ | \cos{\theta_{bP}} | \gg 3\,\ell_{\nabla,\,{\rm kpc}}\,v_{\rm max,\,300}/(\alpha \Delta x_{\rm pc})$. Even with $\alpha=1$ (no reduced speed of light), if there is ISM structure on scales $\ll 100\,$pc, then this cannot be satisfied and the solutions will be incorrect over a wide range of conditions. But even if all the conditions above are met almost everywhere in the domain and $c, \hat{\nu} \rightarrow \infty$ so streaming is a perfect approximation, a major motivation for two-moment approaches (explicitly discussed in both JO+C19) is that the one-moment streaming equation becomes singular where $\bhat \cdot \nabla P$ changes sign (i.e.\ if the magnetic fields rotate through perpendicular $\bhat \cdot \nabla P = 0$ to a larger-scale CR gradient), and even if that occurs in just a single cell in the domain, propagating CRs \textit{through} that point with an incorrect flux equation can non-linearly change the solution in the domain beyond that point (see e.g.\ \citealt{sharma.2010:cosmic.ray.streaming.timestepping,thomas.pfrommer.18:alfven.reg.cr.transport,tsung:2020.sims.cosmicray.modified.shocks,hanasz:2021.cr.propagation.sims.review,gupta:2021.cosmic.ray.numerical.nonuniqueness.with.streaming.only}, some of whom note the analogy between this and multi-fluid shocks without well-defined jump conditions). If such a situation can occur, it is especially important to use the correct transport equations.

Finally, we note for completeness that simply taking JO/C19 and ``binning'' the energy/flux equations to represent a CR energy spectrum (each evolving an independent version of $e_{\rm cr}$ and $F_{\rm cr}$) obviously does not eliminate any of these errors, and it introduces several additional leading-order (large) errors in both flux ($\bar{f}_{1}$ or $f_{\mu}$) and energy ($\bar{f}_{0}$) equations (immediately clear from comparison with \S~\ref{sec:transport.meso} or classic texts).

\,
\\

\end{appendix}

\end{document}